\documentclass[usenames,onecolumn,dvipsnames]{aastex63}

\usepackage{graphicx}	% Including figure files
\usepackage{amsmath}	% Advanced maths commands
\usepackage{amssymb}	% Extra maths symbols
\usepackage{relsize}
\usepackage{mathtools}
\usepackage{bigints}
\usepackage{sidecap}
\usepackage{caption,subcaption}

\DeclareMathOperator{\sech}{sech}

\DeclareMathOperator{\erf}{erf}
\DeclareMathOperator{\sgn}{sgn}
\usepackage{floatrow}
\usepackage{ulem}
\usepackage{soul}
\usepackage{comment}
\usepackage{bm}

\usepackage{macros_ub}
\graphicspath{{./}{figures/}}

\received{XXX}
\revised{YYY}
\accepted{ZZZ}

\begin{document}
\vspace{1mm}

%--------------------------------------------------------

\shorttitle{Phase-space spirals}
\shortauthors{}

\title{A Comprehensive Perturbative Formalism for Phase Mixing in Perturbed Disks. 
II. Phase Spirals in an Inhomogeneous Disk Galaxy with a Non-responsive Dark Matter Halo}

\correspondingauthor{Uddipan Banik}
\email{uddipan.banik@yale.edu}

\author[0000-0002-9059-381X]{Uddipan Banik}
\affiliation{Department of Astronomy, Yale University, PO. Box 208101, New Haven, CT 06520, USA}

\author[0000-0003-3236-2068]{Frank~C.~van den Bosch}
\affiliation{Department of Astronomy, Yale University, PO. Box 208101, New Haven, CT 06520, USA}

\author[0000-0003-2660-2889]{Martin~D.~Weinberg}
\affiliation{Department of Astronomy, University of Massachusetts at Amherst, 710 N. Pleasant St., Amherst, MA 01003}

%\label{firstpage}

\begin{abstract}

We develop a linear perturbative formalism to compute the response of an inhomogeneous stellar disk embedded in a non-responsive dark matter (DM) halo to various perturbations like bars, spiral arms and encounters with satellite galaxies. Without self-gravity to reinforce it, the response of a Fourier mode phase mixes away due to an intrinsic spread in the vertical ($\Omega_z$), radial ($\Omega_r$) and azimuthal ($\Omega_\phi$) frequencies, giving rise to local phase-space spirals. Collisional diffusion due to scattering of stars by structures like giant molecular clouds causes super-exponential damping of the phase-spiral amplitude. The $z-v_z$ phase-spiral is one-armed (two-armed) for vertically anti-symmetric (symmetric) bending (breathing) modes. Only transient perturbations with timescales ($\tau_\rmP$) comparable to the vertical oscillation period ($\tau_z \sim 1/\Omega_z$) can trigger vertical phase-spirals. Each $(n,l,m)$ mode of the response to impulsive ($\tau_\rmP<\tau=1/(n\Omega_z+l\Omega_r+m\Omega_\phi)$) perturbations is power law ($\sim \tau_\rmP/\tau$) suppressed, but that to adiabatic ($\tau_\rmP>\tau$) perturbations, is exponentially weak ($\sim \exp{\left[-\left(\tau_\rmP/\tau\right)^\alpha\right]}$) except resonant ($\tau\to \infty$) modes. Slower ($\tau_\rmP>\tau_z$) perturbations, e.g., distant encounters with satellite galaxies, induce stronger bending modes. Sagittarius dominates the Solar neighborhood response of the Milky Way (MW) disk to satellite encounters. Thus, if the Gaia phase-spiral was triggered by a MW satellite, Sagittarius is the leading contender. However, the survival of the phase-spiral against collisional damping necessitates an impact $\sim 0.6-0.7\Gyr$ ago. We discuss the impact of the detailed galactic potential on the shape of phase-spirals: phase mixing occurs slower and thus phase-spirals are more loosely wound in the outer disk and in presence of an ambient DM halo.
\end{abstract}


\section{Introduction}
\label{sec:intro}

Disk galaxies are characterized by large-scale ordered motion and are therefore highly responsive to perturbations. Following a time-dependent gravitational perturbation, the actions of the disk stars are modified. This in turn causes a perturbation in the distribution function (DF) of the disk known as the response. Over time the response decays away as the system `relaxes' towards a new quasi-equilibrium via collisionless processes that include kinematic processes like phase mixing (loss of coherence in the response due to different oscillation frequencies of stars) and secular/self-gravitating/collective processes like Landau damping \citep[loss of coherence due to wave-particle interactions,][]{LyndenBell.62}. As pointed out by \cite{Sridhar.89} and \cite{Maoz.91}, phase mixing is the key ingredient of all collisionless relaxation and re-equilibration. 

The timescale of collisionless equilibration is typically longer than the orbital periods of stars. Therefore disk galaxies usually harbour prolonged features of incomplete equilibration following a perturbation, e.g., bars, spiral arms, warps and other asymmetries. An intriguing example is the one-armed phase-space spiral, or phase-spiral for short, discovered in the Gaia DR2 data \citep[][]{Gaia_collab.18a} by \cite{Antoja.etal.18} and discussed in more detail in subsequent studies \citep[e.g.,][]{Bland-Hawthorn.etal.19, Laporte.etal.19, Li.Widrow.21, Li.21, Gandhi.etal.22}. \cite{Antoja.etal.18} plotted the density of stars in the Solar neighborhood in the $(z,v_z)$-plane of vertical position, $z$, and vertical velocity, $v_z$, and noticed a faint spiral pattern which became more pronounced when colour-coding the $(z,v_z)$-`pixels' by the median radial or azimuthal velocities. The one-armed spiral shows 2-3 complete wraps like a snail shell, and is interpreted as an indication of vertical phase mixing following a perturbation that is anti-symmetric about the midplane (bending mode) and occurred $\sim 500\Myr$ ago. More recently, \cite{Hunt.etal.22} used the more extensive Gaia DR3 data to study the distributions of stars in $z-v_z$ space at different locations in the MW disk. They found that unlike the one-armed phase-spiral or bending mode at the Solar radius, the inner disk shows a two-armed phase-spiral that corresponds to a breathing mode or symmetric perturbation about the midplane. They inferred that while the one-armed spiral in the Solar neighborhood might have been caused by the impact of a satellite galaxy such as the Sagittarius dwarf, the two-armed spiral in the inner disk could not have been induced by the same since almost all satellite impacts are far too slow/adiabatic from the perspective of the inner disk. Rather, they suggested that the two-armed phase-spiral might haven been triggered by a transient spiral arm or bar.

The phase-spiral holds information about the perturbative history and gravitational potential of the disk and can therefore serve as an essential tool for galacto-seismology \citep[][]{Widrow.etal.14, Johnston.etal.17}. For a given potential, the winding of the spiral is an indication of the time elapsed since the perturbation occurred with older spirals revealing more wraps. A one-armed (two-armed) phase-spiral corresponds to a bending (breathing) mode. Which mode dominates, in turn, depends on the time-scale of the perturbation, with temporally shorter (longer) perturbations (e.g., a fast or slow encounter with a satellite) predominantly triggering breathing (bending) modes \citep[][]{Widrow.etal.14,Banik.etal.22}.

In addition to depending on the nature of the perturbation, the phase-spiral also encodes information about the oscillation frequencies of stars and thus the detailed potential. In particular, the shape of the spiral depends on how the vertical frequencies, $\Omega_z$, vary as a function of the vertical action, $I_z$, which in turn depends on the underlying potential. In \citep[][hereafter Paper~I]{Banik.etal.22} we showed that the amplitude of the phase-spiral can damp away due to lateral mixing, with a damping rate that depends on both the spatio-temporal nature of the perturbation and the frequency structure of the galaxy. This damping, though, only affects the response in the coarse-grained sense, i.e., upon marginalization of the response over the lateral degrees of freedom (the action-angle variables). Damping at the fine-grained level requires collisional diffusion, such as that arising from the gravitational scattering of stars against giant molecular clouds (GMCs), or dark matter (DM) substructure \citep[][]{Tremaine.etal.22}.

Paper~I addresses the problem of inferring the nature of the perturbation from the amplitude and structure of the phase-spiral using a model of an infinite, isothermal slab for the unperturbed disk. This simple, yet insightful, model provides us with essential physical understanding of the perturbative response of disks without the complexity of modelling a realistic, inhomogeneous disk. However it suffers from certain glaring caveats: (i) lateral uniformity leading to an incorrect global structure of the response in the lateral direction, (ii) Maxwellian distribution of velocities in the lateral direction that overpredicts lateral mixing and thereby the rate at which the amplitude of the phase-spiral damps out, (iii) absence of a DM halo and (iv) absence of self-gravity of the response. In this paper we relax the first three assumptions. We consider an inhomogeneous disk characterized by a realistic DF similar to the pseudo-isothermal DF \citep[][]{Binney.10}, that properly captures the orbital dynamics of the disk stars in 3D. In addition, we consider the effect of an underlying DM halo which for the sake of simplicity we consider to be non-responsive. This ambient DM halo alters the potential and thus the frequencies of stars, which can in turn affect the shape of the phase-spiral and its coarse-grained survival. We also consider the impact of small-scale collisionality on the fine-grained survival of the phase-spiral. Since in this paper we are primarily interested in the phase mixing of the disk response that gives rise to phase-spirals, we ignore the self-gravity of the response, which to linear order spawns coherent point mode oscillations of the disk as a whole \citep[for treatments of the self-gravitating response of isothermal slabs, see][]{Mathur.90,Weinberg.91} and somewhat enhances the amplitude of phase-spirals. 

This paper is organized as follows. Section~\ref{sec:linear_theory} describes the standard linear perturbation theory for collisionless systems and its application to a realistic disk galaxy embedded in a DM halo that is exposed to a general perturbation. Sections~\ref{sec:disk_resp_spiral} and \ref{sec:disk_resp_sat} are concerned with computing the disk response for different perturber models. In Section~\ref{sec:disk_resp_spiral} we compute the disk response and phase-spirals induced by bars and spiral arms. We also discuss the impact of collisional diffusion on the fine-grained survivability of the phase-spiral. In Section~\ref{sec:disk_resp_sat} we compute the response to encounters with satellite galaxies. Section~\ref{sec:pot_const} describes how phase-spirals can be used to constrain the galactic potential. We summarize our findings in Section~\ref{sec:concl}.

%\newpage

\section{Linear perturbation theory for galaxies}
\label{sec:linear_theory}

\subsection{Linear perturbative formalism}

A galaxy, to very good approximation, is devoid of star-star collisions. However, there are other potential sources of collisions such as scatterings due to gravitational interactions of stars with giant molecular clouds (GMCs) or DM substructure. The dynamics of stars in such a system is governed by the Boltzmann equation:
\begin{align}
\frac{\partial f}{\partial t}+[f,H]=C[f],
\label{CBE_master}
\end{align}
where $f$ denotes the DF, $H$ denotes the Hamiltonian, square brackets denote the Poisson bracket, and $C[f]$ denotes the collision operator due to small-scale fluctuations, which can be approximated by a Fokker-Planck operator \citep[see Appendix~A of][]{Tremaine.etal.22}:
\begin{align}
C[f] = \frac{1}{2} \frac{\partial}{\partial \xi_i} \left(D_{ij} \frac{\partial f}{\partial \xi_j} \right),
\end{align}
where $\boldsymbol{\xi}=\left(\bq,\bp\right)$ with $\bq$ and $\bp$ denoting the canonically conjugate position and momentum variables, and $D_{ij}$ denotes the diffusion coefficient tensor.

Let the unperturbed steady state Hamiltonian of the galaxy be $H_0$ and the corresponding DF be given by $f_0$, which satisfies the unperturbed Fokker-Planck equation (FPE),
\begin{align}
[f_0,H_0]=C[f_0].
\end{align}
In presence of a small time-dependent perturbation in the potential, $\Phi_\rmP(t)$, the perturbed Hamiltonian can be written as
\begin{align}
H=H_0+\Phi_\rmP(t)+\Phi_1(t),
\end{align}
where $\Phi_1$ is the gravitational potential related to the response density, $\rho_1 = \int f_1 \rmd^3\bv$,
via the Poisson equation,
\begin{align}
\nabla^2\Phi_1=4\pi G\rho_1.
\end{align}
The perturbed DF can be written as
\begin{align}
f=f_0+f_1,
\end{align}
where $f_1$ is the linear order perturbation in the DF. In the weak perturbation limit where linear perturbation theory holds, the time-evolution of $f_1$ is dictated by the following linearized form of the FPE:
\begin{align}
\frac{\partial f_1}{\partial t}+[f_1,H_0]+[f_0,\Phi_\rmP]+[f_0,\Phi_1]=C[f_1].
\label{CBE_perturb}
\end{align}
Throughout this paper we neglect the self-gravity of the disk response, which implies that we set the polarization term, $[f_0,\Phi_1]=0$. The implications of including self-gravity are discussed in Paper~I.

\subsection{Response of a Galactic Disk to a realistic perturbation}
\label{sec:galdisk}

The dynamics of a realistic disk galaxy like the Milky Way (MW) is quasi-periodic, i.e., can be characterized by oscillations in the azimuthal, radial and vertical directions. In close proximity to the mid-plane and under radial epicyclic approximation, the Hamilton-Jacobi equation becomes separable, implying that all stars confined within a few vertical scale heights from the mid-plane of the disk are on regular, quasi-periodic orbits that are characterized by a radial action, $I_R$, an azimuthal action $I_\phi$, and a vertical action $I_z$. Hence, the motion of each star is characterized by three frequencies:
\begin{align}
\Omega_R = \frac{\partial H_0}{\partial I_R}\,,\;\;\; 
\Omega_\phi = \frac{\partial H_0}{\partial I_\phi}\,,\;\;\; 
\Omega_z = \frac{\partial H_0}{\partial I_z}\,.
\label{freqs}
\end{align}
This quasi-periodic nature of the orbits near the mid-plane is approximately preserved even in the presence of a (non-triaxial) DM halo since this preserves the axi-symmetry of the potential. Typically, as discussed in section~\ref{sec:pot_const}, the presence of a halo increases the oscillation frequencies of the disk stars.

In terms of these canonical conjugate action-angle variables, using equation~(\ref{freqs}), the linearized form of the FPE given in Equation~(\ref{CBE_perturb}) becomes
\begin{align}
\frac{\partial f_1}{\partial t}+\Omega_z\frac{\partial f_1}{\partial w_z}+\Omega_R\frac{\partial f_1}{\partial w_R}+\Omega_\phi\frac{\partial f_1}{\partial w_\phi}-\frac{\partial \Phi_\rmP}{\partial w_z}\frac{\partial f_0}{\partial I_z}-\frac{\partial \Phi_\rmP}{\partial w_R}\frac{\partial f_0}{\partial I_R}-\frac{\partial \Phi_\rmP}{\partial w_\phi}\frac{\partial f_0}{\partial I_\phi} &= D_z\frac{\partial}{\partial I_z}\left(I_z\frac{\partial f_1}{\partial I_z}\right) \nonumber \\
&+ \frac{D_z}{4 I_z}\frac{\partial^2 f_1}{\partial w^2_z}+\frac{D_R}{4 I_R}\frac{\partial^2 f_1}{\partial w^2_R}
\label{CBE_perturb_gen}
\end{align}
Here we have performed certain simplifications of the Fokker-Planck operator. Firstly, following \cite{Binney.Lacey.88}, we have assumed that $D_{zz}=D_z I_z$ and $D_{RR}=D_R I_R$, since this preserves the pseudo-isothermal form of the unperturbed DF (equation~[\ref{DF_MW}]) of the disk. Secondly, the $I_R$ diffusion of the response $f_1$ is negligible. This is because the frequencies do not depend on $I_R$ under the radial epicyclic approximation (and only mildly depend on $I_R$ without it) and therefore the response does not develop $I_R$ gradients. Thirdly, following \cite{Binney.Lacey.88}, we have neglected diffusion in $I_\phi$ and $w_\phi$ since the terms involving $D_{\phi\phi}$, $D_{r\phi}$ and $D_{\phi z}$ are smaller than the $I_z$ and $I_R$ diffusion terms by factors of at least $\sigma_R/v_c$ or $\sigma_z/v_c$, which are typically much smaller than unity ($\sigma_R$ and $\sigma_z$ are radial and vertical velocity dispersions respectively, and $v_c$ is the circular velocity along $\phi$). We have retained the $w_z$ and $w_R$ diffusion terms for the sake of completeness, but as we point out later, the diffusion in angles typically occurs over much longer timescales than that in actions and hence is comparatively less important.

Since the stars move along quasi-periodic orbits characterized by actions and angles, we can expand the perturbations, $\Phi_\rmP$ and $f_1$, as discrete Fourier series in the angles as follows
\begin{align}
\Phi_\rmP\left(\bw,\bI,t\right)&=\sum_{n=-\infty}^{\infty} \sum_{l=-\infty}^{\infty} \sum_{m=-\infty}^{\infty} \exp{\left[i (n w_z + l w_R + m w_\phi)\right]}\, \Phi_{nlm}\left(\bI,t\right),\nonumber \\
f_1\left(\bw,\bI,t\right)&=\sum_{n=-\infty}^{\infty} \sum_{l=-\infty}^{\infty} \sum_{m=-\infty}^{\infty} \exp{\left[i (n w_z + l w_R + m w_\phi)\right]}\, f_{1nlm}(\bI,t),
\label{fourier_series_gen}
\end{align}
where $\bw=(w_z,w_R,w_\phi)$ and $\bI=(I_z,I_R,I_\phi)$. Substituting these Fourier expansions in equation~(\ref{CBE_perturb_gen}) yields the following differential equation for the evolution 
of $f_{1nlm}$:
\begin{align}
\frac{\partial f_{1nlm}}{\partial t}+i(n\Omega_z+l\Omega_R+m\Omega_\phi)f_{1nlm}&=i\left(n\frac{\partial f_0}{\partial I_z}+l\frac{\partial f_0}{\partial I_R} + m\frac{\partial f_0}{\partial I_\phi}\right)\Phi_{nlm} \nonumber \\
&+ D_{z}\frac{\partial}{\partial I_z}\left(I_z\frac{\partial f_{1nlm}}{\partial I_z}\right) - \left[\frac{n^2 D_z}{4 I_z} + \frac{l^2 D_R}{4 I_R}\right] f_{1nlm}.
\label{f1nk_de}
\end{align}
This can be solved using the Green's function technique, with the initial condition, $f_{1nlm}(t_\rmi)=0$, to yield the following closed integral form for $f_{1nlm}$:
\begin{align}
f_{1nlm}(\bI,t)&=i\left(n\frac{\partial f_0}{\partial I_z}+l\frac{\partial f_0}{\partial I_R}+m\frac{\partial f_0}{\partial I_\phi}\right) \calI_{nlm}(\bI,t).
\label{f1nk_gensol}
\end{align}
Here, for $D_z \ll \sigma^2_z$ ($\sigma_z$ is the vertical velocity dispersion), which is typically the case, $\calI_{nlm}(\bI,t)$ can be approximately expressed as

\begin{align}
\calI_{nlm}(\bI,t)&\approx \int_{t_\rmi}^{t}\rmd \tau\, \calG_{nlm}(\bI,t-\tau)\, \Phi_{nlm}(\bI,\tau).
\end{align}
Here $\calG_{nlm}(t-\tau)$ is the Green's function (see Appendix~A of \cite{Tremaine.etal.22} for derivation), given by
\begin{align}
\mathcal{G}_{nlm}(\bI,t-\tau) &\approx \exp{\left[-i(n\Omega_z+l\Omega_R+m\Omega_\phi)(t-\tau)\right]} \nonumber \\
&\times \exp{\left[-\left(\frac{n^2 D_{z}}{4 I_z}+\frac{l^2 D_R}{4 I_R}\right) \left(t-\tau\right)\right]}\, \exp{\left[-\frac{{\left(n\Omega_{z1}\right)}^2 D_{z} I_z}{3}{\left(t-\tau\right)}^3\right]},
\end{align}
where $\Omega_{z1}=\partial \Omega_z/ \partial I_z$. The sinusoidal factor represents the oscillations of stars at their natural frequencies which vary with actions, leading to the formation of phase-spirals (see section~\ref{sec:spiral_cless} for details). The first exponential damping factor indicates the damping of the response due to diffusion in angles while the second damping factor manifests the damping of the $I_z$ gradients of the response by diffusion in $I_z$. As discussed in section~\ref{sec:spiral_c}, the diffusion in actions is much more efficient than that in angles.

Each $(n,l,m)$ Fourier coefficient of the response acts as a forced damped oscillator with three different natural frequencies, $n\Omega_z$, $l\Omega_R$ and $m\Omega_\phi$, which is being driven by an external time-dependent perturber potential, $\Phi_{nlm}$, and damped due to collisional diffusion. A similar expression, albeit without allowing for collisionality, for the DF perturbation has been derived by \cite{Carlberg.Sellwood.85} in the context of spiral arm induced perturbations and radial migrations in the galactic disk, and by other previous studies \citep[e.g.,][]{LyndenBell.Kalnajs.72, Tremaine.Weinberg.84, Carlberg.Sellwood.85, Weinberg.89, Weinberg.91, Weinberg.04, Kaur.Sridhar.18, Banik.vdBosch.21a, Kaur.Stone.22} in the context of dynamical friction in spherical systems. To obtain the final expression for $f_{1nlm}$, we need to specify the DF $f_0$ of the unperturbed galaxy, as well as the spatio-temporal behavior of the perturber potential, $\Phi_\rmP$, which is addressed below.

\subsection{The unperturbed galaxy}
\label{sec:disk_model}

Under the radial epicyclic approximation (small $I_R$), the unperturbed DF, $f_0$, for a rotating MW-like disk galaxy can be well approximated as a pseudo-isothermal DF, i.e., written as a nearly isothermal separable function of the azimuthal, radial and vertical actions. Following \cite{Binney.10}, we write
\begin{align}
f_0 \approx \frac{\sqrt{2}}{\pi^{3/2} \, \sigma_z h_z} {\left(\frac{\Omega_\phi \Sigma}{\kappa\, \sigma^2_R}\right)}_{\Rc} \, \exp{\left[-\frac{\kappa I_R}{\sigma^2_R}\right]} \, \exp{\left[-\frac{E_z(I_z)}{\sigma^2_z}\right]} \, \Theta(L_z)\,,
\label{DF_MW}
\end{align}
The vertical structure of this disk is isothermal, while the radial profile is pseudo-isothermal. Here $\Sigma = \Sigma(R) = \int_{-\infty}^\infty dz\, \rho(R,z)$ is the surface density of the disk, $L_z$ is the $z$-component of the angular momentum, which is equal to $I_\phi$, $\Rc = \Rc(L_z)$ is the guiding radius, $\Omega_\phi$ is the circular frequency, and $\kappa = \kappa(\Rc) = \lim_{I_R \to 0}{\Omega_R}$ is the radial epicyclic frequency \citep[][]{Binney.Tremaine.87}. $\Theta(x)$ is the Heaviside step function. Thus we assume that the entire galaxy is composed of prograde stars with $L_z>0$. 

The density profile, $\rho(R,z)$, of the disk corresponding to the above DF is the product of a radially exponential profile with scale radius $h_r$ and a vertically isothermal ($\sech^2$) profile with scale height $h_z$ (equation~[\ref{disk_exp_iso_rho_app}]).
As shown by \cite{Smith.etal.15}, this density profile is accurately approximated by a sum of three \cite{Miyamoto.Nagai.75} disks\footnote{the 3MN profile as implemented in the {\tt Gala Python package} \citep[][]{gala, gala_code_adrian}.}, which has a simple, analytical form for the associated potential. Throughout, we therefore use this 3MN approximation for our disk since this drastically simplifies the computation of orbital frequencies. The disk is assumed to be embedded in an extended DM halo characterized by a spherical NFW \citep[][]{Navarro.etal.97} density profile, with virial mass $\Mvir$, concentration $c$, scale radius $r_s$ and the corresponding potential $\Phi_\rmh$ given by equation~(\ref{Phi0_halo_app}). Throughout, for the purpose of computing the disk response, we assume typical MW like parameters for the various quantities, i.e., $\Rsun=8$ kpc, disk mass $M_\rmd=5\times 10^{10}\Msun$, $h_R=2.2 \kpc$, $\sigma_R(\Rsun) = \sigma_{R,\odot} = 35\, {\rm km}/\rms$, $h_z=0.4 \kpc$ and $\sigma_z(\Rsun) = \sigma_{z,\odot} = \sqrt{2\pi G h_z \Sigma(\Rsun)} = 23\, {\rm km}/\rms$ \citep[][]{McMillan.11, Bovy.Rix.13}. For the NFW DM halo, we adopt $\Mvir=9.78\times 10^{11}\Msun$, $r_s=16$ kpc, and $c=15.3$ \citep[][]{Bovy.15}.

The combined potential experienced by the disk stars is simply the sum of disk and halo potentials, i.e.,
\begin{align}
\Phi_0(R,z)=\Phi_\rmd(R,z)+\Phi_\rmh(R,z).
\end{align}
The total energy of a disk star under the radial epicyclic approximation is $E= L^2_z/2 R^2_c + \Phi_0(\Rc,0) + \kappa I_R + E_z$, where the vertical part of the energy is given by $E_z=v^2_z/2+\Phi_z(\Rc,z)$, with $\Rc(L_z)$ the guiding radius given by $L^2_z/\Rc^3=\partial \Phi_0/\partial R|_{R=\Rc}$. The vertical potential, $\Phi_z(\Rc,z)$, is given by
\begin{align}
\Phi_z(\Rc,z)=\Phi_0(\Rc,z)-\Phi_0(\Rc,0).
\end{align}
The vertical action, $I_z$, can be obtained from $E_z$ as follows
\begin{align}
I_z = \frac{1}{2\pi} \oint v_z \, d z= \frac{2}{\pi} \int_0^{z_{\rm max}} \sqrt{2[E_z - \Phi_z(\Rc,z)]} \, d z,
\end{align}
where $\Phi_z(\Rc,z_{\rm max})=E_z$. This implicit equation can be inverted to obtain $E_z(\Rc,I_z)$. The time period of vertical oscillation can then be obtained using
\begin{align}
T_z(\Rc,I_z) = \oint \frac{d z}{v_z} = 4\int_0^{z_{\rm max}} \frac{d z} {\sqrt{2\left[E_z(\Rc,I_z)-\Phi_z(\Rc,z)\right]}},
\label{T_z}
\end{align}
which yields the vertical frequency, $\Omega_z(\Rc,I_z) = 2\pi/T_z(\Rc,I_z)$.

Substituting the expression for $f_0$ given by Equation~(\ref{DF_MW}) in Equation~(\ref{f1nk_gensol}), we obtain the following integral form for $f_{1nlm}$,
\begin{align}
f_{1nlm}(\bI,t) & \approx -\frac{2i}{\pi\sigma^2_R}\, \frac{1}{\sqrt{2\pi}h_z\sigma_z} \exp{\left[-\frac{\kappa I_R}{\sigma^2_R}\right]} \exp{\left[-\frac{E_z(I_z)}{\sigma^2_z}\right]} \nonumber \\ \nonumber \\ 
&\times \left[\left\{\left(\frac{n\Omega_z}{\sigma^2_z} + \frac{l\kappa}{\sigma^2_R}\right) {\left(\frac{\Omega_\phi \Sigma}{\kappa}\right)} - m\frac{\rmd}{\rmd L_z}\left(\frac{\Omega_\phi \Sigma}{\kappa}\right)\right\} \Theta(L_z) - m\frac{\Omega_\phi \Sigma}{\kappa} \delta(L_z)\right] \calI_{nlm}(\bI,t).
\label{f1nk_gensol_f0}
\end{align}
As we shall see, the first order disk response expressed above phase mixes away and gives rise to phase-spirals due to oscillations of stars with different frequencies except when they are resonant with the frequency of the perturber. However this `direct' response of the disk does not include certain effects. First of all, we ignore the self-gravity of the response. As discussed in Paper~I, to linear order self-gravity gives rise to point mode oscillations of the disk that are decoupled from the phase mixing component of the response which is what we are interested in.  Secondly, for the sake of simplicity, we consider the ambient DM halo to be non-responsive and therefore ignore the indirect effect of the halo response on disk oscillations. We leave the inclusion of these two effects in the computation of the disk response for future work.

The spatio-temporal nature of the perturbing potential dictates the disk response. In this paper we explore two different types of perturbation to which realistic disc galaxies can be exposed, and which are thus of general astrophysical interest. The first is an in-plane spiral/bar perturbation with a vertical structure, either formed as a consequence of secular evolution, or triggered by an external perturbation. We consider both short-lived (transient) and persistent spirals. The second type of perturbation that we consider is that due to an encounter with a massive object, e.g., a satellite galaxy or DM subhalo.

\section{Disk response to spiral arms and bars}
\label{sec:disk_resp_spiral}

We model the potential of a spiral arm perturbation as one with a vertical profile and a sinusoidal variation along radial and azimuthal directions,
\begin{align}
\Phi_\rmP(R,\phi,z) &= -\frac{2\pi G \Sigma_\rmP}{k_R}\, \left[\alpha\,\calM_\rmo(t)\,\calF_\rmo(z) + \calM_\rme(t)\,\calF_\rme(z)\right] \sum_{m_\phi=0,2} \sin{\left[k_R R + m_\phi \left(\phi-\Omega_\rmP t\right)\right]}\,.
\label{Phip_spiral}
\end{align}
Here $\Omega_\rmP$ is the pattern speed and $k_R$ is the horizontal wave number of the spiral perturbation. The long wavelength limit, $k_R\to 0$, corresponds to a bar. We consider the in-plane part of $\Phi_\rmP$ to be a combination of an axisymmetric ($m_\phi=0$) and a 2-armed spiral mode ($m_\phi=2$), and the vertical part to be a combination of anti-symmetric/odd and symmetric/even perturbations respectively denoted by $\calF_\rmo$ and $\calF_\rme$, that are modulated by growth functions, $\calM_\rmo(t)$ and $\calM_\rme(t)$, capturing the growth and/or decay of the spiral strength over time. The ratio of the maximum strengths of the anti-symmetric and symmetric parts of the perturbation is $\alpha$. We consider the following two functional forms for $\calM_j(t)$ (where the subscript $j=\rmo$ or $\rme$):
\begin{align}
\calM_j(t) &= 
\begin{cases}
\frac{1}{\sqrt{\pi}}\exp{\left[-\omega^2_j t^2\right]}, & \text{Transient spiral/bar} \\
\exp{\left[\gamma_j t\right]} + \left(1-\exp{\left[\gamma_j t\right]}\right)\,\Theta(t), & \text{Persistent spiral/bar.}
\end{cases}
\label{modulation}
\end{align}
The first option describes a transient spiral/bar that grows and decays like a Gaussian pulse with a characteristic life-time $\tau_{\rmP j} \sim 1/\omega_j$ \citep[][]{Banik.etal.22}. The second form describes a persistent spiral perturbation that grows exponentially on a timescale $\tau_{\rmG j} \sim 1/\gamma_j$ and then saturates to a constant amplitude. We shall see shortly that these two kinds of spiral perturbations perturb the disk in very different ways.

The vertical part of the perturbation consists of an anti-symmetric function, $\calF_\rmo(z)$, and a symmetric function, $\calF_\rme(z)$, which, for the sake of simplicity, we take to be the following trigonometric functions: 
\begin{align}
\calF_\rmo(z) &= \sin{\left(k_z^{(\rmo)} z\right)}, \nonumber \\
\calF_\rme(z) &= \cos{\left(k_z^{(\rme)} z\right)}.
\end{align}
Here $k^{(\rmo)}_z$ and $k^{(\rme)}_z$ denote the vertical wave-numbers of the anti-symmetric and symmetric perturbations, respectively. Since the above functions form a complete Fourier basis in $z$, any (vertical) perturber profile can be expressed as a linear superposition of $\calF_\rmo$ and $\calF_\rme$. The disk response involves the Fourier coefficients of the perturbing potential, $\Phi_{nlm}$, which can be obtained by taking the Fourier transform of $\Phi_\rmP$ given in Equation~(\ref{Phip_spiral}) with respect to the angles, $w_R$, $w_\phi$ and $w_z$, as detailed in Appendix~\ref{App:fourier_spiral}.

\begin{figure*}
\centering
\begin{subfigure}{0.49\textwidth}
  \centering
  \includegraphics[width=1\textwidth]{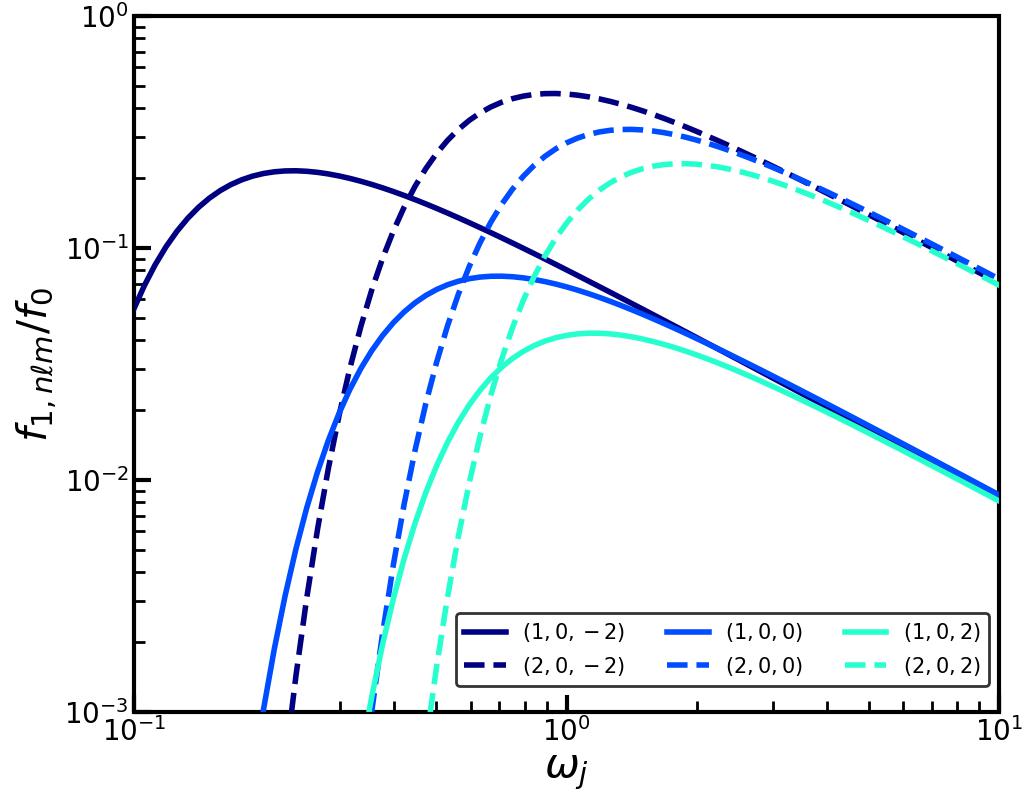}
  \label{fig:f1_vs_omega_gauss}
\end{subfigure}
\begin{subfigure}{0.49\textwidth}
  \centering
  \includegraphics[width=1\textwidth]{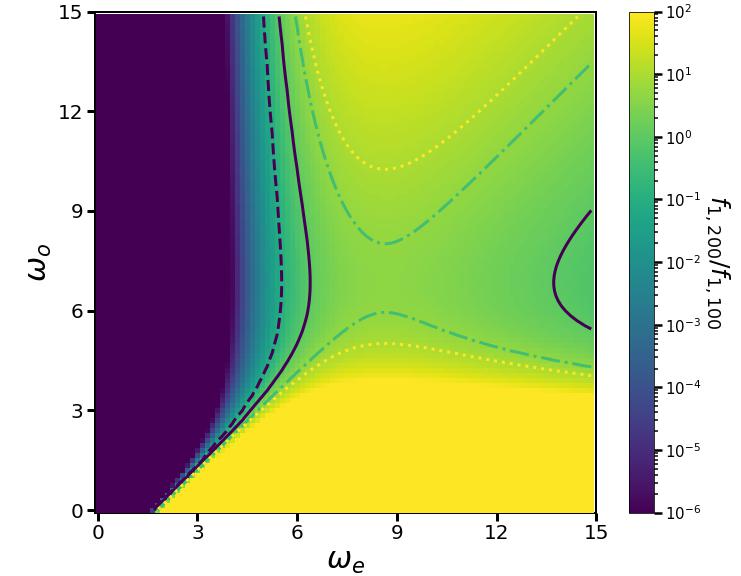}
  \label{fig:BB_gauss}
\end{subfigure}
\caption{MW disk response to transient bars/2-armed spirals with Gaussian temporal modulation in absence of collisional diffusion: Left panel shows the steady state ($t\to \infty$) amplitude of the disk response, $f_{1,nlm}/f_0$, in the Solar neighborhood, computed using equations~(\ref{trans_spiral_Resp}) and (\ref{Pnlm_trans_spiral}) in presence of an ambient DM halo, as a function of the pulse frequency, $\omega_j$, where the subscript $j=\rmo$ and $\rme$ for vertically anti-symmetric (odd $n$) and symmetric (even $n$) perturbations. Solid (dashed) lines indicate the $n=1$ bending ($n=2$ breathing) modes and different colors denote $(l,m)=(0,-2)$, $(0,0)$ and $(0,2)$ respectively. We consider $I_z = I_{z,\odot} \equiv h_z\sigma_{z,\odot}$ and marginalize the response over $I_R$. Note that the response peaks at intermediate values of $\omega_j$, which is different for different modes, and is suppressed like a power law in the impulsive (large $\omega_j$) limit and super-exponentially in the adiabatic (small $\omega_j$) limit. Right panel shows the breathing-to-bending ratio, $f_{1,200}/f_{1,100}$, as a function of $\omega_\rme$ and $\omega_\rmo$, the pulse frequencies of the bending and breathing mode perturbations respectively. The dashed, solid, dot-dashed and dotted contours correspond to breathing-to-bending ratios of $0.1,1,5$ and $10$ respectively. The breathing-to-bending ratio rises and falls with increasing $\omega_\rme$ at fixed $\omega_\rmo$, while the reverse occurs with increasing $\omega_\rmo$ at fixed $\omega_\rme$, leading to a saddle point at $(\omega_\rme,\omega_\rmo)\approx (9,7)$.}
\label{fig:trans_spiral}
\end{figure*}

\subsection{Computing the disk response}

The expression for the disk response to bars or spiral arms can be obtained by substituting the Fourier coefficient of the perturber potential given in Equation~(\ref{spiral_fourier_app}) in Equation~(\ref{f1nk_gensol_f0}) and performing the $\tau$ integration with the initial time, $t_\rmi\to -\infty$. This yields the modal response, $f_{1nlm}$ (Equation~[\ref{f1nk_gensol_f0}]), with $\calI_{nlm}(\bI,t)$ given by
\begin{align}
\calI_{nlm}(\bI,t) &= \alpha\,\Psi^{(\rmo)}_{nlm}(\bI)\, \calP^{(\rmo)}_{nlm}(\bI,t) + \Psi^{(\rme)}_{nlm}(\bI)\, \calP^{(\rme)}_{nlm}(\bI,t),
\label{trans_spiral_Resp}
\end{align}
where $\Psi^{(\rmo)}_{nlm}$ and $\Psi^{(\rme)}_{nlm}$ respectively denote the time-independent parts of the odd and even terms in the expression for $\Phi_{nlm}$, and
\begin{align}
\calP^{(j)}_{nlm}(\bI,t) &= \exp{\left[-i\,m\Omega_\rmP\,t\right]} \nonumber \\
&\times \int_0^{\infty} \rmd \tau \exp{\left[-i\,\Omega_{\rm res}\,\tau\right]}\, \exp{\left[-\left(\frac{n^2 D_{z}}{4 I_z}+\frac{l^2 D_R}{4 I_R}\right)\tau\right]}\, \exp{\left[-\frac{{\left(n\Omega_{z1}\right)}^2 D_z I_z}{3}{\tau}^3\right]} \, \calM_j(t-\tau),
\label{Pnlm_spiral}
\end{align}
which characterizes the temporal evolution of the response. Here the subscript $j=\rmo$ or $\rme$, and the resonance frequency, $\Omega_{\rm res}$, is given by
\begin{align}
\Omega_{\rm res} = n\Omega_z + l\kappa + m(\Omega_\phi-\Omega_\rmP).
\end{align}

\subsubsection{Collisionless limit}\label{sec:spiral_cless}

First we examine the response in the limit of zero diffusion, i.e., $D_z=0$, where each star acts as a forced oscillator. 

\paragraph{\ul{Transient spirals and bars}}

First we consider the case of transient spiral arm or bar perturbations that grow and decay in strength over time, i.e., the temporal modulation $\calM_j(t)$ is given by the first of equations~(\ref{modulation}). In this case,
\begin{align}
\calP_{nlm}^{(j)}(\bI,t) &= 
\frac{1}{2\,\omega_j}\,\exp{\left[-\frac{\Omega^2_{\rm res}}{4\omega^2_j}\right]} \left[1+\erf{\left(\omega_j t - i\frac{\Omega_{\rm res}}{2\omega_j}\right)}\right] \exp{\left[-i(n\Omega_z+l\kappa+m\Omega_\phi)t\right]}\nonumber \\
&\xrightarrow{t\to\infty} \frac{1}{\omega_j}\,\exp{\left[-\frac{\Omega^2_{\rm res}}{4\omega^2_j}\right]}\, \exp{\left[-i(n\Omega_z+l\kappa+m\Omega_\phi)t\right]}.
\label{Pnlm_trans_spiral}
\end{align}
The error function describes the growth and transient oscillations of the response amplitude; over time the transients die away, and in the limit $t\to\infty$ the response saturates to a constant amplitude (in the absence of collisional diffusion).

The left-hand panel of Fig.~\ref{fig:trans_spiral} plots the amplitude of the steady state disk response to transient spiral/bar perturbations, relative to the unperturbed DF, as a function of the modulation/pulse frequency, $\omega_j$ ($j=\rmo$ and $\rme$ for bending and breathing modes respectively), for different modes indicated in different colors. Solid and dashed lines correspond to the $n=1$ bending modes and the $n=2$ breathing modes, respectively. We adopt $\Sigma_\rmP=5.5\Msun \pc^{-2}$, $\Omega_\rmP=12 \kms\kpc^{-1}$, $k^{(o)}_z=k^{(e)}_z=1\kpc$, $k_R=10\kpc$, and $I_z = I_{z,\odot} \equiv h_z\sigma_{z,\odot}=9.2\kms$, and marginalize the response over $I_R$. We set $\alpha=1$, implying equal maximum strengths for the bending and breathing modes. As evident from this figure, and also from equation~(\ref{Pnlm_trans_spiral}), the long-term strength of the disk response (after the initial transients have died out like $e^{-\omega^2_j t^2}$) scales as $\sim 1/\omega_j$ in the impulsive (large $\omega_j$) limit, but is super-exponentially suppressed ($\sim \exp\left[-\Omega^2_{\rm res}/4 \omega^2_j\right]$) in the adiabatic (small $\omega_j$) limit away from resonances, i.e., for $\Omega_{\rm res}\neq 0$. The adiabatic suppression scales differently with $\omega_j$ for other functional forms of $\calM_j(t)$, e.g., for $\calM_j(t)=1/\sqrt{1+\omega^2_j t^2}$ the response strength is exponentially suppressed $(\sim \exp[-\Omega_{\rm res}/\omega_j])$. The response of resonant modes ($\Omega_{\rm res}=0$) however does not undergo adiabatic suppression and scales as $\sim 1/\omega_j$ throughout, becoming non-linear in the adiabatic regime. Since there are many resonance modes, the cumulative response in the adiabatic limit of {\it all} modes combined is only suppressed as a power-law, rather than an exponential, in $\omega_j$ \citep[][]{Weinberg.94a,Weinberg.94b}.

The sinusoidal factor, $\exp{\left[-i(n\Omega_z+l\kappa+m\Omega_\phi)t\right]}$, in $\calP_{nlm}^{(j)}$ describes the oscillations of stars with three different frequencies, $\Omega_z$, $\kappa$ and $\Omega_\phi$, along the vertical, radial and azimuthal directions, respectively. Due to the dependence of these frequencies on the actions, that of $\Omega_z$ on $I_z$ and of $\kappa$ and $\Omega_\phi$ on $I_\phi=L_z$, the response integrated over actions eventually phase mixes away. This manifests as phase-spirals in the $I_z\cos{w_z}-I_z\sin{w_z}$ and $I_\phi\cos{\phi}-I_\phi\sin{\phi}$ phase-spaces, which are proxies for the $z-v_z$ and $\phi-\dot{\phi}$ phase-spaces, respectively. As is evident from equation~(\ref{Pnlm_spiral}), $\calP_{nlm}^{(j)} \sim \exp{\left[-i m \Omega_\rmP t\right]}$ in the adiabatic limit ($\omega_j\to 0$); hence, in this limit the  sinusoidal factor, $\exp{\left[-i(n\Omega_z+l\kappa+m\Omega_\phi)t\right]}$ is absent from the response, which implies that phase-spirals only occur for sufficiently impulsive perturbations. As shown in Paper~I, $n=1$ bending modes involve a dipolar perturbation in the vertical phase-space ($I_z\cos{w_z}-I_z\sin{w_z}$) distribution immediately after the perturbing pulse reaches its maximum strength. This dipolar distortion is subsequently wound up into a one-armed phase-spiral since $\Omega_z$ is a function of $I_z$. Breathing modes, on the other hand, involve an initial quadrupolar perturbation in the phase-space distribution which is subsequently wrapped up into a two-armed phase-spiral. Since $\Omega_z$, $\Omega_\phi$ and $\Omega_R$ all depend on $L_z$, the amplitude of the $I_z\cos{w_z}-I_z\sin{w_z}$ phase-spiral damps out over time due to mixing between stars with different $L_z$. The modal response, $f_{1nlm}$, when marginalized in a narrow bin of size $\Delta L_z$ around $L_z$, damps out as follows:
\begin{align}
\left<f_{1nlm}\right>(\bI,t) &= \frac{1}{\Delta L_z} \int_{L_z-\Delta L_z/2}^{L_z+\Delta L_z/2} \rmd L_z\, f_{1nlm} (\bI,t) \nonumber \\ \nonumber \\
&\approx \dfrac{\;\;\;\sin{\left[\left(\dfrac{\partial}{\partial L_z}\left(n\Omega_z+l\kappa+m\Omega_\phi\right)\right)\dfrac{\Delta L_z}{2} t\right]}\;\;\;} {\;\;\;\;\;\left(\dfrac{\partial}{\partial L_z}\left(n\Omega_z+l\kappa+m\Omega_\phi\right)\right)\dfrac{\Delta L_z}{2} t}\, f_{1nlm}(\bI,t).
\label{lateral_mixing_disk}
\end{align}
Since the frequencies vary with $L_z$, marginalizing over $L_z$ mixes phase-spirals that differ slightly in phases, giving way to a $\sim 1/t$ damping accompanied by a beat-like modulation with a characteristic lateral mixing timescale,

\begin{align}
\tau_\rmD^{(\rm{LM})} &= \dfrac{1} {\left(\dfrac{\partial}{\partial L_z}\left(n\Omega_z+l\kappa+m\Omega_\phi\right)\right)\dfrac{\Delta L_z}{2}}.
\end{align}
This explains why the density-contrast of the Gaia phase-spiral is enhanced upon color-coding by $v_\phi$ or, equivalently, $L_z$ \citep[][]{Antoja.etal.18, Bland-Hawthorn.etal.19}. Radial phase mixing is also present, but is typically much weaker because none of the frequencies depend on $I_R$ under the radial epicyclic approximation and only mildly depend on $I_R$ without it. Hence, due to ordered motion, the phase-spiral amplitude in a realistic disk galaxy damps out at a much slower rate, as $\sim 1/t$ (in absence of collisional diffusion), than the lateral mixing damping in the isothermal slab case considered in Paper~I, which arises from the unconstrained lateral velocities of the stars and exhibits a Gaussian temporal behavior.

It is worth emphasizing that not all frequencies undergo phase mixing. In fact the resonant frequencies, for which
\begin{align}
\Omega_{\rm res}=n\Omega_z+l\kappa+m(\Omega_\phi-\Omega_\rmP)=0,
\end{align}
do not phase mix away. Hence, parts of the phase-space closer to a resonance take longer to phase-mix away. Moreover, as manifest from the adiabatic suppression factor, $\exp[-\Omega^2_{\rm res}/4\omega^2_j]$, the near-resonant modes with $\Omega_{\rm res} \ll 2\omega_j$ have much larger amplitude than those with $\Omega_{\rm res} \gg 2\omega_j$ that are far from resonance. Therefore the long-term disk response consists of stars in (near) resonance with the perturbing bar or spiral arm. Most of the strong resonances are confined to the disk-plane, including the co-rotation resonance $(n,l,m)=(0,0,m)$, the Lindblad resonances $(0,\pm 1, \pm 2)$, the ultraharmonic resonances $(0,\pm 1,\pm 4)$, and so on. For thin disks with $h_z \ll h_R$ , the vertical degrees of freedom are generally not in resonance with the radial or azimuthal ones since $\Omega_z$ is much larger than $\Omega_\phi$ or $\kappa$. Hence the vertical oscillation modes ($n\neq 0$) such as the $n=1$ bending or $n=2$ breathing modes undergo phase mixing and give rise to phase-spirals. However, if the disk has significant thickness, then the vertical degrees of freedom can be in resonance with the horizontal ones, e.g., banana orbits ($\Omega_z=2\Omega_r$) in barred disks.

The excitability of the bending and breathing modes is dictated by the perturbation timescale, or more precisely by the ratio of the pulse frequency, $\omega_j$, and the resonant frequency, $\Omega_{\rm res}$. The right panel of Fig.~\ref{fig:trans_spiral} shows the breathing-to-bending ratio, $f_{1,200}/f_{1,100}$, as a function of $\omega_\rme$ and $\omega_\rmo$, with blue (yellow) shades indicating low (high) values. In general, the breathing-to-bending ratio rises steeply and falls gradually with $\omega_\rme$ at fixed $\omega_\rmo$ while the trend is reversed as a function of $\omega_\rmo$ at fixed $\omega_\rme$, resulting in a saddle point at $(\omega_\rme,\omega_\rmo)\approx (9,7)$. This owes to the super-exponential suppression in the adiabatic ($\omega_j \ll \Omega_{\rm res}$) limit and the power-law suppression in the impulsive ($\omega_j \gg \Omega_{\rm res}$) limit. Along the $\omega_\rmo=\omega_\rme$ line, the bending (breathing) modes dominate in the adiabatic (impulsive) limit, as evident from the left panel of Fig.~\ref{fig:trans_spiral}. All this suggests that bending modes dominate over breathing modes when (i) the anti-symmetric perturbation is more impulsive, i.e., evolves faster than the symmetric one, or (ii) both symmetric and anti-symmetric perturbations occur over comparable timescales but slower than the stellar vertical oscillation period.

\paragraph{\ul{Persistent spirals and bars}}

Next we consider perturbations caused by a persistent spiral arm or bar that grows exponentially until it saturates at a constant strength. The corresponding temporal modulation $\calM_j(t)$ is given by the second of equations~(\ref{modulation}). In this case, as shown by equation (19) of \cite{Banik.vdBosch.21a},
\begin{align}
\calP_{nlm}^{(j)}(\bI,t) = \frac{\exp{\left[\gamma_j t\right]} \exp{\left[-i m \Omega_\rmP t\right]}}{\gamma_j+i\Omega_{\rm res}}\left[1-\Theta(t)\right]+i\left[\frac{\gamma_j \exp{\left[-i(n\Omega_z+l\kappa+m\Omega_\phi)t\right]}}{\Omega_{\rm res}(\gamma_j+i\Omega_{\rm res})}-\frac{\exp{\left[-i m \Omega_\rmP t\right]}}{\Omega_{\rm res}}\right]\Theta(t).
\end{align}
Up to $t=0$ when the perturber amplitude stops growing, the response from all modes oscillates with the pattern speed $\Omega_\rmP$ and grows hand in hand with the perturber. Subsequently, as the perturbation attains a steady strength, the disk response undergoes temporary phase mixing due to the oscillations of stars at different frequencies, giving rise to phase-spirals. These transients, however, are quickly taken over by long term oscillations driven at the forcing frequency $\Omega_\rmP$. 

For a slowly growing spiral/bar, i.e., in the `adiabatic growth' limit ($\gamma \to 0$), the entire disk oscillates at the driving frequency, $\Omega_\rmP$, i.e.,
\begin{align}
\calP_{nlm}^{(j)}(\bI,t) &\xrightarrow{\gamma_j\to 0} \exp{\left[-im\Omega_\rmP t\right]} \left[\pi\delta(\Omega_{\rm res})-\frac{i}{\Omega_{\rm res}}\right].
\end{align}
This has two major implications. First of all, since all stars, both resonant and non-resonant, are driven at the pattern speed of the perturbing spiral/bar, transient phase mixing does not occur and thus no phase-spiral arises. Secondly, the response is dominated by the resonances, $\Omega_{\rm res}=0$. In fact the resonant response diverges, reflecting the failure of (standard) linear perturbation theory near resonances. The adiabatic invariance of actions is partially broken near these resonances, causing the stars to get trapped in librating near-resonant orbits. A proper treatment of the near-resonant response can be performed by working with `slow' and `fast' action-angle variables \citep[][]{Tremaine.Weinberg.84, Lichtenberg.Lieberman.92, Chiba.Schonrich.22, Banik.vdBosch.22, Hamilton.etal.22}, which are uniquely defined for each resonance as linear combinations of the original action-angle variables. The fast actions remain nearly invariant while the fast angles oscillate with periods comparable to the unperturbed orbital periods of stars. The slow action-angle variables, on the other hand, undergo large amplitude oscillations about their resonance values over a libration timescale that is typically much longer than the orbital periods. For example, at co-rotation resonance ($n=l=0$), angular momentum behaves as the slow action while the radial and vertical actions behave as the fast ones.

Based on the above discussion, we infer that phase-spirals can only be excited in the galactic disk by transient spiral/bar perturbations whose amplitude changes over a timescale comparable to the vertical oscillation periods of stars. Persistent spirals or bars rotating with a fixed pattern speed cannot give rise to phase-spirals. Rather they trigger stellar oscillations at the pattern speed itself, which manifests in phase-space as a steadily rotating dipole or quadrupole depending on whether the $n=1$ or $2$ mode dominates the response. Thus, a phase-spiral is necessarily always triggered by a transient perturbation.

\subsubsection{Impact of collisions on the disk response}\label{sec:spiral_c}

\begin{figure*}
  \centering
  \includegraphics[width=0.9\textwidth]{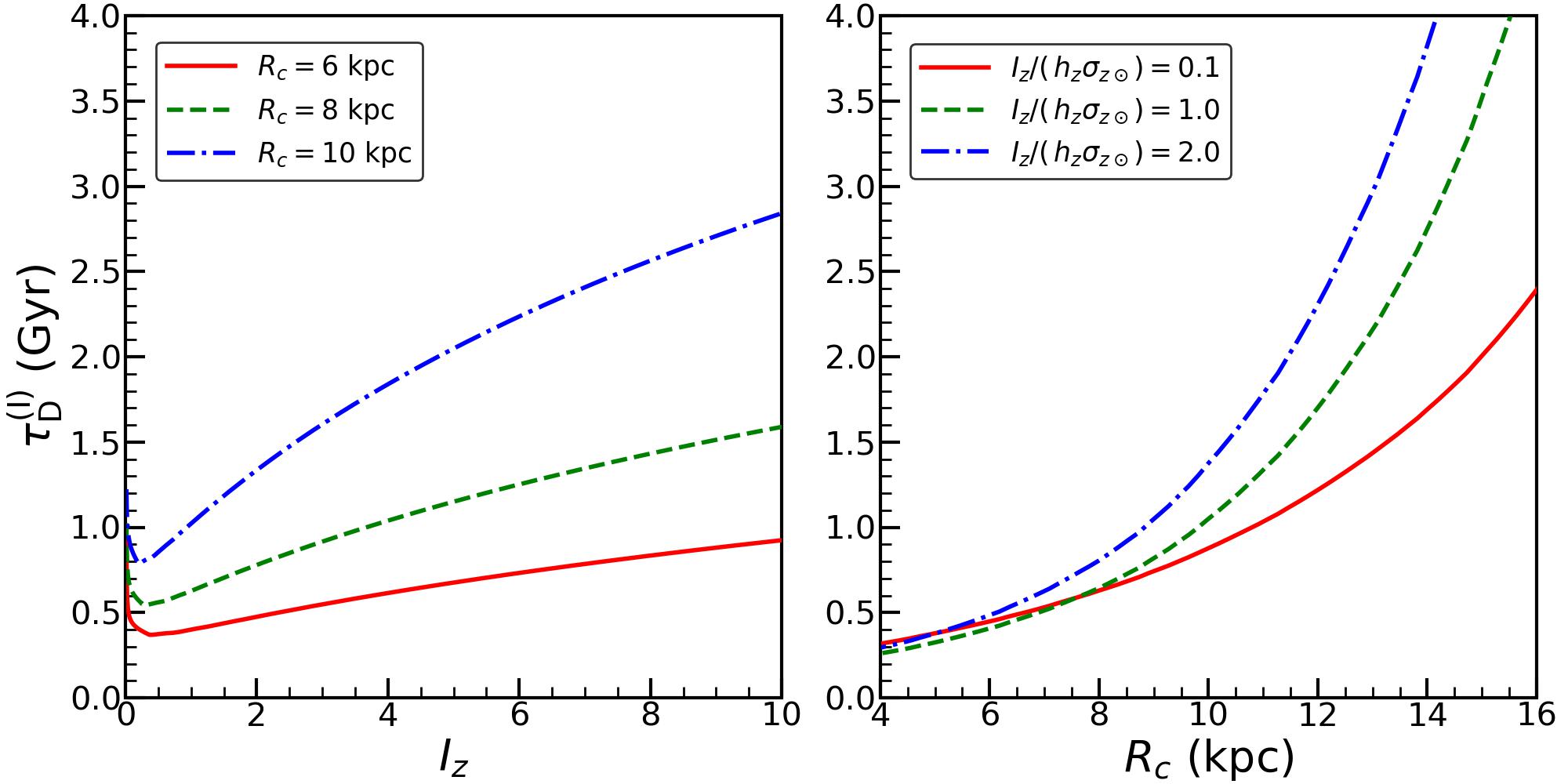}
  \caption{Timescale at which the disk response damps away due to collisional diffusion, i.e., small-scale scatterings of stars with structures like GMCs, is plotted as a function of $I_z$ ($R_c$) for three different values of $R_c$ ($I_z$) as indicated, in the left (right) panel. Typically, collisional diffusion occurs faster for smaller $I_z$ and smaller $R_c$.}
  \label{fig:diffusion_timescale}
\end{figure*}

In the above section we discussed the characteristics of the disk response in the absence of collisions. However, in a real galaxy like the MW disk, small-scale collisionality can potentially damp away any coherent response to a perturbation. Collisional diffusion arises not from star-star collisions, which is typically negligible, but from gravitational scattering with other objects, such as GMCs, DM substructure, etc. As discussed in Section~\ref{sec:galdisk}, the impact of collisional diffusion is mainly captured by the diffusion coefficients $D_z$ and $D_R$. Following \cite{Tremaine.etal.22} we assume that the disk stars have gained their mean vertical and radial actions over the age of the disk, $T_{\rm disk}=10\Gyr$, due to collisional heating, which implies that $D_a=\left<I_a\right>/T_{\rm disk}$ where $a$ is either $z$ or $R$ and $\left<I_a\right>=\int \rmd I_a \, I_a \, f_0 \, /\int \rmd I_a \, f_0$.

\begin{figure*}
\centering
\begin{subfigure}{1\textwidth}
  \centering
  \includegraphics[width=1\textwidth]{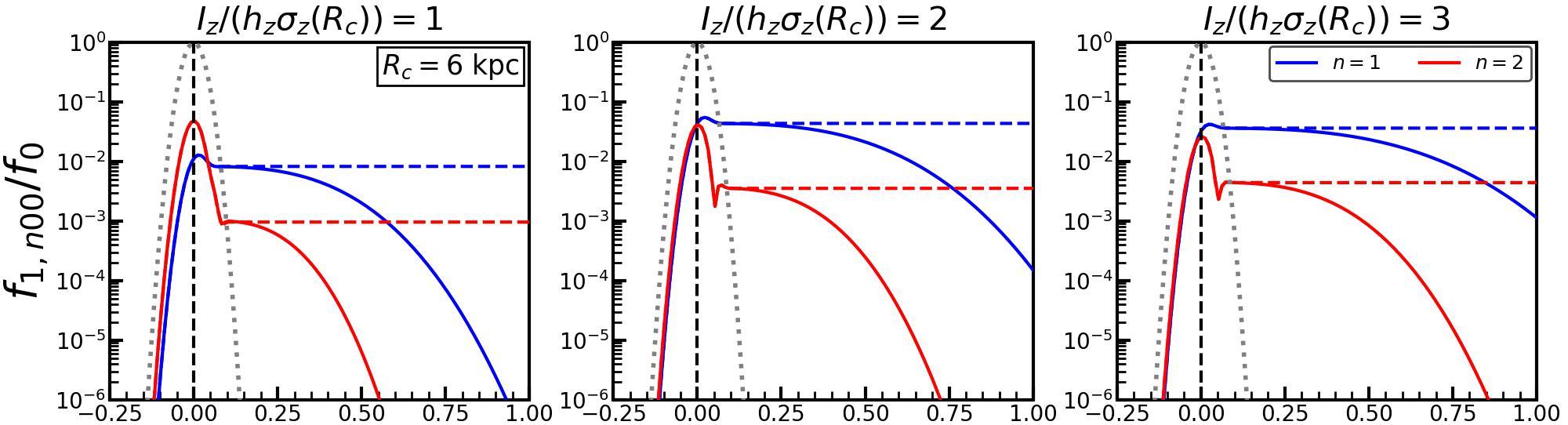}
  \label{fig:f1_vs_t_1}
\end{subfigure}\\
\begin{subfigure}{1\textwidth}
  \centering
  \includegraphics[width=1\textwidth]{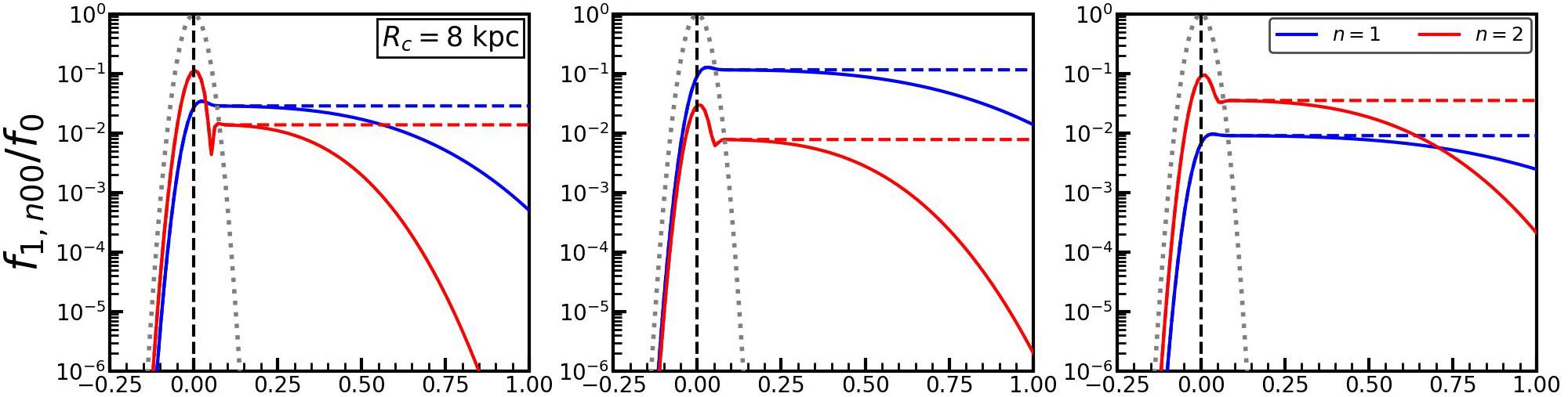}
  \label{fig:f1_vs_t_2}
\end{subfigure}\\
\begin{subfigure}{1\textwidth}
  \centering
  \includegraphics[width=1\textwidth]{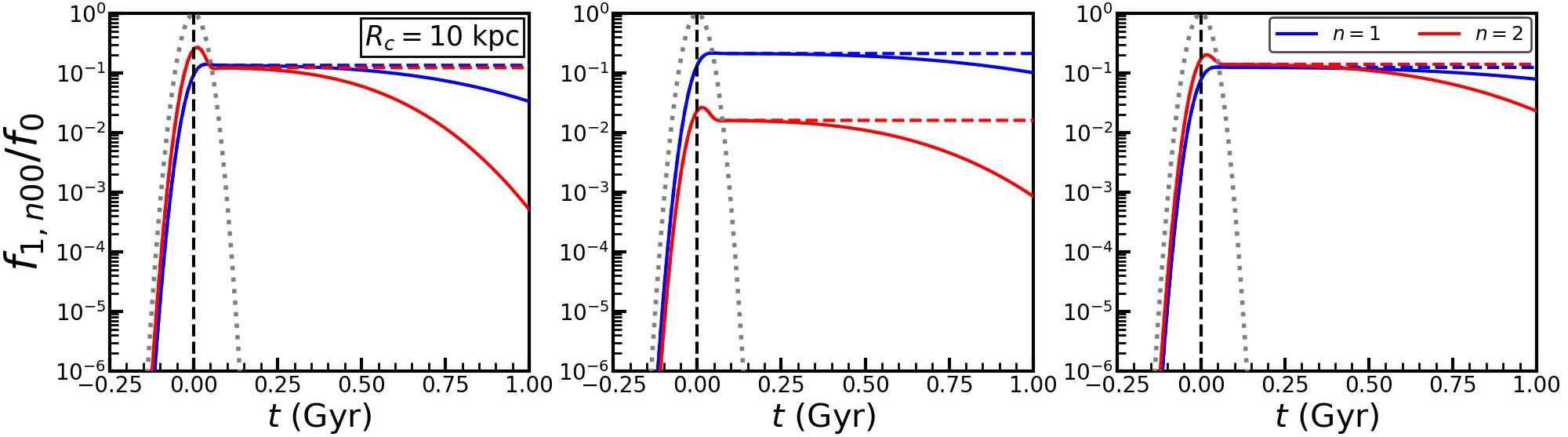}
  \label{fig:f1_vs_t_3}
\end{subfigure}
\caption{MW disk response to transient bars/2-armed spirals with Gaussian temporal modulation of pulse frequency, $\omega_\rmo=\omega_\rme=0.5\, \sigma_{z,\odot}/h_z$: the amplitude of the disk response, $f_{1n00}/f_0$, is plotted as a function of time. The rows and columns respectively denote different values of $R_c$ and $I_z$ as indicated. Blue and red lines indicate the $n=1$ and $2$ modes, while the solid and dashed lines respectively denote the cases with and without collisional diffusion (due to interactions of stars with structures like GMCs). The disk response initially rises and falls hand in hand with the perturbing pulse (indicated by the grey dotted line), before saturating to a steady state in the collisionless case and undergoing super-exponential damping in the collisional case. Note that collisional damping is faster for smaller $I_z$, smaller $R_c$ and larger $n$ modes.}
\label{fig:trans_spiral_damp}
\end{figure*}

For a transient bar/spiral with pulse frequency $\omega_j$, $\calP^{(j)}_{nlm}$ is given by equation~(\ref{Pnlm_spiral}). In the impulsive limit ($\omega_j \to \infty$), we have that $\calM_j(t-\tau) \to \omega_j \delta(t-\tau)$. Upon absorbing $\omega_j$ in the prefactor, the expression for $\calP^{(j)}_{nlm}$ then simplifies to \citep[c.f. Appendix A of][]{Tremaine.etal.22}
\begin{align}
\calP^{(j)}_{nlm} (\bI,t) &\approx \Theta(t)\, \exp{\left[-i\,(n\Omega_z+l\kappa+m\Omega_\phi)\,t\right]} \, \exp{\left[-\left(\frac{n^2 D_{z}}{4 I_z}+\frac{l^2 D_R}{4 I_R}\right)t\right]} \, \exp{\left[-\frac{{\left(n\Omega_{z1}\right)}^2 D_z I_z}{3}t^3\right]}.
\end{align}
This demonstrates that, in the impulsive limit, the disk response instantaneously grows and spawns phase-spirals whose amplitude decays due to collisional diffusion, manifest in the exponential damping terms. The first and second exponential factors, respectively, characterize the diffusion in vertical angle and action, which occur over the following timescales:
\begin{align}
\tau_\rmD^{(\rmw)} &= {\left[\frac{n^2 D_z}{4 I_z}+\frac{l^2 D_R}{4 I_R}\right]}^{-1}, \;\;\;\; \tau_\rmD^{(\rmI)} = {\left[\frac{3}{{\left(n\Omega_{z1}\right)}^2 D_z I_z}\right]}^{1/3}.
\label{diffusion_timescale}
\end{align}
Of these, the timescale for the diffusion in angles, $\tau_\rmD^{(\rmw)}$, typically exceeds that for the diffusion in actions, $\tau_\rmD^{(\rmI)}$, by at least an order of magnitude, implying that angle diffusion is negligible. Hence collisional diffusion mainly causes the abatement of action gradients in the phase-space structure of the response (arising from the action dependence of the frequencies, i.e., $\Omega_{z1}\neq 0$). The left (right) panel of Fig.~\ref{fig:diffusion_timescale} plots the diffusion timescale, $\tau_\rmD^{(\rmI)}$, as a function of $I_z$ ($\Rc$) for three different values of $\Rc$ ($I_z$) as indicated. Note that $\tau_\rmD^{(\rmI)}$ diverges in the small $I_z$ limit, attains a minimum around $I_z\sim 0.2-0.5\, h_z\sigma_{z,\odot}$, and increases as $I_z^\beta$ with $\beta<1$ at large $I_z$. As a function of $R_\rmc$, $\tau_\rmD^{(\rmI)}$ shows an approximately exponential rise. This owes to the fact that $\left<D_z\right> \sim h_z \, \sigma_z(R_\rmc) / T_{\rm disk} \sim \exp{\left[-R_\rmc/2h_R\right]} / T_{\rm disk}$ for the 3MN profile adopted for the MW disk. At $R_\rmc=\Rsun=8\kpc$, $\tau_\rmD^{(I)}\sim 0.6-0.7\Gyr$, in agreement with \cite{Tremaine.etal.22}. Hence, we see that collisional diffusion in action space is fairly efficient, and thus that phase-spirals are short-lived features.

Fig.~\ref{fig:trans_spiral_damp} plots the amplitude of the disk response (for $I_R=0$) to a transient spiral of pulse frequency, $\omega_\rmo=\omega_\rme=0.5 \sigma_{z,\odot}/h_z$, computed using equations~(\ref{f1nk_gensol_f0}), (\ref{trans_spiral_Resp}) and (\ref{Pnlm_spiral}), as a function of time. Dashed and solid lines show the results with and without collisional diffusion, respectively. The rows correspond to different values of $\Rc$ while the columns denote different values of $I_z/(h_z\sigma_{z,\odot})$ as indicated. The blue and red lines denote the response for the $(n,l,m)=(1,0,0)$ and $(2,0,0)$ modes, respectively, and the dotted grey line represents the Gaussian pulse strength. The response for both bending and breathing modes initially grows hand in hand with the perturbing pulse. Following the point of maximum pulse strength, the response follows the decaying pulse strength before saturating to the steady state amplitude given in equation~(\ref{Pnlm_trans_spiral}) in the collisionless limit. In the presence of collisional diffusion, however, the response continues to damp out as $\sim \exp{[-(t/\tau_\rmD^{(\rmI)})^3]}$ after temporarily saturating at the collisionless steady state. Note that the collisional damping is faster for smaller $\Rc$ and smaller $I_z$. In addition, $n=2$ breathing modes damp out faster than the $n=1$ bending modes due to the $n^{-2/3}$ dependence of $\tau_\rmD^{(\rmI)}$.

To summarize, we have shown that phase-spirals can be triggered by impulsive perturbations resulting from transient spiral arms or bars, but are subject to super-exponential damping due to collisional diffusion that is likely to be dominated by scattering against GMCs. This collisional damping is more efficient in the inner disk, for stars with smaller $I_z$, and for modes of larger $n$.

\section{Disk response to satellite encounter}
\label{sec:disk_resp_sat}

In addition to the spiral arm/bar perturbations considered above, we also consider disk perturbations triggered by encounters with a satellite galaxy. For the sake of brevity, we only compute the disk response in the collisionless limit. In the case of impulsive encounters, the impact of collisional diffusion is simply expressed by multiplying the collisionless response by the collisional damping factor $\exp[-(t/\tau_\rmD^{(\rmI)})^3]$, with $\tau_\rmD^{(\rmI)}$ given by equation~(\ref{diffusion_timescale}).

\begin{SCfigure}
  \centering
  \includegraphics[width=0.5\textwidth]{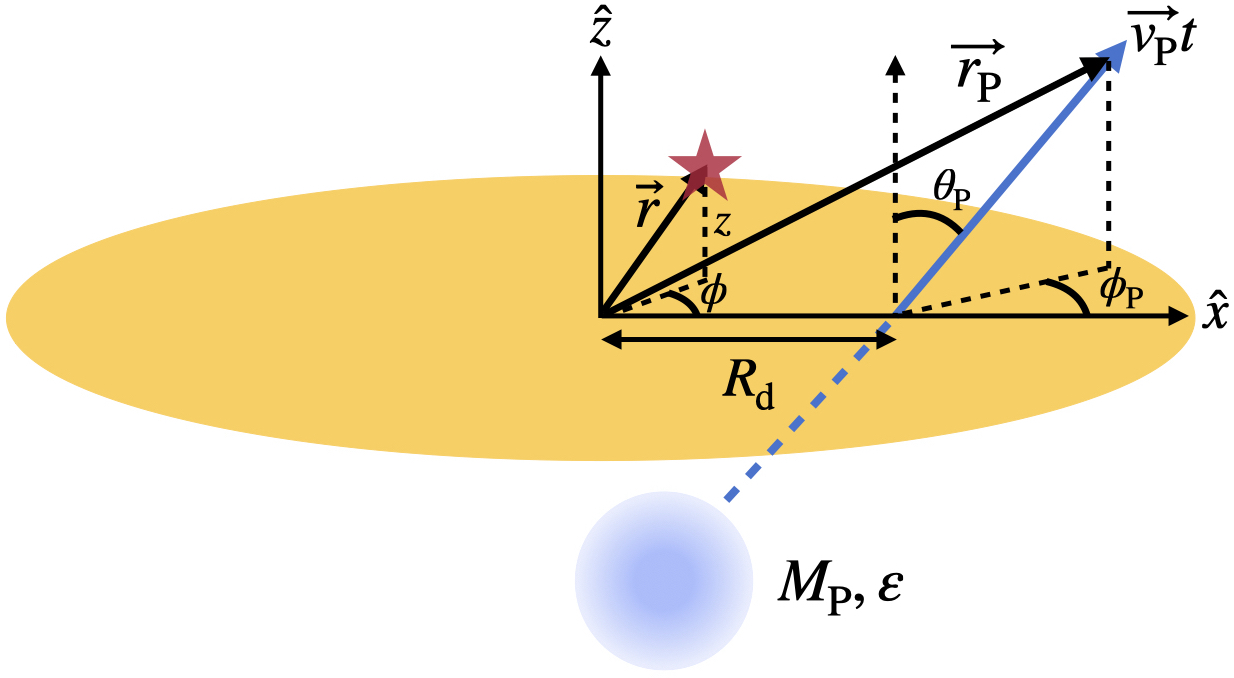}
  \caption{Illustration of the geometry of a satellite galaxy with mass $M_\rmP$ impacting a disk galaxy with uniform velocity $\vp$ along a straight line. The impact occurs at a galactocentric distance $\rd$. The orientation of $\bvp$ is specified by $\thetap$, the angle between $\bvp$ and the $z$-axis, and $\phip$, the angle between the projection of $\bvp$ on the mid-plane and the $x$-axis.}
  \label{fig:sat_enc_orient}
\end{SCfigure}

For simplicity, we assume that the satellite is moving with uniform velocity $\vp$ along a straight line, impacting the disk at a galactocentric distance $\rd$ with an arbitrary orientation, specified by the angles, $\thetap$ and $\phip$, which are respectively defined as the angles between $\bvp$ and the $z$-axis, and between the projection of $\bvp$ on the mid-plane and the $x$-axis (see Fig.~\ref{fig:sat_enc_orient}). Thus the position vector of the satellite with respect to the galactic center can be written as
\begin{align}
\brp = (\rd + \vp\sin{\thetap}\cos{\phip}\,t)\,\hat{\bx} + \vp\sin{\thetap}\sin{\phip}\,t\,\hat{\by} + \vp\cos{\thetap}\,t\,\hat{\bz},
\label{rp_sat}
\end{align}
while that of a star is given by
\begin{align}
\br = R(\cos{\phi}\,\hat{\bx} + \sin{\phi}\,\hat{\by}) + z\,\hat{\bz}.
\label{r_sat}
\end{align}
We consider the satellite to be a Plummer sphere of mass $M_\rmP$ and size $\varepsilon$, such that its gravitational potential at location $\br$ is given by
\begin{align}
\Phi_\rmP &= G M_\rmP \left[-\frac{1}{\sqrt{{\left|\br-\brp\right|}^2+\varepsilon^2}} + \frac{\br \cdot \brp}{{\left(\brp^2+\varepsilon^2\right)}^{3/2}}\right]\,.
\label{sat_pot}
\end{align}
Here the first term is the `direct' term and the second is the `indirect' term that accounts for the reflex motion of the disk and the fact that the disk center is accelerated by the satellite and is thus non-inertial. Typically, the first one dominates over the second.

In order to compute the disk response to this external perturbation, we need to compute its Fourier coefficients, which is challenging. Rather, we first evaluate the $\tau$-integral in Equation~(\ref{f1nk_gensol_f0}), setting $t_\rmi\to -\infty$, and then compute the Fourier transform of the result, as worked out in Appendix~\ref{App:sat_disk_Resp}. For $I_R\approx 0$ (this is justified since we adopt the radial epicyclic approximation in this paper), this yields a modal response, $f_{1nlm}$ (Equation~[\ref{f1nk_gensol_f0}]), with $\calI_{nlm}(\bI,t)$ given by

\begin{align}
&\calI_{nlm}(\bI,t)\approx-\frac{2G M_\rmP}{\vp} \exp{\left[-i\Omega t\right]} \times \exp{\left[-i\frac{\Omega\sin{\thetap}\cos{\phip}}{\vp} \rd\right]} \nonumber \\ 
&\times \frac{1}{{\left(2\pi\right)}^2} \int_0^{2\pi} d w_z \exp{\left[-in w_z\right]} \exp{\left[i\frac{\Omega \cos{\thetap}}{\vp} z\right]} \int_0^{2\pi} d \phi\,\exp{\left[-im\phi\right]}\, \exp{\left[i\frac{\Omega \sin{\thetap} \cos{\left(\phi-\phip\right)}}{\vp} \Rc\right]} \nonumber \\
&\times K_{0i}\left(\frac{\Omega\sqrt{\calR^2_\rmc+\varepsilon^2}}{\vp},\frac{\vp t-\calS_\rmc}{\sqrt{\calR^2_\rmc+\varepsilon^2}}\right),
\label{sat_gen}
\end{align}
where $\Omega$ is given by
\begin{align}
\Omega = n\Omega_z + l\kappa + m\Omega_\phi.
\end{align}
Here $\calR_\rmc=\calR(\Rc)$ and $\calS_\rmc=\calS(\Rc)$ with $\calR$ and $\calS$ given by equation~(\ref{sat_RS_app}). $K_{0i}$ is given by equation~(\ref{K0i_app}), which asymptotes to the modified Bessel function of the second kind, $K_0\left(\left|\Omega\right|\sqrt{\calR^2_\rmc+\varepsilon^2}/\vp\right)$, in the large time limit. A more precise expression for $\calI_{nlm}$ that is valid for higher values of $I_R$ is given by equation~(\ref{sat_gen_IR_app}) of Appendix~\ref{App:sat_disk_Resp}.

The expression for $\calI_{nlm}$ given in equation~(\ref{sat_gen}) exhibits several key features of the disk response to satellite encounters. The $\exp{\left[-i\Omega t\right]}$ factor encodes the phase mixing of the response due to oscillations at different frequencies, giving rise to phase-spirals. The $\exp{\left[i\left(\Omega \cos{\thetap}/\vp\right)z\right]}$ and $\exp{\left[i\left(\Omega \sin{\thetap} \cos{(\phi-\phip)}/\vp\right) \Rc\right]}$ factors respectively indicate that the satellite induces wave-like perturbations in the disk with two characteristic wave-numbers: the vertical wave-number, $k_z\approx \Omega\cos{\thetap}/\vp$ and the horizontal wave-number, $k_R\approx \Omega\sin{\thetap}/\vp$. Therefore, the disk response will be vertically (horizontally) stratified in case of a perpendicular (planar) impact of the satellite. As shown in Appendix~\ref{App:special}, expressions~(\ref{f1nk_gensol_f0}) and (\ref{sat_gen}), which are complicated to compute, yield the correct response in the impulsive limit of a satellite having a face-on, perpendicular encounter through the center of the disk.

\subsection{Asymptotic behaviour of the response}

It is instructive to study the two extreme cases of encounter speed, the impulsive limit (large $\vp$) and the adiabatic limit (small $\vp$). Using the asymptotic form of the $K_0$ Bessel function that appears in equation~(\ref{sat_gen}), we obtain the following approximate asymptotic behaviour of $f_{1nlm}$ at large time:
\begin{align}
f_{1nlm} &\sim \frac{G M_\rmP}{\vp} \exp{\left[-i\Omega t\right]} \times
\begin{cases} 1, & \vp\to \infty \\ \\
\sqrt{\vp/\Omega b}\, \exp{\left[-\Omega b/\vp\right]}, & \vp \to 0,
\end{cases}
\label{sat_asymptote}
\end{align}
where $b$ is the impact parameter of the encounter, defined as the perpendicular distance of the nearest star on the mid-plane from the satellite's (straight) orbit, and expressed as
\begin{align}
b = \left|\rd - \Rc \right| \, \sqrt{1-\sin^2{\thetap}\cos^2{\phip}}.
\label{impact_parameter}
\end{align}
It is clear from these limits that the disk response is most pronounced for intermediate velocities, $\vp\sim\Omega b$. For impulsive encounters, the response is suppressed as a power law in $\vp$, whereas in the adiabatic limit the response is exponentially suppressed, except at resonances, $\Omega=n\Omega_z+l\kappa+m\Omega_\phi=0$. In this limit, far from the resonances, the perturbation timescale, $b/\vp$, is much larger than $\Omega^{-1}$, and the net response is washed away due to many oscillations during the perturbation (i.e., the actions are adiabatically invariant), a phenomenon known as adiabatic shielding \citep[][]{Weinberg.94a,Weinberg.94b,Gnedin.Ostriker.99}.

\begin{figure*}
  \centering
  \includegraphics[width=1\textwidth]{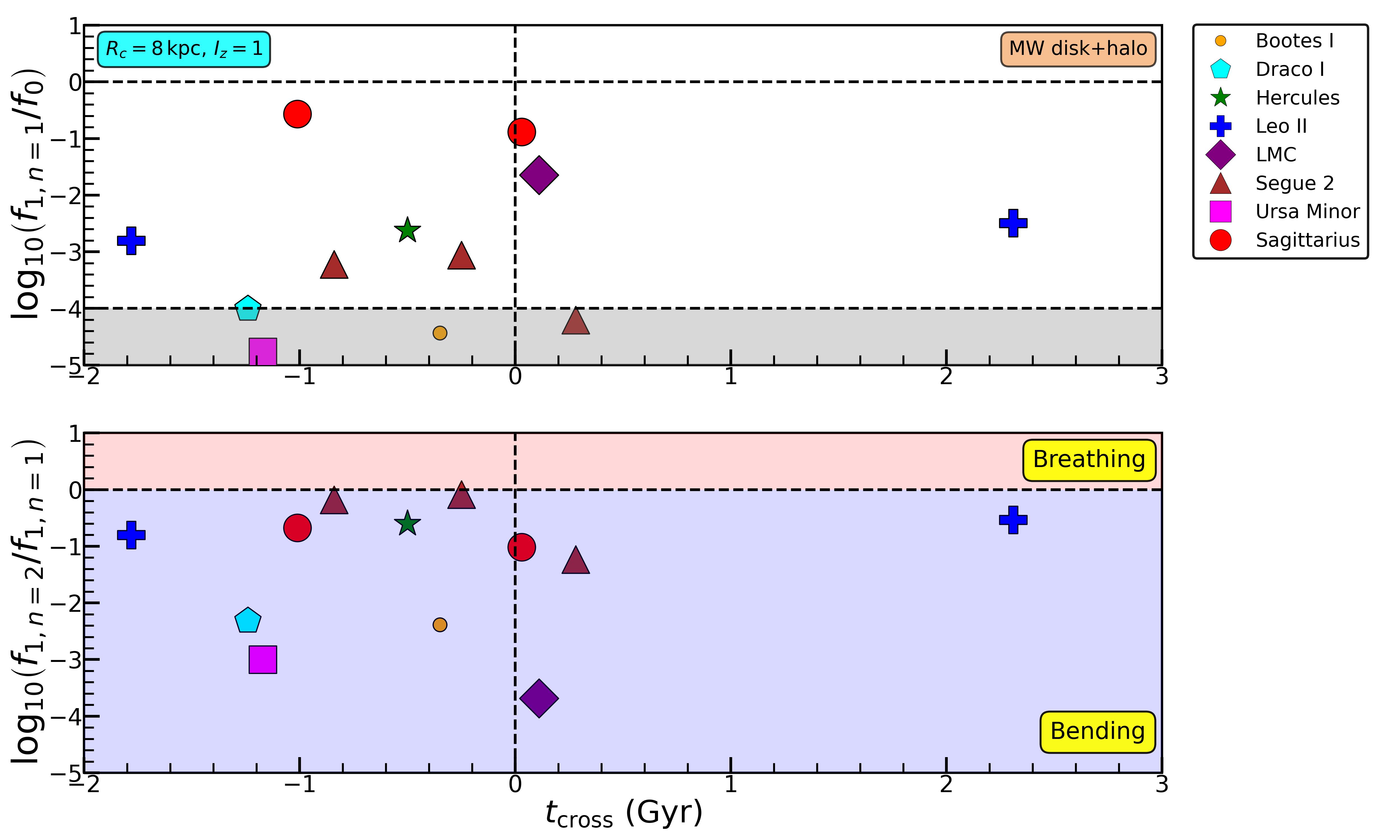}
  \caption{Steady state MW disk response to satellite encounter in the collisionless limit: bending mode strength, $f_{1,n=1}/f_0$ (upper panel), and the corresponding breathing vs bending ratio, $f_{1,n=2}/f_{1,n=1}$ (lower panel) for the $(l,m)=(0,0)$ modes, in the Solar neighborhood for the MW satellites, as a function of the disk crossing time, $t_{\rm cross}$, in Gyr, where $t_{\rm cross}=0$ marks today. The previous two and the next impacts are shown. Here we consider $I_z=h_z \sigma_{z,\odot}$, with fiducial MW parameters, and marginalize over $I_R$. The effect of the (non-responsive) ambient DM halo on the stellar frequencies is taken into account. The estimates of $t_{\rm cross}$ are very sensitive to the detailed potential of the MW system, while the response estimates are fairly robust (see text for details). In the upper panel, the region with bending mode response, $f_{1,n=1}/f_0<10^{-4}$, has been grey-scaled, indicating that the response from the satellites in this region is far too weak and adiabatic to be detected by Gaia. Note that the response is dominated by that due to Sgr, followed by Hercules, Leo II, Segue 2 and the Large Magellanic Cloud (LMC). Also note that the previous two and next impacts of all the satellites excite bending modes in the Solar neighborhood.}
  \label{fig:MW_sat_Resp}
\end{figure*}
\begin{figure*}[t!]
  \centering
  \begin{subfigure}{0.43\textwidth}
    \centering
    \includegraphics[width=1\textwidth]{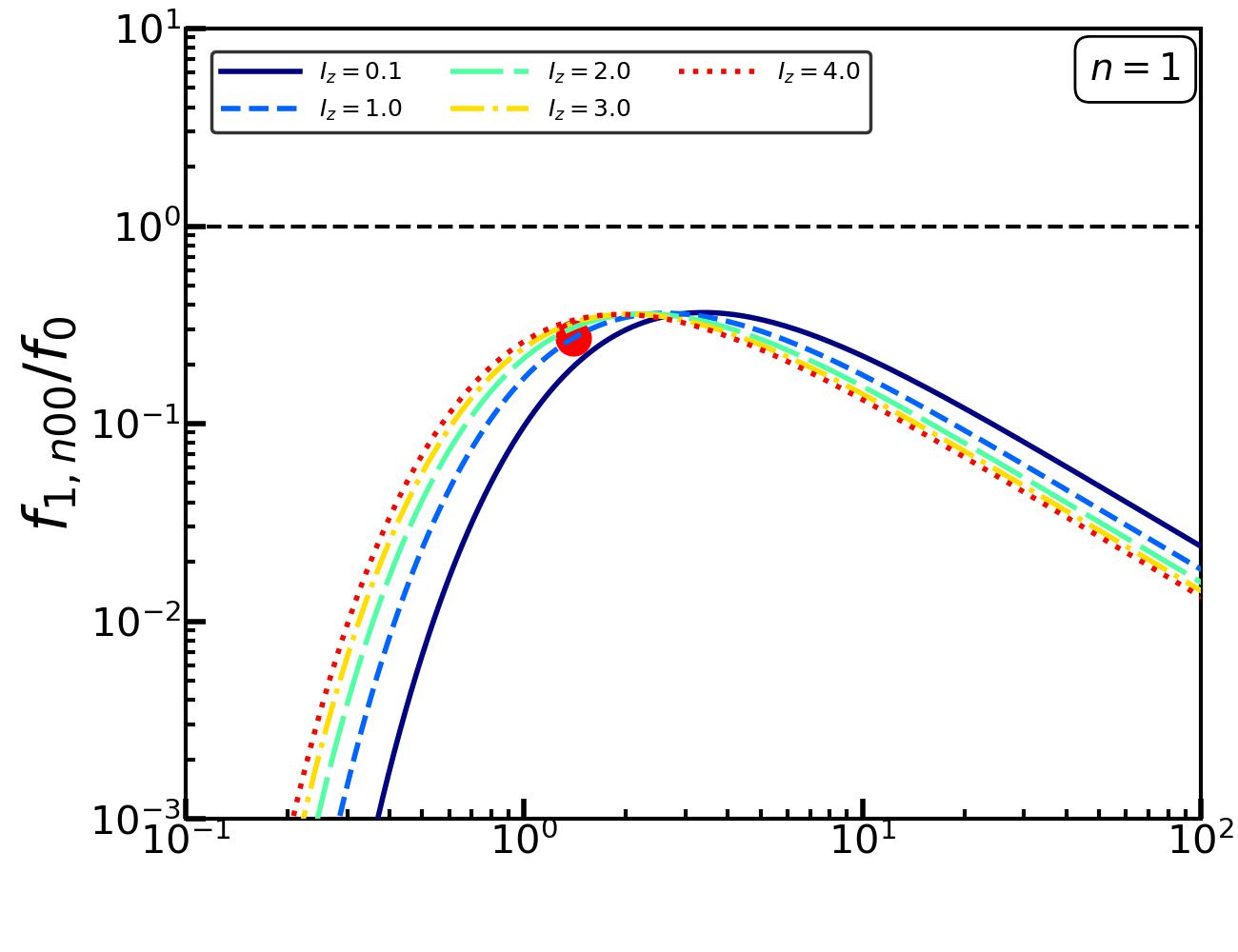}
    \label{disk_Resp_n1_0.5pi}
  \end{subfigure}
  \begin{subfigure}{0.43\textwidth}
    \centering
    \includegraphics[width=1\textwidth]{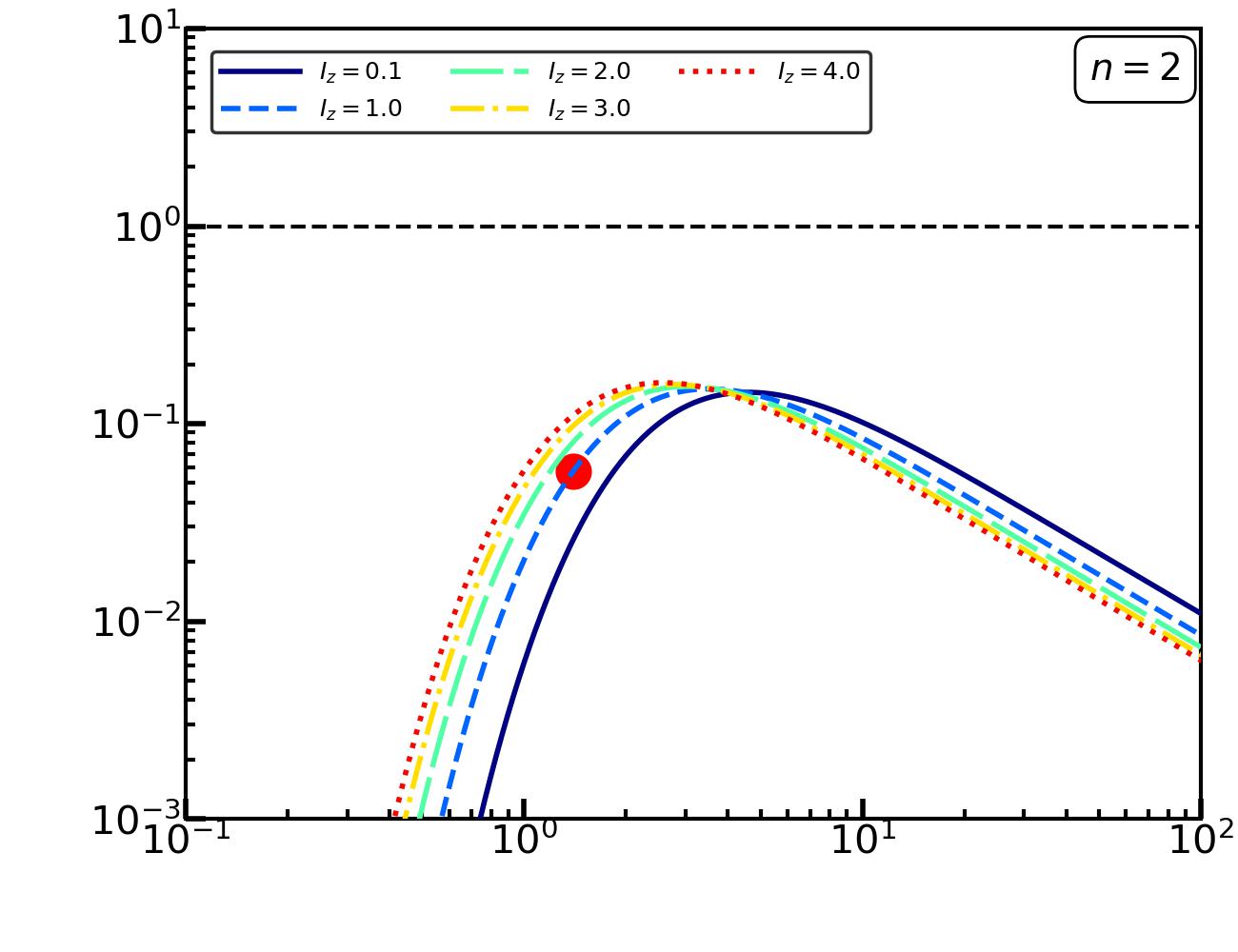}
    \label{disk_Resp_n1_0.5pi}
  \end{subfigure}
  \\
  \begin{subfigure}{0.43\textwidth}
    \centering
    \includegraphics[width=1\textwidth]{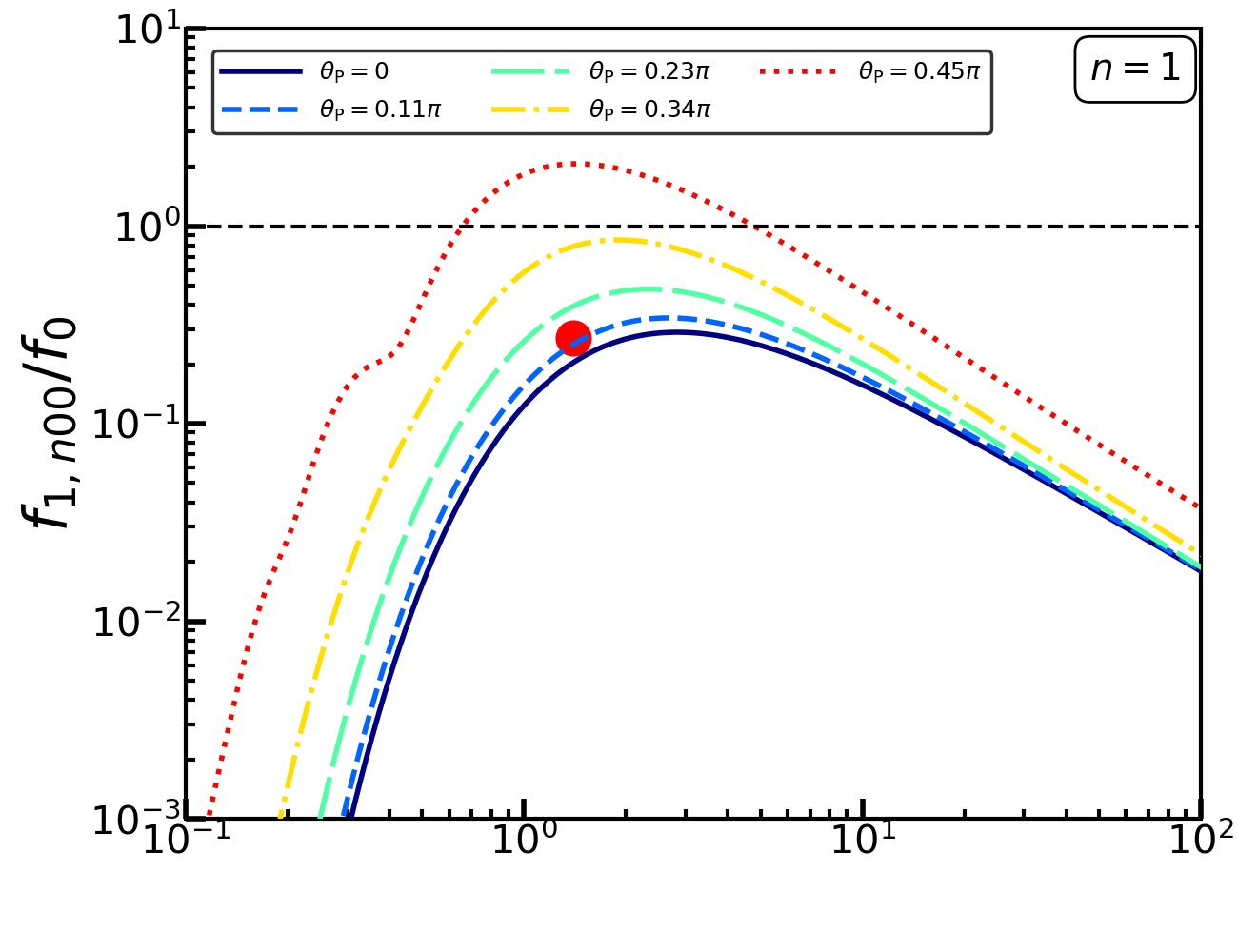}
    \label{disk_Resp_n2_0.5pi}
  \end{subfigure}
  \begin{subfigure}{0.43\textwidth}
    \centering
    \includegraphics[width=1\textwidth]{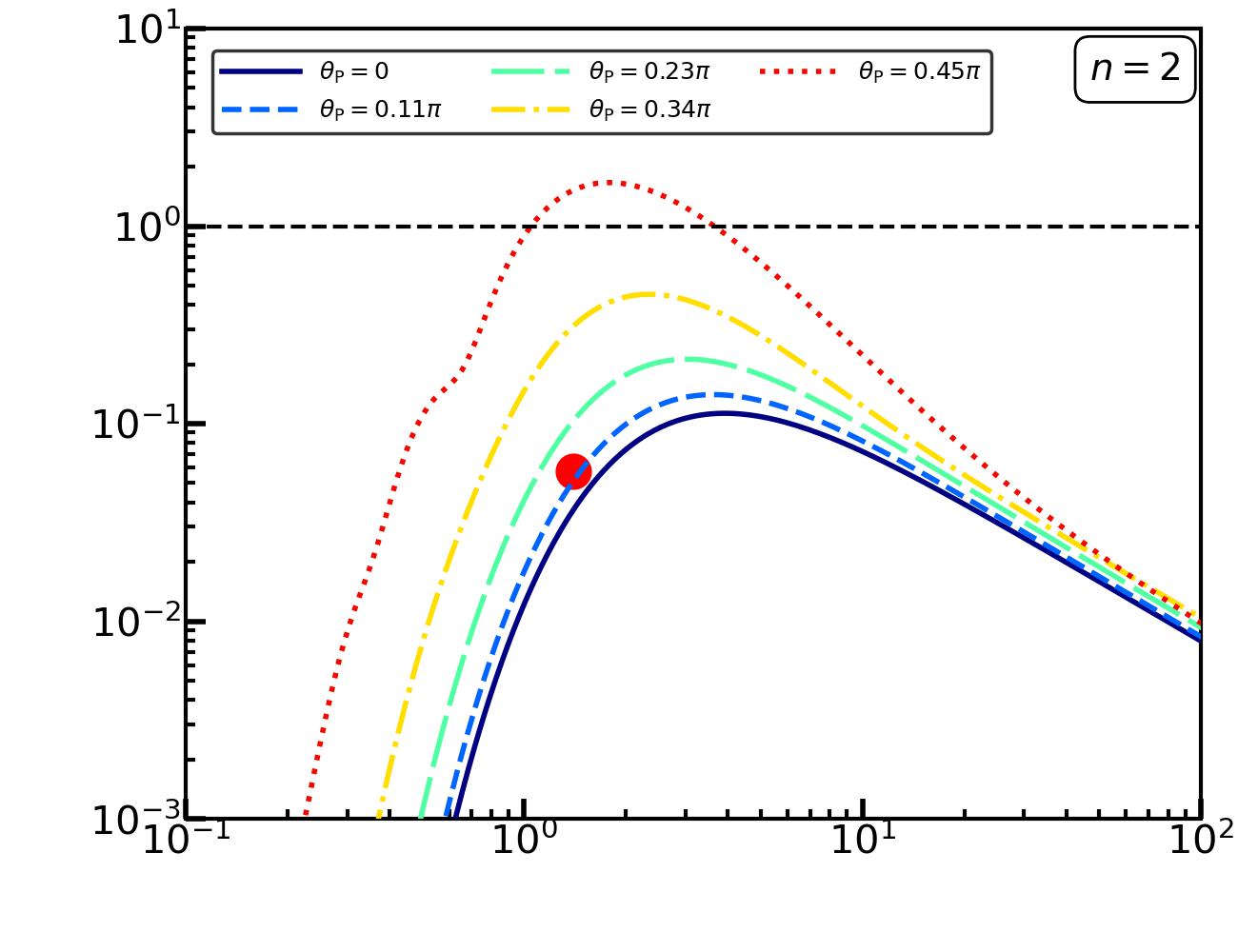}
    \label{disk_Resp_n2_0.5pi}
  \end{subfigure}
  \\
  \begin{subfigure}{0.43\textwidth}
    \hspace{0.001cm}
    %\centering
    \includegraphics[width=1\textwidth]{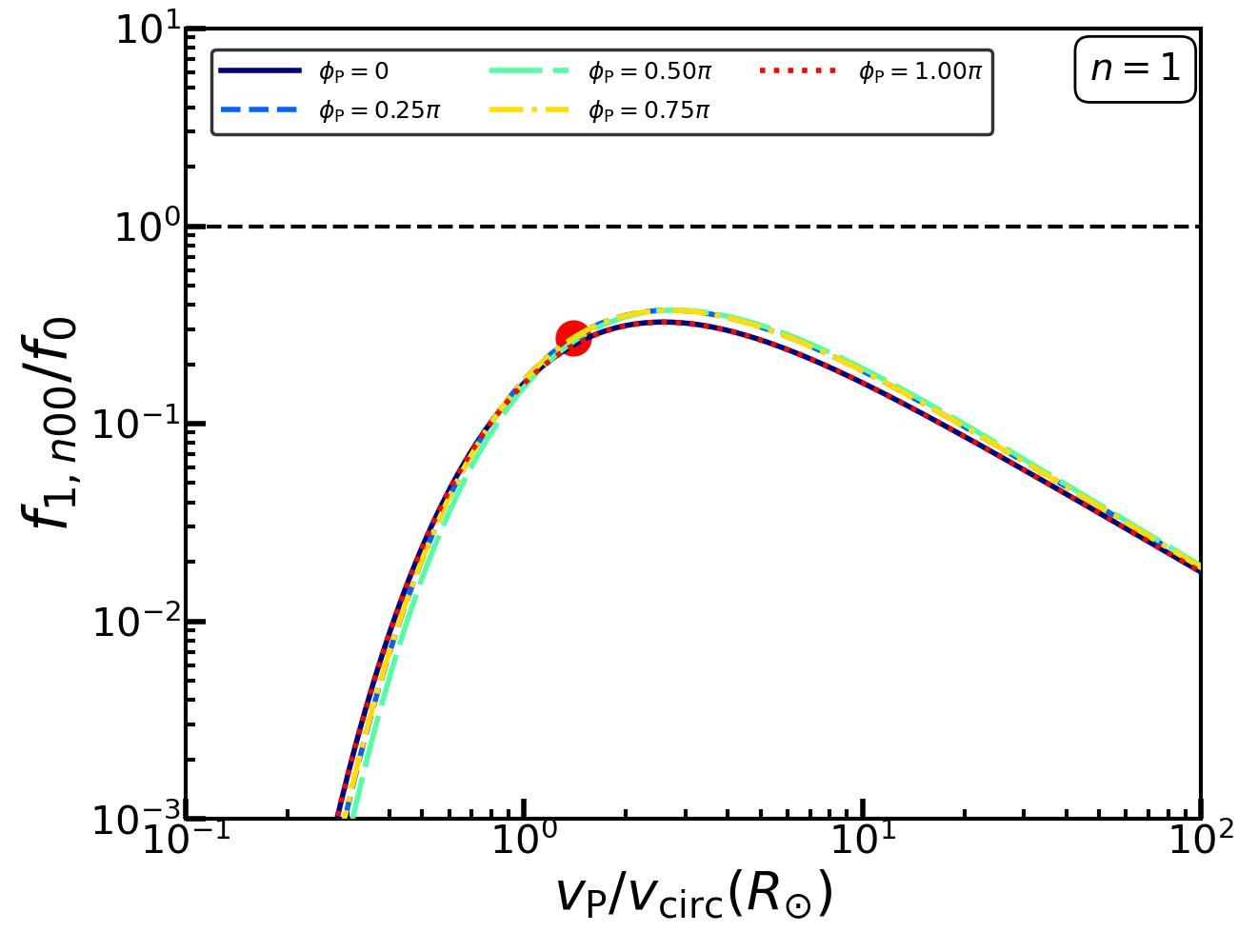}
    \label{disk_Resp_n2_0.25pi}
  \end{subfigure}
  \begin{subfigure}{0.43\textwidth}
    \hspace{0.001cm}
    %\centering
    \includegraphics[width=1\textwidth]{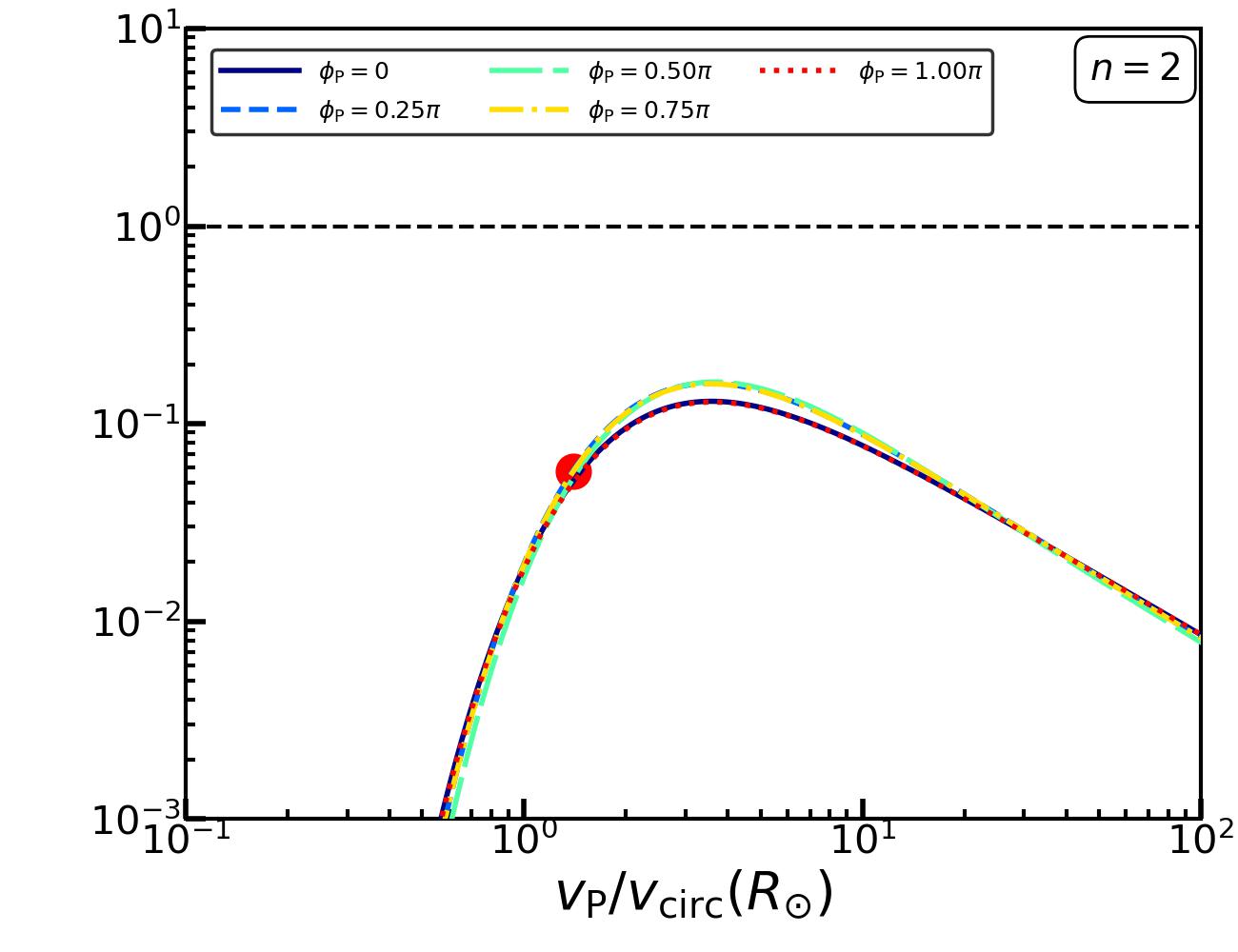}
    \label{disk_Resp_n2_0.25pi}
  \end{subfigure}
  \caption{Steady state MW disk response to satellite encounter in the collisionless limit: each panel shows the behaviour of the disk response amplitude, $f_{1,n00}/f_0$ (evaluated using equations~[\ref{f1nk_gensol_f0}] and [\ref{sat_gen_IR_app}]) and marginalized over $I_R$), as a function of the impact velocity, $\vp$, in the Solar neighborhood, i.e., $\Rc = \Rsun = 8\kpc$, in presence of an ambient DM halo. The left and right columns respectively indicate the response for the $n=1$ bending and $n=2$ breathing modes. The top, middle and bottom rows show the same for different values of $I_z$ (in units of $I_{z,\odot}$), $\thetap$ and $\phip$ respectively as indicated, with the fiducial parameters corresponding to $I_{z,\odot}$ and the parameters for Sgr impact, the response amplitude for which is indicated by the red circle. Note that the response is suppressed as $\vp^{-1}$ in the impulsive (large $\vp$) limit but exponentially suppressed in the adiabatic (small $\vp$) regime, and peaks at an intermediate velocity, $\vp\sim 2-3\, v_{\rm circ}(\Rsun)$ (which is very similar to the encounter speed of Sgr). The peak of the response shifts to smaller $\vp$ for larger $I_z$, since $\Omega_z$ decreases with $I_z$. The response depends only very weakly on $\phip$ but is quite sensitive to $\thetap$; more planar encounters, i.e., increasing $\thetap$ triggers stronger responses.}
  \label{fig:disk_Resp_vp}
\end{figure*}

\subsection{Response of the MW disk to satellites}\label{sec:MW_disk_resp_sat}

The MW halo harbors several fairly massive satellite galaxies that repeatedly perturb the MW disk. Here we use existing data on the phase-space coordinates of those MW satellites to compute the disk response of satellite encounters that occurred in the past few hundred Myr, which are those for which we may expect phase-spirals that were triggered to have survived to the present day.

To compute the disk response to the MW satellites, we proceed as follows. As in Paper~I, we adopt the galactocentric coordinates and velocities computed and documented by \cite{Riley.etal.19} \citep[table A.2, see also][]{Li.etal.20} and \cite{Vasiliev.Belokurov.20} as initial conditions for the MW satellites. We then simulate their orbits in the combined gravitational potential of the MW halo, disk plus bulge\footnote{The bulge is modelled as a spherical \cite{Hernquist.90} profile with mass $M_b=6.5\times 10^9\Msun$ and scale radius $r_b=0.6$ kpc.} using a second order leap-frog integrator. For each individual orbit, we record the times, $t_{\rm cross}$, and the galactocentric radii, $\rd$, corresponding to disk crossings. We also register the corresponding impact velocities, $\vp=\sqrt{v^2_z+v^2_R+v^2_\phi}$, and the angles of impact, $\thetap=\cos^{-1}{(v_z/\vp)}$ and $\phip=\tan^{-1}{(v_\phi/v_R)}$. We substitute these quantities in equation~(\ref{sat_gen_IR_app}) and compute the disk response (integrated over $I_R$) following the satellite encounter, using equations~(\ref{f1nk_gensol_f0}) and (\ref{sat_gen}). Results are summarized in Table~\ref{tab:MW_sat_resp} of Appendix~\ref{App:sat_disk_Resp}. Fig.~\ref{fig:MW_sat_Resp} plots the amplitude of the Solar neighborhood (for which $\Rc(L_z)=\Rsun=8\kpc$) bending mode response, $f_{1,n=1}/f_0$ (top panel), and breathing-to-bending ratio, $f_{1,n=2}/f_{1,n=1}$ (bottom panel), as a function of $t_{\rm cross}$. Here we only show the responses for $(l,m)=(0,0)$ modes, and consider stars with $I_z = I_{z,\odot} = 9.2\kpc\kms$.

It is noteworthy that the responses in the realistic MW disk computed here are $\sim 1-2$ orders of magnitude larger than those evaluated for the isothermal slab model shown in Fig.~7 of Paper~I. This owes to the reduced damping of the phase-spiral amplitude due to lateral mixing, which is more pronounced in the isothermal slab with unconstrained lateral velocities than in the realistic disk with constrained, ordered motion. From the lower panel of Fig.~\ref{fig:MW_sat_Resp} it is evident that, as in the isothermal slab case, almost all satellites trigger a bending mode response in the Solar neighborhood, resulting in a one-armed phase-spiral in qualitative agreement with the Gaia snail. However, as is evident from the upper panel, only five of the satellites trigger a detectable response in the disk, with $f_{1,n=1}/f_0 > \delta_{\rm min} \equiv 10^{-4}$ (see Appendix~C of Paper~I for a derivation of this approximate detectability criterion for Gaia). The response to encounters with the other satellites is weak either because they have too low mass or because the encounter with respect to the Sun is too slow and adiabatically suppressed. Sgr excites the strongest response by far; its bending mode response, $f_{1,n=1}/f_0$, is at least $1-2$ orders of magnitude above that for any other satellite. Its penultimate disk crossing, about the same time as its last pericentric passage $\sim 1\Gyr$ ago, triggered a strong response of $f_{1,n=1}/f_0 \sim 0.3$ in the Solar neighborhood. For comparison, the response from its last disk crossing, which nearly coincides with its last apocentric passage about $350\Myr$ ago, triggered a very weak, adiabatically suppressed response ($\sim 5\times 10^{-8}$) that falls below the lower limit of Fig.~\ref{fig:MW_sat_Resp}. Its next disk crossing in about $30\Myr$ is estimated to trigger a strong response with $f_{1,n=1}/f_0\sim 0.1$. Besides Sgr, the satellites that excite a detectable response, $f_{1,n=1}/f_0 > \delta_{\rm min}$ are Hercules, Segue 2, Leo II and the LMC. The imminent crossing of LMC is estimated to trigger $f_{1,n=1}/f_0 \sim 2\times10^{-2}$, which is an order of magnitude below Sgr. Only for $I_z/I_{z,\odot} \gtrsim 4.5$ ($z_{\rm max}\gtrsim 3.4 h_z$), the LMC response dominates over Sgr. This exercise therefore suggests that Sgr is the leading contender, among the MW satellites considered here, for triggering the Gaia snail in the Solar neighborhood, in agreement with several previous studies \citep[][]{Antoja.etal.18,Binney.Schonrich.18, Laporte.etal.18, Laporte.etal.19, Darling.Widrow.19b, Bland-Hawthorn.etal.19, Hunt.etal.21, Bland-Hawthorn.Garcia.21, Bennett.etal.22}.

We caution that the estimates of the disk response computed above ignore dynamical friction. Moreover, the disk crossing times are sensitive to the satellite orbits and therefore to the detailed MW potential and the current phase-space coordinates of the satellites. For example, a heavier MW model with a total mass of $1.5\times 10^{12}\Msun$ leaves the relative amplitudes of the satellite responses (in the collisionless limit) nearly unchanged, but makes the satellites more bound, bringing most of the disk crossing times closer to the present day. In particular, the last disk passage of Sgr that triggers a significant response now occurs $\sim 600\Myr$ ago (as opposed to $1 \Gyr$ ago in the fiducial case) which is closer to the winding time of $\sim 500\Myr$ inferred from the phase-spiral observed in the Solar neighborhood \citep[][]{Bland-Hawthorn.etal.19}.

In this section, we have computed the responses in the collisionless limit. In reality, collisional diffusion due to interactions of stars with GMCs, etc. would damp away the response super-exponentially over a timescale that is $\sim 0.6-0.7\Gyr$ in the Solar neighborhood (see section~\ref{sec:spiral_c}). This would almost completely wash away the response to any satellite encounter that occurred $\gtrsim 1\Gyr$ ago. For example, the present day response to the last pericentric passage of Leo II that occurred $\sim 1.8\,\Gyr$ ago would be completely erased. If the last disk crossing of Sgr that induced a strong response occurred $\sim 1\Gyr$ ago as in the fiducial MW model, the response would have been damped out by $\sim 2$ orders of magnitude by today, deeming Sgr unlikely to be the agent behind the Gaia snail. However, as discussed above, the disk crossing times are sensitive to the satellite orbits. The heavier MW model with a total mass of $1.5\times 10^{12}\Msun$ implies a Sgr crossing time of $\sim 0.6\Gyr$ instead of $1\Gyr$. In this case the response would only have been damped by a factor of $\sim 0.4$. Therefore, the collisionality argument suggests that if the Gaia snail was indeed triggered by Sgr, the impact causing it must have happened within $\sim 0.6-0.7\Gyr$ from the present day.

\subsection{Exploring parameter space}

\begin{figure*}[t!]
\centering
\begin{subfigure}{0.49\textwidth}
  \hspace{0.5mm}
  %\centering
  \includegraphics[width=1\textwidth]{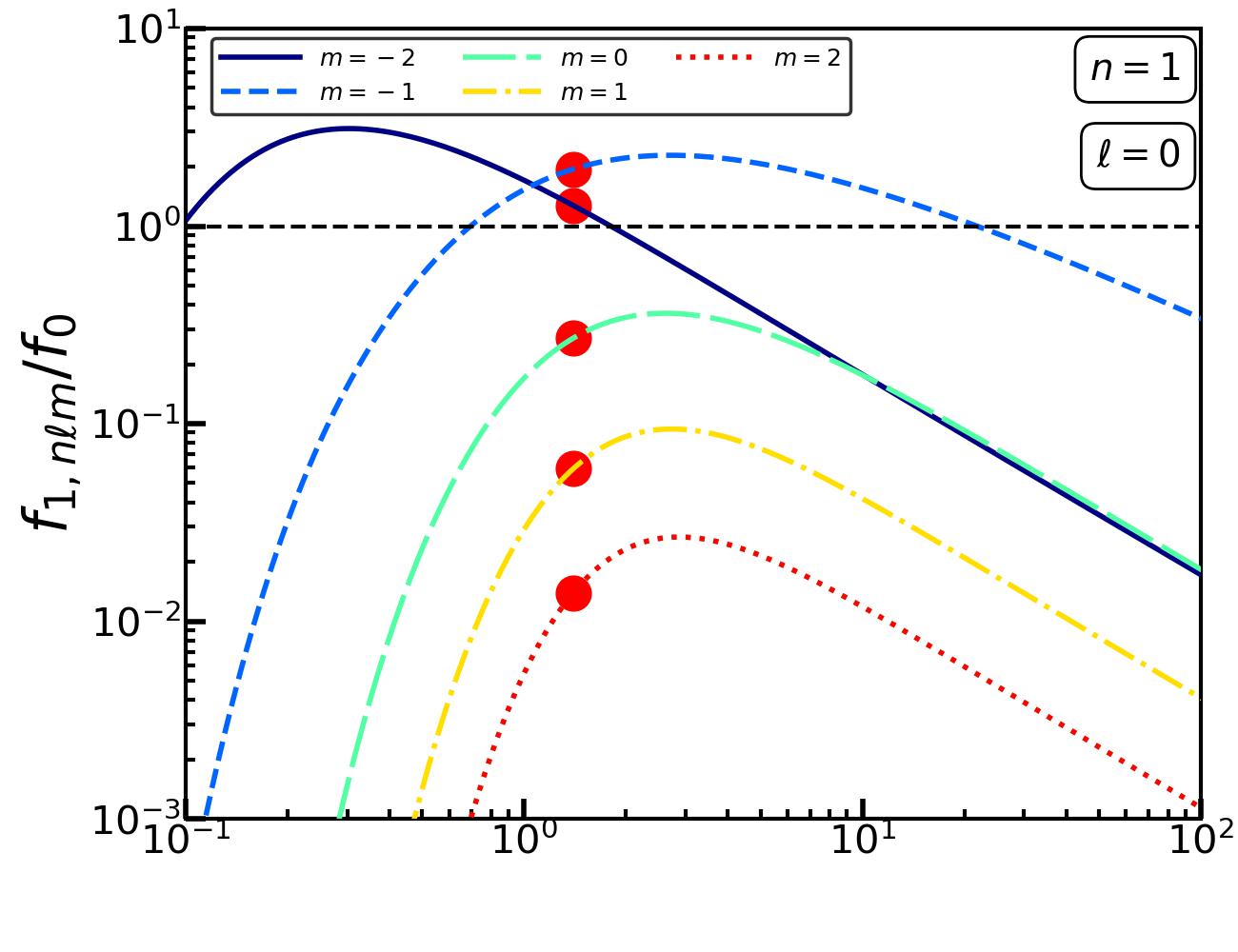}
  \label{disk_Resp_n1_l0}
\end{subfigure}
\begin{subfigure}{0.49\textwidth}
  \hspace{0.5mm}
  %\centering
  \includegraphics[width=1\textwidth]{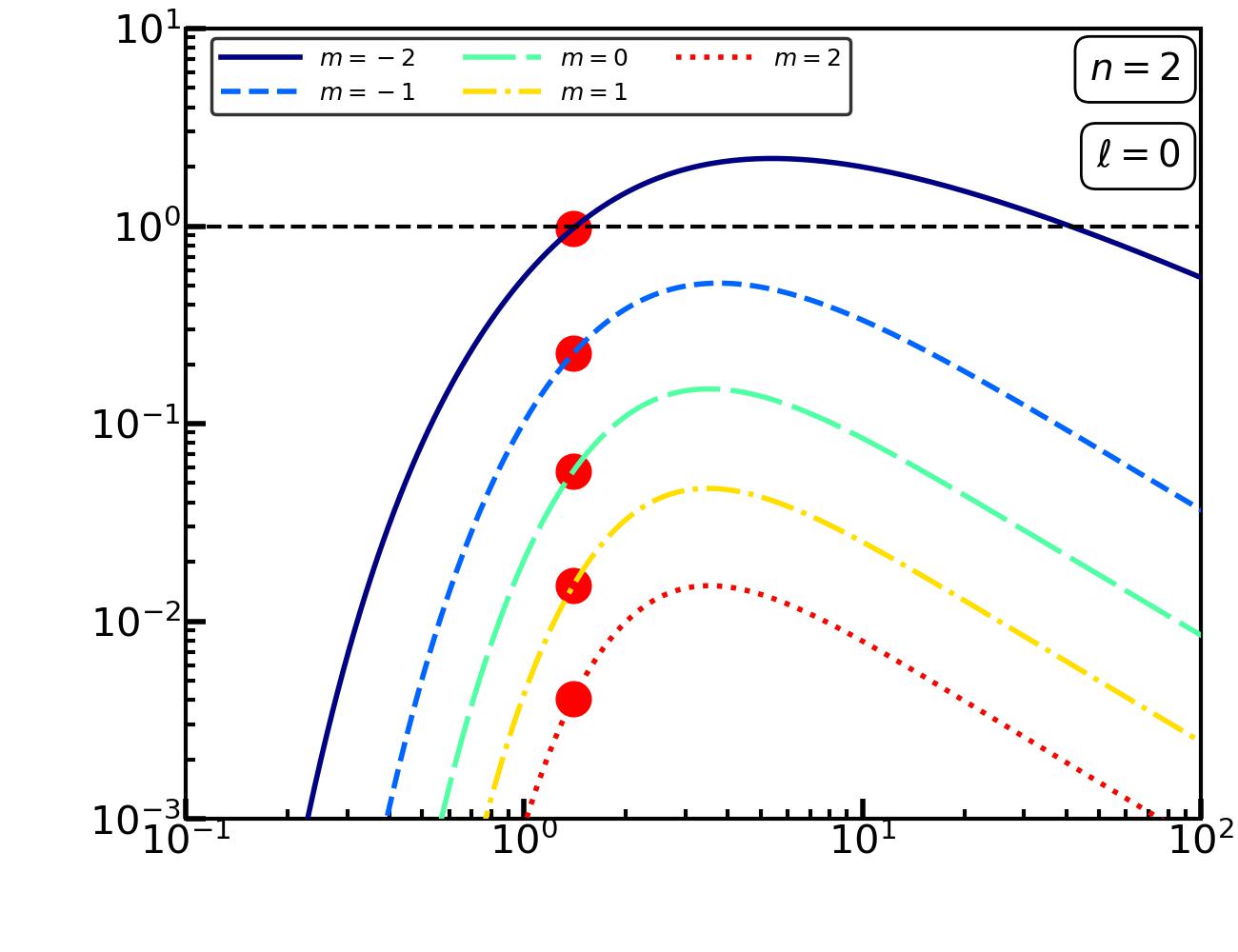}
  \label{disk_Resp_n2_l0}
\end{subfigure}
\begin{subfigure}{0.49\textwidth}
  \hspace{0.5mm}
  %\centering
  \includegraphics[width=1\textwidth]{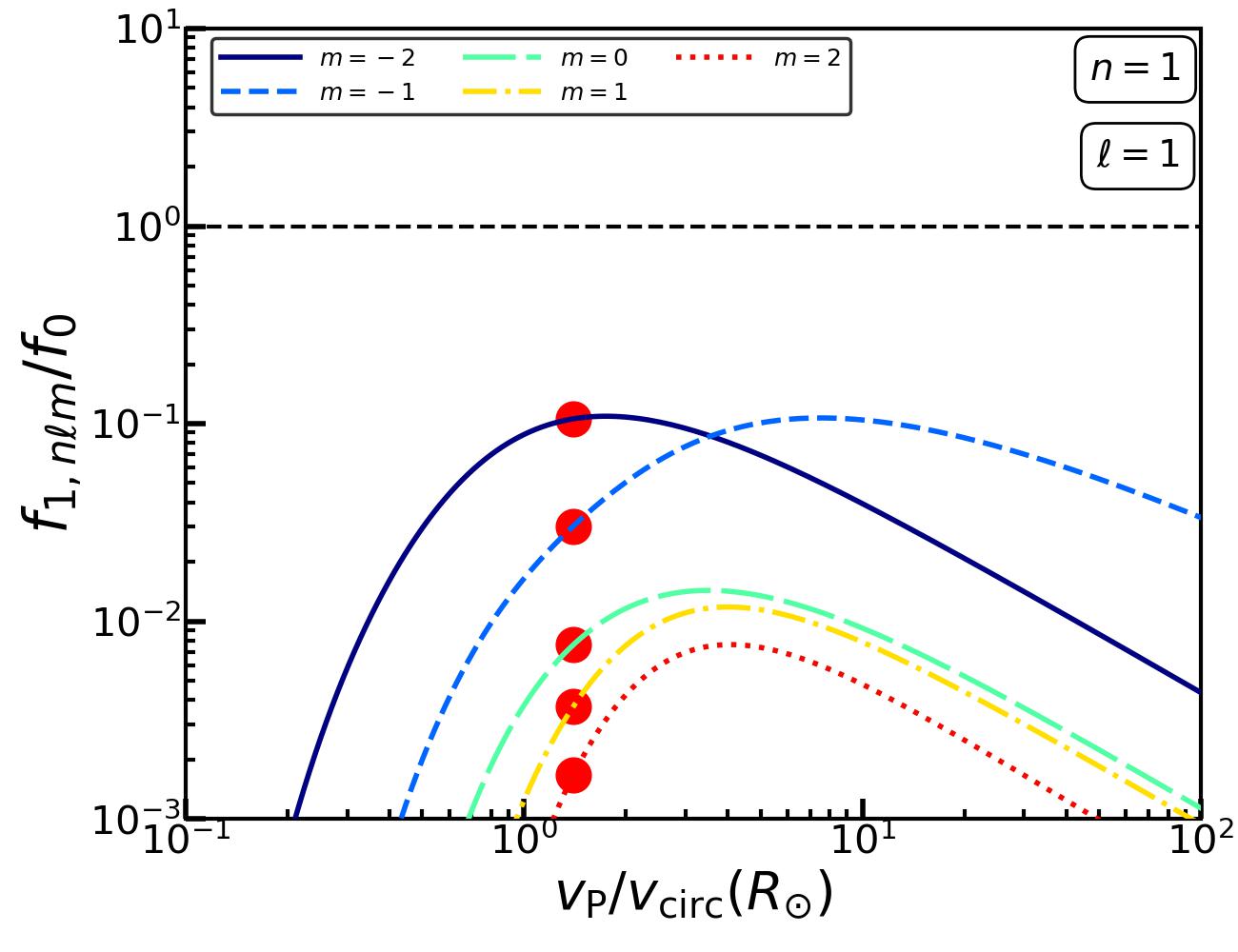}
  \label{disk_Resp_n1_l1}
\end{subfigure}
\begin{subfigure}{0.49\textwidth}
  \hspace{0.5mm}
  %\centering
  \includegraphics[width=1\textwidth]{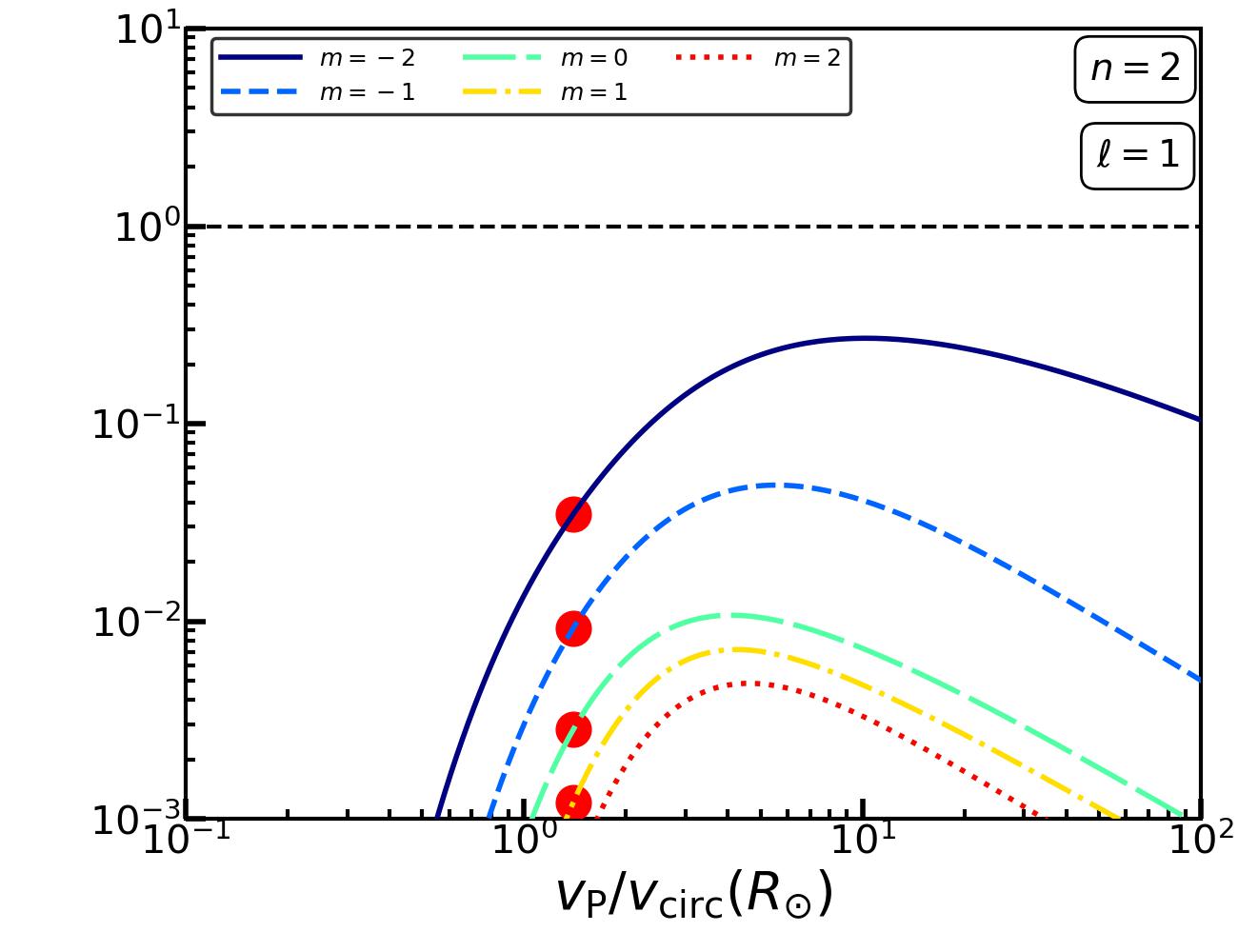}
  \label{disk_Resp_n2_l1}
\end{subfigure}
\caption{Steady state MW disk response to satellite encounter in the collisionless limit: each panel shows the behaviour of the disk response amplitude, $f_{1,nlm}/f_0$ (marginalized over $I_R$), as a function of the impact velocity, $\vp$, in the Solar neighborhood, in presence of an ambient DM halo. Different lines correspond to different $m$ modes as indicated. The top and bottom rows show the response for $l=0$ and $1$ while the left and right columns indicate it for the $n=1$ bending and $n=2$ breathing modes. The fiducial parameters correspond to $I_z=I_{z,\odot}$ and the parameters for Sgr impact, the response amplitudes for which are indicated by the red circles in each panel. The response is dominated by the $(n,l,m)=(1,0,-2)$ mode or the two-armed warp at small $\vp$ and the $(2,0,-2)$ mode or the two-armed spiral at large $\vp$. Typically, the $m=-2$ and $-1$ responses dominate over $m=0,1$ and $2$, while the $l=0$ response is more pronounced than $l=1$.}
\label{fig:disk_Resp_lm}
\end{figure*}

\begin{figure*}[t!]
\centering
\begin{subfigure}{0.49\textwidth}
  \centering
  \includegraphics[width=1\textwidth]{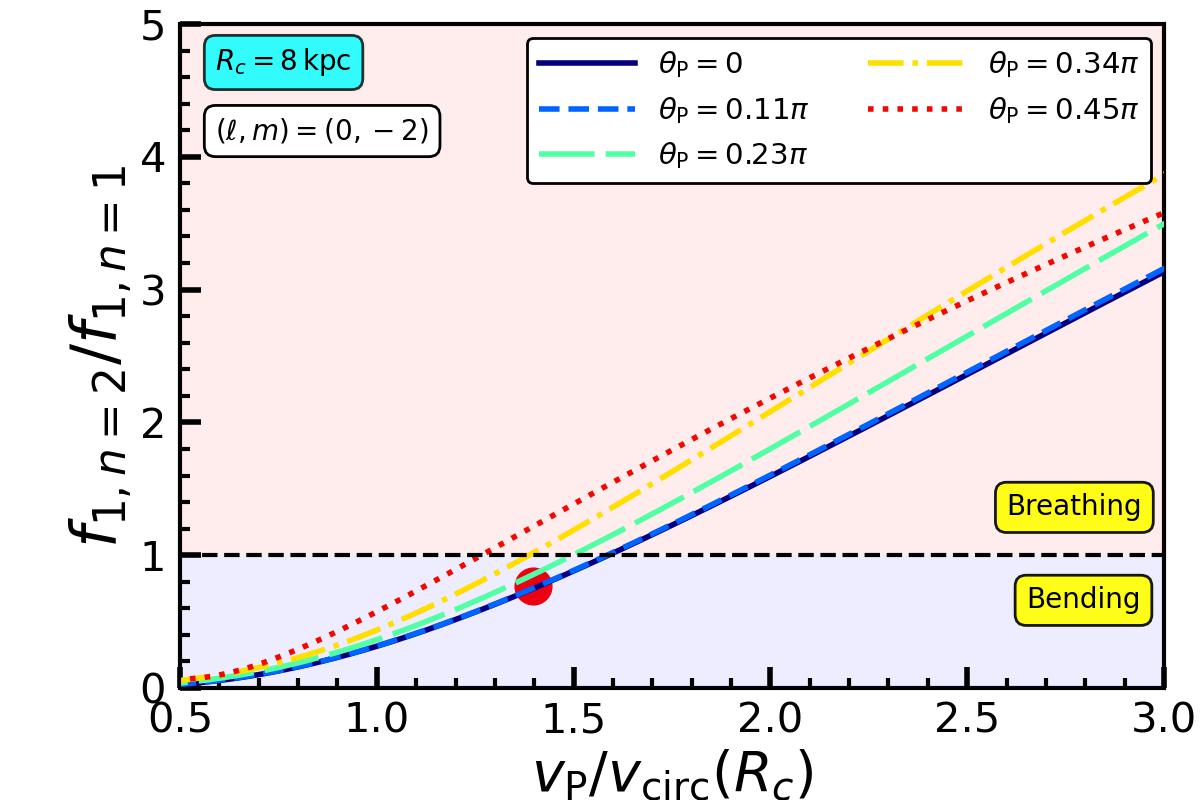}
  \label{disk_Resp_Ratio_thetap}
\end{subfigure}
\begin{subfigure}{0.49\textwidth}
  \hspace{-0.5mm}
  %\centering
  \includegraphics[width=1\textwidth]{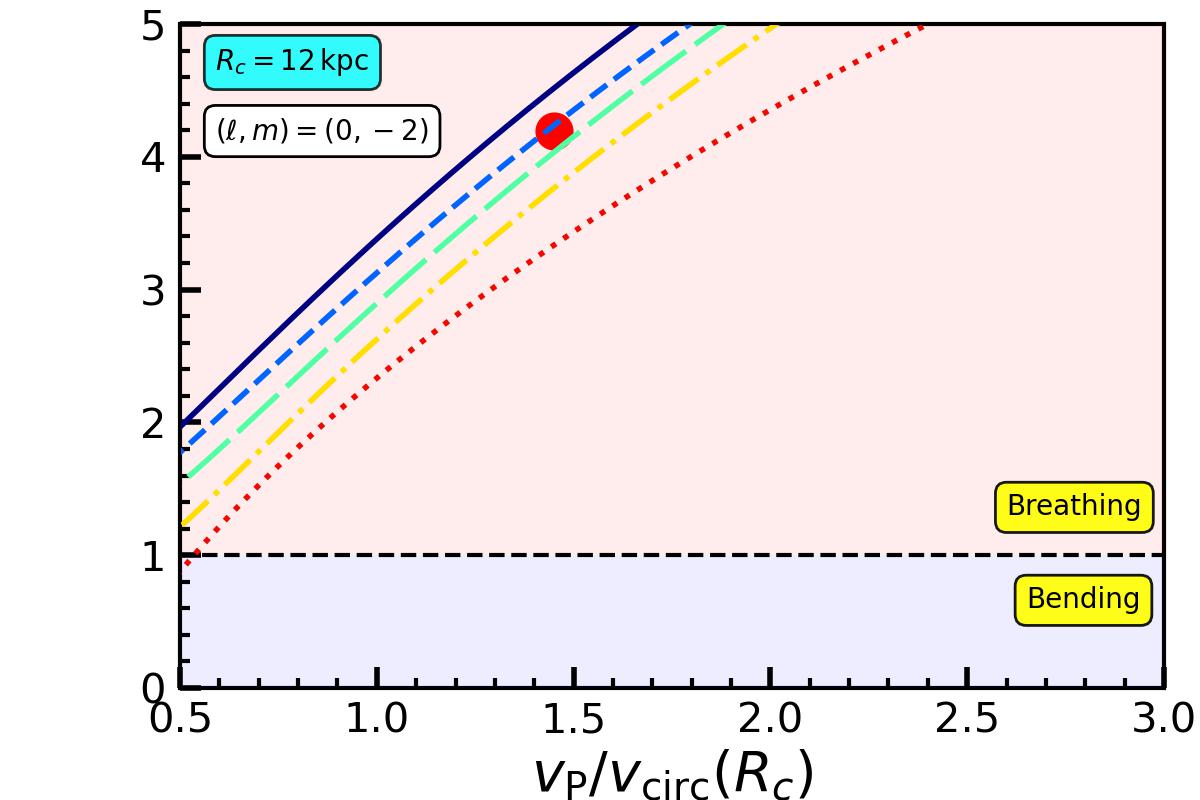}
  \label{disk_Resp_Ratio_phip}
\end{subfigure}
\caption{MW disk response to satellite encounter: breathing-to-bending ratio or the relative strength of the $n=2$ and $n=1$ modes of disk response to a Sgr-like impact is plotted as a function of the impact velocity, $\vp$, at $\Rc=\Rsun=8\kpc$ and $\Rc=1.5\Rsun=12\kpc$ shown in the left and right columns respectively, for the $(l,m)=(0,-2)$ mode which typically dominates the response. Different lines correspond to different values of $\thetap$ as indicated. We consider $I_z=I_{z,\odot}$ and the fiducial parameters to correspond to those for Sgr encounter, for which the breathing-to-bending ratio is denoted by the red circle. Bending modes dominate over breathing modes at small $\vp$ and vice versa at large $\vp$. Breathing modes are relatively more pronounced than bending modes in the outer disk, closer to the Sgr impact radius, $\rd=17\kpc$. More planar (perpendicular) encounters trigger larger breathing-to-bending ratios farther away from (closer to) the impact radius.}
\label{fig:disk_Resp_Ratio}
\end{figure*}

Having computed the MW disk response to its satellites, we now investigate the sensitivity of the response to the various encounter parameters. In Fig.~\ref{fig:disk_Resp_vp} we plot the amplitude of the Solar neighborhood response, $f_{1,nlm}/f_0$ (marginalized over $I_R$), as a function of the impact velocity, $\vp$ (in units of the circular velocity at $\Rc=\Rsun$), for the $(n,l,m)=(1,0,0)$ bending and $(n,l,m)=(2,0,0)$ breathing modes, shown in the left and right columns respectively. The top, middle and bottom rows show the results for varying $I_z$, $\thetap$ and $\phip$ respectively, assuming the fiducial parameters to be those for Sgr (mass $M_\rmP=10^9 \Msun$, scale radius $\varepsilon=1.6$ kpc) during its penultimate disk crossing (most relevant for the Gaia snail), i.e., impact radius $\rd=17\kpc$, impact velocity $\vp=340$ km/s, and angles of impact, $\thetap={21}^\circ$ and $\phip={150}^\circ$. In Fig.~\ref{fig:disk_Resp_lm} we plot the bending and breathing mode response amplitudes (in the Solar neighborhood) as a function of $\vp$ for different $(l,m)$ modes, with the fiducial parameters again corresponding to Sgr. The left and right columns respectively indicate the $n=1$ bending and $n=2$ breathing modes, while the top and bottom rows correspond to $l=1$ and $l=2$ respectively. The different lines in each panel denote the responses for $m=-2,-1,0,1$ and $2$. Fig.~\ref{fig:disk_Resp_Ratio} shows the ratio of the bending and breathing response amplitudes as a function of $\vp$ for the dominant mode $(l,m)=(0,-2)$. Different lines indicate breathing-to-bending ratios for different values of $\thetap$, while the left and right columns respectively indicate the ratios observed at $\Rc=8$ and $12\kpc$.

From Figs.~\ref{fig:disk_Resp_vp} and \ref{fig:disk_Resp_lm} it is evident that, as shown in equation~(\ref{sat_asymptote}), the disk response is suppressed like a power law ($\sim v^{-1}_\rmP$) in the high velocity/impulsive limit and exponentially ($\sim \exp{\left[-\Omega b/\vp\right]}$) suppressed in the low velocity/adiabatic limit. The response is the strongest for intermediate velocities, $\vp\sim 2-3\, v_{\rm circ}(\Rsun)$, where the time periods of the vertical, radial and azimuthal oscillations of the stars are nearly commensurate with the encounter timescale, $\sqrt{b^2+\varepsilon^2}/\vp$. The $v^{-1}_\rmP$ and $K_{0i}$ factors in equation~(\ref{sat_gen}) conspire to provide the near-resonance condition for maximum response,

\begin{align}\label{rescond}
n\Omega_z + l\kappa + m\Omega_\phi \approx \frac{0.6\,\vp}{\sqrt{b^2 + \varepsilon^2}},
\end{align}
where $b$ is the impact parameter of the encounter, given by equation~(\ref{impact_parameter}). From the top panels of Fig.~\ref{fig:disk_Resp_vp}, it is clear that the peak response shifts to smaller $\vp$ with increasing $I_z$. This is easy to understand from the fact that the corresponding vertical frequency, $\Omega_z$, decreases with increasing $I_z$, making the encounter more impulsive for larger actions. The middle and bottom panels show that the response depends strongly on the polar angle of the encounter, $\thetap$, but very mildly on the azimuthal angle, $\phip$. Moreover, the middle panels indicate that more planar encounters (larger $\thetap$) induce stronger responses. 

The in-plane structure of the disk response depends on the relative contribution of the different $(l,m)$ modes. From Fig.~\ref{fig:disk_Resp_lm} it is evident that a typical Sgr-like encounter predominantly excites $(l,m)=(0,-1)$ and $(l,m)=(0,-2)$ in the Solar neighborhood. The dominant mode for slower encounters is $(n,l,m)=(1,0,-2)$ while that for faster ones is $(n,l,m)=(2,0,-2)$. Since $f_{1,nlm}/f_0\gtrsim 1$ in these cases, the response to the impact by Sgr is in fact non-linear in the Solar neighborhood. Either way, a satellite encounter is typically found to excite strong $m=-2$ modes, i.e., $2$-armed warps ($n=1$) and spirals ($n=2$). This is due to a quadrupolar tidal distortion of the disk by the satellite, which manifests as a stretching of the disk in the direction of the impact and a compression perpendicular to it.

Fig.~\ref{fig:disk_Resp_Ratio} elucidates that the bending mode response dominates for slower encounters, i.e., smaller $\vp$, and at guiding radii far from the impact radius, $\rd$. More planar impacts trigger larger breathing-to-bending ratios farther away from the impact radius while this trend reverses closer to it. This is because more planar encounters cause more vertically symmetric perturbations farther away from the impact radius. The predominance of bending modes for low $\vp$ encounters while that of breathing modes for high $\vp$ ones has been observed by \cite{Widrow.etal.14} and \cite{Hunt.etal.21} in their N-body simulations of satellite-disk encounters. As demonstrated by \cite{Widrow.etal.14}, slower encounters provide energy to the stars near one of the vertical turning points while drain energy from those near the other turning point, thereby driving bending wave perturbations that are asymmetric about the mid-plane. On the other hand, fast satellite passages are impulsive and impart energy to the stars near both the turning points, thus triggering symmetric breathing waves. 

The predominance of breathing (bending) modes closer to (farther away from) the impact radius is qualitatively similar to the observation by \cite{Hunt.etal.21} in their simulations of MW-Sgr encounter that the outer part of the MW disk which is closer to the impact radius shows a preponderance of two-armed phase-spirals or breathing modes. This can be understood within the framework of our formalism by noting that the impact parameter, $b$, and therefore the encounter timescale $\sim \sqrt{b^2+\varepsilon^2}/\vp$ decreases with increasing proximity to the point of impact; hence the impact is faster than the vertical oscillations of stars near the point of impact, driving stronger breathing mode perturbations. However, contrary to these predictions for the MW-Sgr encounter, \cite{Hunt.etal.22}, using Gaia DR3 data, revealed two-armed phase-spirals, and therefore breathing modes, in the inner disk ($\Rc \sim 6-7\kpc$). Our analysis suggests that none of the MW satellites could have caused this. Using N-body simulations of an isolated MW system, \cite{Hunt.etal.22} suggested that a transient spiral arm or bar could be a potential trigger for breathing modes in the inner disk. However, such a transient perturbation would have to be sufficiently impulsive, i.e., occur over a timescale that is comparable to or smaller than the vertical oscillation timescale in the inner disk (see section~\ref{sec:spiral_cless}), in order to produce two-armed phase-spirals with density contrast as strong as in the data. Such short timescales are unlikely to arise from the secular evolution of the disk alone and may instead require forcing of the inner disk by perturbations in the MW halo. Another possible trigger of this feature is the recent passage of dark satellite(s) through the inner disk. The true origin of this feature is however unclear. Hence, we conclude that the presence of two-armed phase-spirals in the inner disk is rather unexpected, and that its origin poses an intriguing conundrum.

\section{phase-spirals and the Galactic potential}
\label{sec:pot_const}

Thus far we mainly focused on how the nature of the perturbation dictates the vertical (i.e., bending and breathing modes) as well as the in-plane (various $(l,m)$ modes) structure of the disk response. However, the detailed structure, in particular the winding, of the phase-spiral not only depends on the triggering agent but also holds crucial information about the underlying potential in which the stars move, and can thus be used to constrain the potential of the combined disk plus halo system \citep[see also][]{Widmark.etal.22a,Widmark.etal.22b}.

The winding of the vertical phase-spiral can be characterized by the pitch-angle, $\phi_\rmI$, along the ridge of maximum density. It is defined as the angle between the azimuthal direction and the tangent to the line of constant density \citep[][]{Binney.Tremaine.87}. It is related to the local dependence of the vertical frequency on the vertical action according to:
\begin{align}
\phi_\rmI = \cot^{-1}{\left[\left|I_z \frac{\rmd \Omega_z }{\rmd I_z}\right| t\right]} = \cot^{-1}{\left[\left|\frac{\rmd \Omega_z }{\rmd \ln{I_z}}\right| t\right]}.
\label{pitch_angle}
\end{align}
Following a perturbation, the pitch angle decreases with time, asymptoting towards zero, as the spiral winds up as a consequence of the ongoing phase mixing. Based on the above expression for $\phi_\rmI$, we can define the following timescale of phase mixing:
\begin{align}
\tau_\phi = \left|\frac{\rmd \ln{I_z}}{\rmd \Omega_z}\right|\,.
\label{phase_mix_timescale}
\end{align}
This timescale, which determines the rate of winding of the spiral, is a function of both the guiding radius, $\Rc$, and the action, $I_z$, and is ultimately dictated by the (unperturbed) potential of the disk+halo system, which sets $\rmd \Omega_z/\rmd I_z$. Hence, the detailed shape of the phase-spiral at a given location in the disk is sensitive to the local disk+halo potential, thereby opening up interesting avenues for constraining the detailed potential of the MW by examining phase-spirals throughout the disk.

\begin{figure}
\centering
\begin{subfigure}{0.32\textwidth}
\centering
\includegraphics[width=1\textwidth]{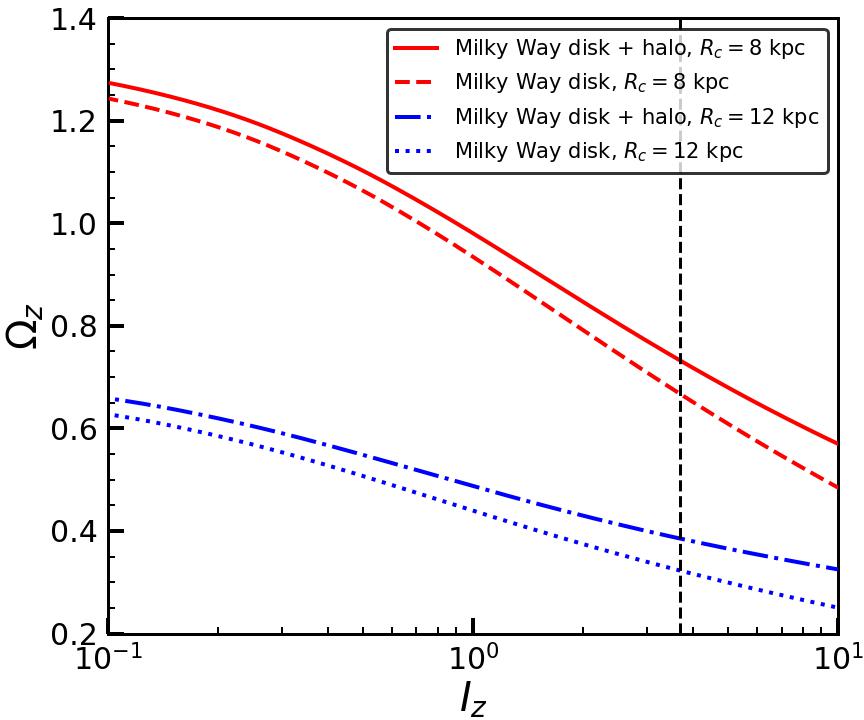}
\label{freq_Iz}
\end{subfigure}
\begin{subfigure}{0.32\textwidth}
\centering
\includegraphics[width=1\textwidth]{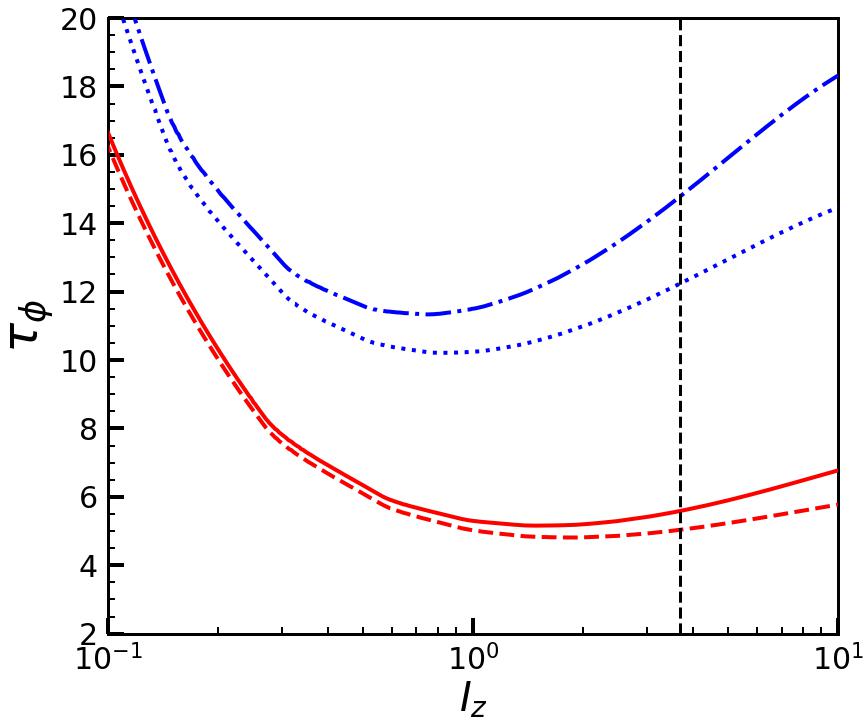}
\label{freq_Iz}
\end{subfigure}
\begin{subfigure}{0.32\textwidth}
\centering
\includegraphics[width=1\textwidth]{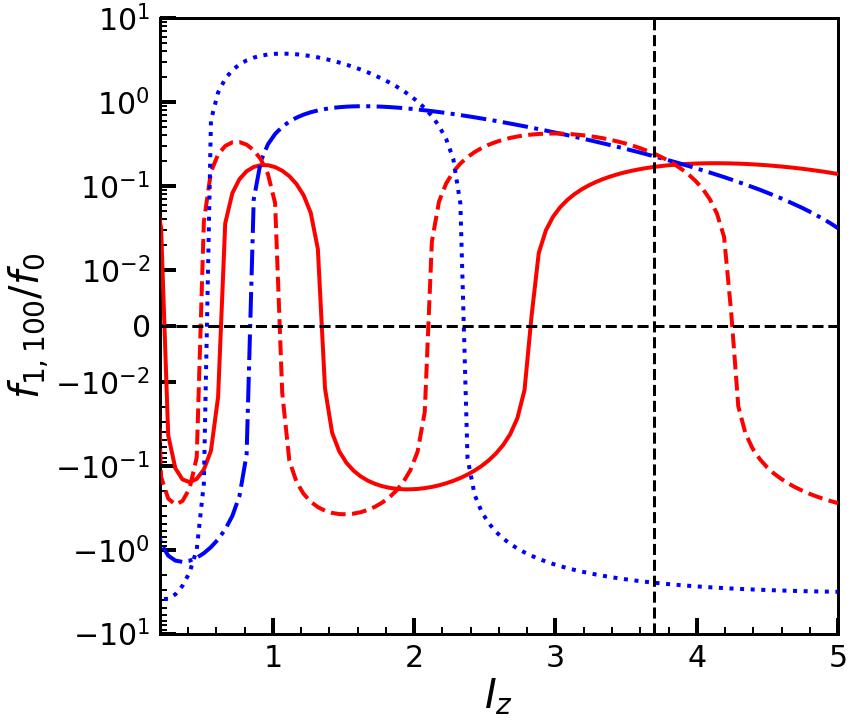}
\label{f1_Iz}
\end{subfigure}
\caption{Impact of DM halo on vertical phase mixing: the panels from left to right respectively indicate the vertical frequency, $\Omega_z$ (units of $\sigma_{z,\odot}/h_z$), the vertical phase mixing timescale, $\tau_\phi$ (given by equation~[\ref{phase_mix_timescale}]), and the $w_z=0$ cuts of the phase-spirals shown in Fig.~\ref{fig:snailn1} as a function of the vertical action, $I_z$ (units of $h_z\sigma_{z,\odot}$). The solid and dashed red lines denote the cases with and without a halo for $\Rc=\Rsun=8\kpc$ while the dot-dashed and dotted blue lines show the same for $\Rc=12\kpc$. The vertical dashed line indicates roughly the maximum $I_z$ for which a phase-spiral is discernible in the Gaia data. Note that phase mixing occurs the fastest for $I_z\sim1$ and that the inner disk phase mixes faster than the outer disk. Also note that the presence of a DM halo increases $\Omega_z$ as well as $\tau_\phi$, leading to slower phase mixing and therefore slower wrapping of the phase-spiral. This effect is more pronounced in the outer disk.}
\label{fig:phase_mix}
\end{figure}
\begin{figure*}
  \centering
  \begin{subfigure}{0.5\textwidth}
    \centering
    \includegraphics[width=1\textwidth]{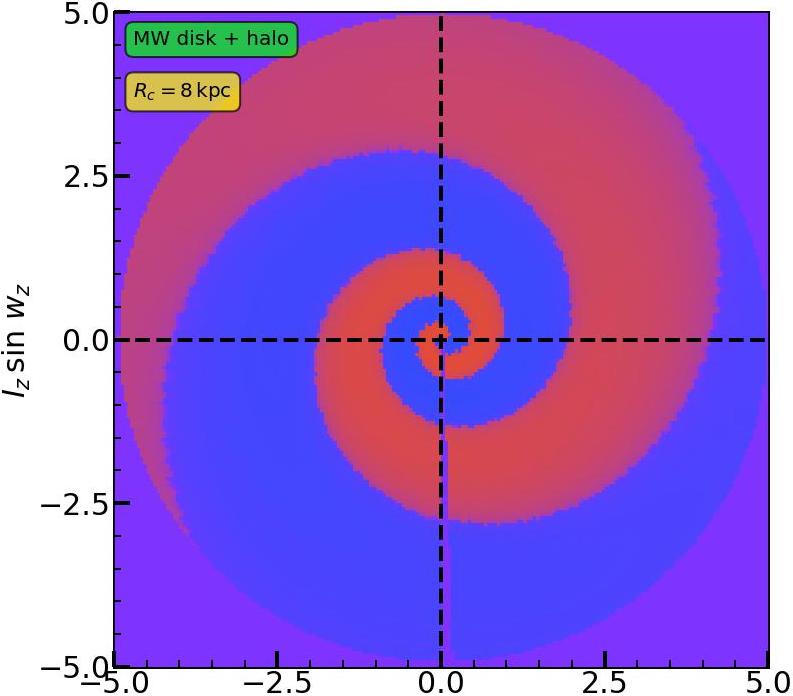}
    \label{snailn1a}
  \end{subfigure}
  \begin{subfigure}{0.475\textwidth}
    \centering
    \includegraphics[width=1\textwidth]{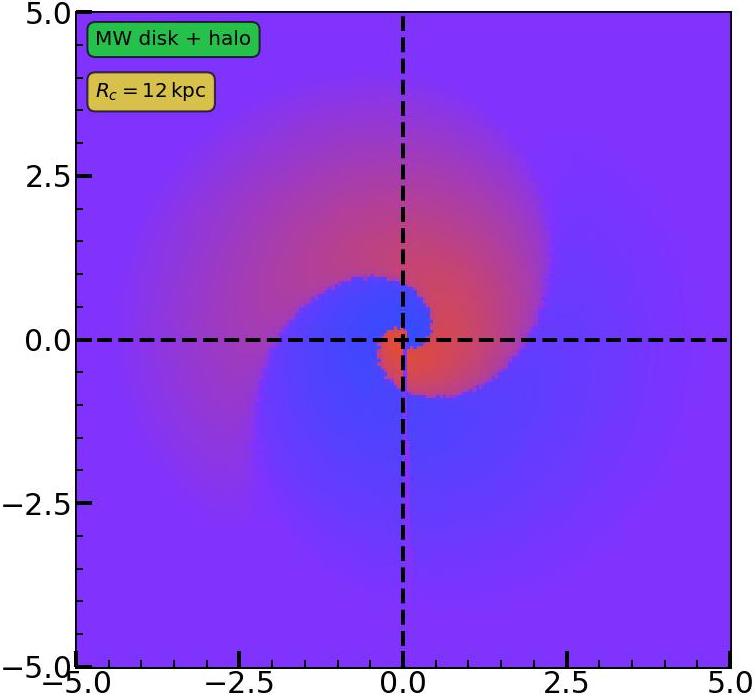}
    \label{snailn1b}
  \end{subfigure}
  \\
  \begin{subfigure}{0.5\textwidth}
    \hspace{0.008cm}
    %\centering
    \includegraphics[width=1\textwidth]{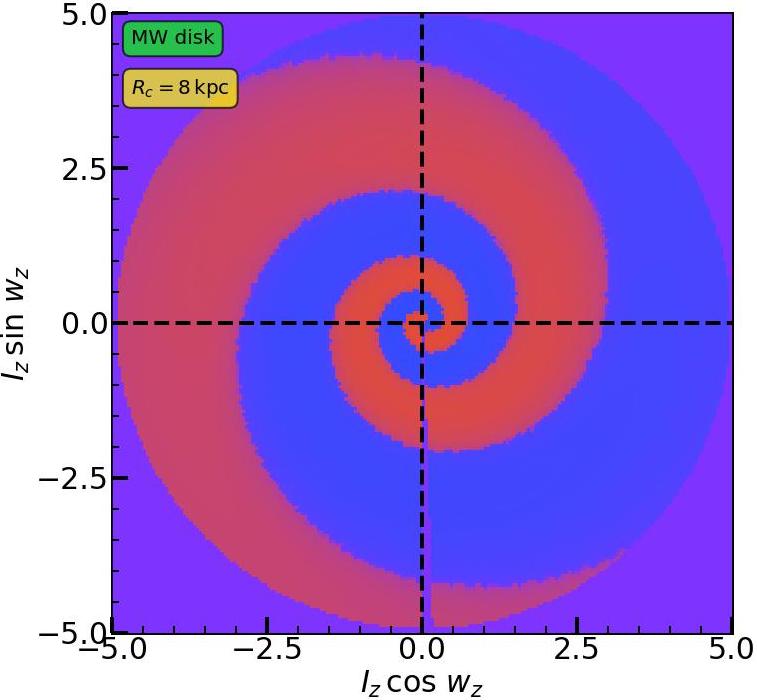}
    \label{snailn1a}
  \end{subfigure}
  \begin{subfigure}{0.476\textwidth}
    \hspace{0.018cm}
    %\centering
    \includegraphics[width=1\textwidth]{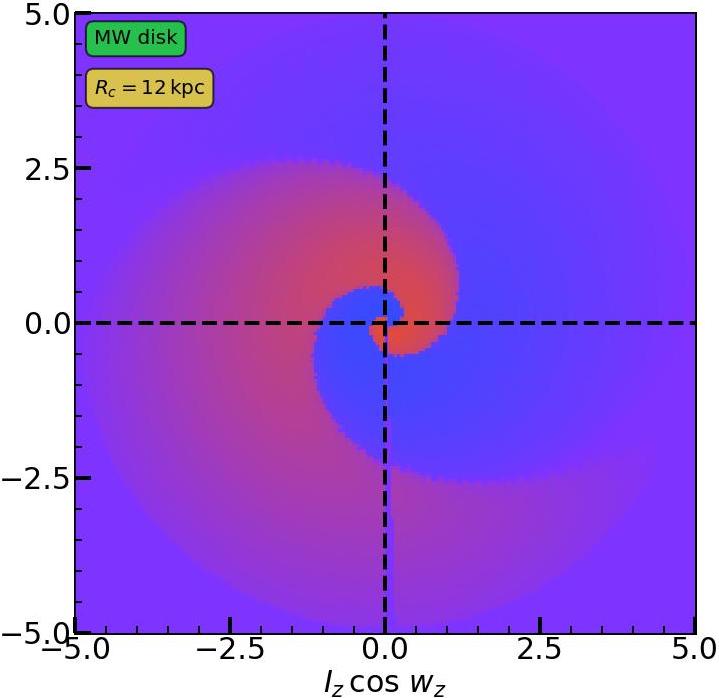}
    \label{snailn1b}
  \end{subfigure}
  \caption{Vertical phase mixing: one-armed phase-spiral corresponding to $n=1$ bending mode excited by the encounter with Sgr for MW disk+halo and MW disk models (rows) at $\Rc=8\kpc$ and $12\kpc$ (columns). The presence of DM halo slows down the rate of phase mixing, leading to more loosely wrapped phase-spirals. Phase mixing occurs more rapidly in the inner disk than in the outer disk.}
  \label{fig:snailn1}
\end{figure*}

The left panel of Fig.~\ref{fig:phase_mix} plots the vertical frequency, $\Omega_z$, as a function of the logarithm of the action, $I_z$, for the MW potential with and without the halo and at guiding radii, $\Rc=8$ (red) and $12\kpc$ (blue). The middle panel shows the behaviour of the corresponding phase mixing timescale, $\tau_\phi$, as a function of $I_z$. Fig.~\ref{fig:snailn1} shows the $(n,l,m)=(1,0,0)$ phase-spirals $400\Myr$ after the penultimate disk crossing of Sagittarius, color coded by the MW disk response, $f_{1,100}$, with blue (red) indicating higher (lower) phase-space density. Results for the cases with and without the halo are shown in rows and for $R_c=8$ and $12\kpc$ are shown in columns. Finally, the right panel of Fig.~\ref{fig:phase_mix} shows the $w_z=0$ cuts of the normalized response, $f_{1,100}/f_0$, as a function of $I_z$, for the four different phase-spirals shown in Fig.~\ref{fig:snailn1}. The vertical frequency, $\Omega_z$, is a decreasing function of $\ln{I_z}$ in all cases, indicating that stars with larger actions (i.e., larger vertical excursion amplitudes) oscillate slower. Note that $\left|\rmd\Omega_z/\rmd \ln{I_z}\right|$ is an increasing (decreasing) function of $I_z$ at small (large) $I_z$, reaching a maximum at intermediate $I_z$. Consequently, the phase mixing timescale, $\tau_\phi$, which is the inverse of $\left|\rmd\Omega_z/\rmd \ln{I_z}\right|$, attains its minimum at $I_z/(h_z\sigma_z)\sim 1$. Thus phase mixing occurs the fastest at intermediate actions and slows down at larger actions, causing the spiral to become more loosely wound (larger pitch angle) farther away from its origin. 

The rate of phase mixing is different in the four different cases. Closer to the galactic center where the potential is deeper and steeper, stars have a larger range of $\Omega_z$, or in other words $\Omega_z$ falls off more steeply with $\ln{I_z}$ in the inner disk than in the outskirts. This leads to faster phase mixing and therefore a much more tightly wound phase-spiral in the inner disk (left panels of Fig.~\ref{fig:snailn1}) as opposed to the outer disk (right panels). The difference in the phase mixing rates is also manifest in the $w_z=0$ response shown in the right panel of Fig.~\ref{fig:phase_mix}; note the longer oscillation wavelengths of the blue lines (outer disk) as opposed to the red lines (inner disk). Hence, in agreement with expectations, the inner part of the disk equilibrates much faster than the outer part. 

The presence of a DM halo deepens the potential well and thus boosts the oscillation frequencies. But the halo also steepens the potential such that the range of frequencies is reduced, i.e., $\Omega_z$ falls off more mildly with $\ln{I_z}$ than in the disk only case. This leads to slower phase mixing and therefore more loosely wound phase-spirals in the presence of the halo (upper panels of Fig.~\ref{fig:snailn1}) than in its absence (lower panels), the effect being more pronounced in the outer (right panels) than in the inner (left panels) disk. Equivalently, the $w_z=0$ response in the right panel of Fig.~\ref{fig:phase_mix} shows longer wavelength wiggles in presence of the halo.

The above sensitivity of the phase mixing timescale to the detailed galaxy potential implies that one can use phase-spirals to constrain it. One can unwind the observed phase-spiral by adopting a form for the galactic potential. Only for the correct potential will the spiral be properly unwound, i.e., the pitch-angle, $\phi_\rmI$, go to zero for all $I_z$ (modulo measurement errors) at the same time, $t_0$, in the past. This $t_0$ then corresponds to the time elapsed since the maximum strength of the perturbation that triggered the phase-spiral. However, this method to constrain the total potential (disk plus dark matter) of the MW relies on the assumption of a single, impulsive perturbation as the trigger. In reality, the phase-spirals may have been impacted by multiple, overlapping perturbations and/or by large-scale temporal fluctuations in the overall potential, which would severely hamper this technique \citep[][]{Tremaine.etal.22}. We intend to investigate the promise of phase-spirals as probes of the galactic potential for different kinds of perturbation in future work.

\section{Conclusion}
\label{sec:concl}

In this paper, we have developed a linear perturbative formalism to analyze the response of a realistic disk galaxy (characterized by a pseudo-isothermal DF) embedded in an ambient spherical DM halo (modelled by an NFW profile) to perturbations of diverse spatiotemporal nature: bars, spiral arms, and encounters with satellite galaxies. Adopting the radial epicyclic approximation, we perturb the FPE up to linear order (in action-angle space) in presence of a perturbing potential, $\Phi_\rmP$, to compute the post-perturbation linear response in the DF, $f_1$. Without self-gravity to reinforce the response, the oscillations in the response phase mix away due to an intrinsic spread in the frequencies of stars, giving rise to spiral features in the phase-space distribution known as phase-spirals. Depending on the timescale of $\Phi_\rmP$, different modes of disk oscillation, corresponding to different phase-spiral structures, are excited. We summarize our conclusions as follows:

\begin{itemize}
    \item Following an impulsive perturbation, the $(n,l,m)$ mode of the disk response consists of stars oscillating with frequencies, $n\Omega_z$, $l\Omega_r\approx l\kappa$ and $m\Omega_\phi$, along vertical, radial and azimuthal directions respectively. Since the frequencies depend on the actions, primarily on the vertical action $I_z$ and the angular momentum $L_z$, the response phase mixes away, spawning phase-spirals. The dominant modes of vertical oscillation are the anti-symmetric bending ($n=1$) and symmetric breathing ($n=2$) modes, which induce initial dipolar and quadrupolar perturbations in the $z-v_z$ or $I_z\cos{w_z}-I_z\sin{w_z}$ phase-space. Over time these features are phase-wrapped into one- and two-armed phase-spirals, respectively, due to the variation of $\Omega_z$ with $I_z$.
    
    \item Since $\Omega_z$ and $\Omega_\phi$ both depend on $L_z$, the amplitude of the $I_z\cos{w_z}-I_z\sin{w_z}$ phase-spiral damps away over time, typically as $\sim 1/t$ (equation~\ref{lateral_mixing_disk}), at a coarse-grained level, i.e., upon marginalization over $L_z$. Therefore, in a realistic disk with ordered motion, lateral mixing causes phase-spirals to damp out much slower than in the isothermal slab with unconstrained lateral velocities discussed in Paper~I, where it occurs like a Gaussian in time.
    
    \item Collisional diffusion due to scatterings of stars by GMCs, DM substructure, etc. damps away the disk response to a perturbation, and therefore the phase-spiral amplitude, at a fine-grained level. Typically, the diffusion in actions is much more efficient than that in angles. The action gradients of the response, which predominantly arise from the action dependence of the oscillation frequencies, are erased by collisional diffusion, causing a super-exponential damping of the response over a timescale, $\tau_\rmD^{(\rmI)}$, which is $\sim 0.6-0.7\Gyr$ in the Solar neighborhood. The diffusion timescale is shorter in the inner disk, for stars with smaller $I_z$, and for higher-$n$ modes.
    
    \item The response to a bar or spiral arm with a fixed pattern speed, $\Omega_\rmP$, is dominated by the near-resonant stars ($\Omega_{\rm res}=n\Omega_z+l\kappa+m(\Omega_\phi-\Omega_\rmP) \approx 0$), especially in the adiabatic regime (slowly evolving perturber amplitude). Moreover, phase-mixing occurs gradually in the near-resonant parts of phase-space. Most of the strong resonances are confined to the disk-plane, such as the co-rotation ($n=l=0$) and Lindblad ($n=0,l=\pm 1,m=\pm 2$) resonances. For a {\it transient} bar or spiral arm whose amplitude varies over time as $\sim \exp{\left[-\omega^2_0 t^2\right]}$, the response is maximal when $\omega_0 \sim \Omega_{\rm res}$. In the impulsive limit ($\omega_0 \gg \Omega_{\rm res}$), the response is power-law suppressed, while in the adiabatic limit ($\omega_0 \ll \Omega_{\rm res}$) it is suppressed (super)-exponentially. 
    
    \item For a thin disk, since $\Omega_z$ is very different from $\Omega_\phi$ and $\kappa$, the vertical modes ($n\neq 0$) are generally not resonant with the radial and azimuthal ones and thus undergo phase mixing. The strength of a vertical mode primarily depends on the nature of the perturbing potential, most importantly its timescale. Slower pulses trigger mainly bending $(n=1)$ modes, while faster pulses excite more pronounced breathing $(n=2)$ modes. Therefore, a transient bar or spiral arm with amplitude $\sim \exp{\left[-\omega^2_0 t^2\right]}$ triggers a bending (breathing) mode when the pulse-frequency, $\omega_0$, is smaller (larger) than $\Omega_z$. The response to very slow perturbations ($\omega_0 \ll \Omega_z$) is however heavily suppressed (adiabatic shielding).
    
    \item For a persistent bar or spiral arm with a fixed pattern speed, $\Omega_\rmP$, that grows and saturates over time, the response initially develops a phase-spiral. However, this transient response is quickly taken over by coherent oscillations at the driving frequency, $\Omega_\rmP$, which manifest in phase-space as a steadily rotating dipole (quadrupole) for the bending (breathing) mode. Therefore, a transient (pulse-like) perturbation, such as a bar or spiral arm whose amplitude varies over a timescale comparable to the vertical oscillation period, $T_z\sim h_z/\sigma_z$, is essential for the formation of a phase-spiral in $z$-$v_z$ space.
    
    \item The above analysis suggests that if the recently discovered two-armed Gaia phase-spiral (breathing mode) in the inner disk of the MW was indeed induced by a spiral arm/bar as suggested by \citet{Hunt.etal.22} using N-body simulations, the spiral arm/bar was probably a transient one with a predominantly symmetric vertical profile whose amplitude varied over a timescale comparable to the vertical oscillation period. However, it remains to be seen whether such a rapid excitation and decay of a spiral arm/bar perturbation is realistic.

    \item We have computed the response of the MW disk, embedded in an extended DM halo, to disk-crossing perturbations by several of its satellite galaxies. We find that the response in the Solar neighborhood is dominated by the perturbations due to Sgr, followed by those due to the LMC, Hercules and Leo II. This implies that, if the Gaia snail near the Solar radius was indeed triggered by a MW satellite (which is still subject to debate), Sgr is the leading contender \citep[see also][]{Banik.etal.22}. However, if that is the case, then the impact (disk crossing) must have happened within the last $\sim 0.6-0.7\Gyr$ in order for the response to have survived damping due to collisional diffusion.
    
    \item The amplitude of the response (at a fixed guiding radius $\Rc$) to satellite encounters scales as $\vp^{-1}$ in the impulsive (large $\vp$) limit, but is exponentially suppressed in the adiabatic (small $\vp$) limit, a phenomenon known as adiabatic shielding. The resonant modes with $n\Omega_z+l\kappa+m\Omega_\phi=0$ are not suppressed but rather become non-linear in the adiabatic regime. The peak response of a mode (with $n\Omega_z+l\kappa+m\Omega_\phi \neq 0$) is achieved at intermediate velocities for which the encounter frequency is commensurate with the oscillation frequencies of the stars, i.e., the near-resonance condition given by equation~(\ref{rescond}) is satisfied.

    \item The response of a disk to an encounter with a satellite galaxy depends primarily on three parameters:  (i) impact velocity $\vp$, (ii) polar angle of impact $\thetap$, and (iii) position on the disk relative to the point of impact where the satellite crossed the disk. Slower (faster) encounters excite predominantly $n=1$ bending ($n=2$ breathing) modes. More planar encounters (those with larger $\thetap$) typically result in larger breathing-to-bending ratios farther away from the impact radius while this trend gets reversed closer to it. In general, breathing modes dominate over bending modes closer to the point of impact, in agreement with $N$-body simulations of the MW-Sgr encounter \citep[][]{Hunt.etal.21}. Since the impact velocities of the MW satellites are all fairly similar to the local circular velocity, the decisive factor for breathing {\it vs.} bending modes is not so much the impact velocity, but rather the distance from the point of impact.
    
    \item The $(n,m)=(1,-2)$ and $(-1,2)$ modes generally dominate the response for slower satellite encounters, e.g., that of Sgr with respect to the Solar neighborhood, due to the tidal distortion of the disk by the satellite. The in-plane spatial structure of the disk response therefore generally resembles a two-armed warp ($n=1$) or spiral ($n=2$).

    \item The presence of an extended DM halo causes phase mixing to occur slower, and modifies the structural appearance of the phase-spirals (i.e., the pitch angle as function of vertical action). Hence, provided that the phase-spiral was triggered by a single, impulsive perturbation, the detailed shape of the phase-spiral can in principle be used to constrain not only the time elapsed since the perturbation \citep[][]{Darragh-Ford.etal.23} but also the total (disk$+$halo) potential. If the phase-spiral has been triggered and/or impacted by multiple, overlapping perturbations the situation is less clear. In future work we intend to investigate the constraining power of phase-spirals for different kinds of perturbation.
\end{itemize}

This paper focused on the analysis of the phase mixing component (phase-spirals) of the `direct' disk response to various perturbations such as bars, spiral arms and satellite galaxies. However this leaves out some other potentially important features of the disk response. First of all, we considered the ambient DM halo to be non-responsive. In reality, the DM halo will also be perturbed, for example by an impacting satellite, and this halo response, which can be enhanced by self-gravity, can indirectly perturb the disk. A preliminary, perturbative analysis based on the $N$-body simulation of the MW-Sgr encounter by \cite{Hunt.etal.21} suggests that the indirect disk response to halo perturbations (triggered by Sgr) is comparable, but sub-dominant, to the direct response to Sgr. However, a more detailed analysis is warranted, which we leave for future work. Secondly, we have neglected the self-gravity of the disk response. As discussed in Paper~I, the dominant effect of self-gravity is to cause coherent point mode oscillations \citep[][]{Mathur.90, Weinberg.91, Darling.Widrow.19a} of the disk, which in linear theory are decoupled from the phase-spirals. However, self-gravity can enhance the amplitude of the phase-spiral. Although recent developments \citep[][]{Dootson.Magorrian.22} have shed some light on the self-gravitating response of razor-thin disks to bar perturbations, a more generic theoretical description of the self-gravitating response of inhomogeneous, thick disks to general perturbations (bars, spiral arms, satellite galaxies, etc) is still lacking. We hope to include the effects of self-gravity on disk perturbations in future work.

\section*{Acknowledgments}

The authors are grateful to Kathryn Johnston, Jason Hunt, Adrian Price-Whelan, Chris Hamilton, Elise Darragh-Ford, James Binney, John Magorrian and Elena D'Onghia for insightful discussions and valuable suggestions. FCvdB is supported by the National Aeronautics and Space Administration through Grant No. 19-ATP19-0059 issued as part of the Astrophysics Theory Program. MDW is supported by the National Science Foundation through Grant No. AST-1812689.

%%%%%%%%%%%%%%%%%
% Bibliography
%%%%%%%%%%%%%%%%%

\bibliography{references_banik}{}
\bibliographystyle{aasjournal}

\appendix

\section{The unperturbed galaxy}
\label{sec:disk_model}

Under the radial epicyclic approximation (small $I_R$), the unperturbed DF, $f_0$, for a rotating MW-like disk galaxy can be well approximated as a pseudo-isothermal DF, i.e., written as a nearly isothermal separable function of the azimuthal, radial and vertical actions. Following \cite{Binney.10}, we write
\begin{align}
f_0 = \frac{1}{\pi} {\left(\frac{\Omega_\phi \Sigma}{\kappa\, \sigma^2_R}\right)}_{R_\rmc} \left(1+\tanh{\frac{L_z}{L_0}}\right)\times \exp{\left[-\frac{\kappa I_R}{\sigma^2_R}\right]} \times \frac{1}{\sqrt{2\pi}h_z\sigma_z} \exp{\left[-\frac{E_z(I_z)}{\sigma^2_z}\right]}\,.
\end{align}
The vertical structure of this disk is isothermal, while the radial profile is pseudo-isothermal. Here $\Sigma=\Sigma(R)$ is the surface density of the disk, $L_z$ is the $z$-component of the angular momentum, which is equal to $I_\phi$, $R_\rmc = R_\rmc(L_z)$ is the guiding radius, $\Omega_\phi$ is the circular frequency, and $\kappa = \kappa(\Rc) = \lim_{I_R \to 0}{\Omega_R}$ is the radial epicyclic frequency \citep[][]{Binney.Tremaine.87}. If $L_0$ is sufficiently small, then we can further approximate the above form for $f_0$ as
\begin{align}
f_0 \approx \frac{\sqrt{2}}{\pi^{3/2} \, \sigma_z h_z} {\left(\frac{\Omega_\phi \Sigma}{\kappa\, \sigma^2_R}\right)}_{\Rc} \, \exp{\left[-\frac{\kappa I_R}{\sigma^2_R}\right]} \, \exp{\left[-\frac{E_z(I_z)}{\sigma^2_z}\right]} \, \Theta(L_z)\,,
\label{DF_MW_app}
\end{align}
where $\Theta(x)$ is the Heaviside step function. Thus we assume that the entire galaxy is composed of prograde stars with $L_z>0$. 

The corresponding density profile can be written as a product of an exponential radial profile and an isothermal ($\sech^2$) vertical profile, i.e.,
\begin{align}
\rho(R,z) = \rho_\rmc \, \exp{\left[-\frac{R}{h_R}\right]} \, \sech^2{\left(\frac{z}{h_z}\right)},
\label{disk_exp_iso_rho_app}
\end{align}
where $h_R$ and $h_z$ are the radial and vertical scale heights, respectively. Throughout we adopt the thin disk limit, i.e., $h_z \ll h_R$. The surface density profile is given by
\begin{align}
\Sigma(R) = \int_{-\infty}^{\infty} \rmd z\, \rho(R,z) = \Sigma_\rmc \exp{\left[-\frac{R}{h_R}\right]},
\label{disk_exp_iso_Sigma_app}
\end{align}
where $\Sigma_\rmc=\rho_\rmc h_z$ is the central surface density of the disk. We assume a radially varying vertical velocity dispersion, $\sigma_z$, satisfying $\sigma^2_z(R) = 2\pi G h_z \Sigma(R)$ \citep[][]{Binney.Tremaine.08}. We assume a similar profile for $\sigma^2_R$ such that the ratio, $\sigma_R/\sigma_z$ is constant throughout the disk \citep[][]{Binney.10} and equal to the value at the Solar vicinity.

Throughout this paper, for the ease of computation of the frequencies (because of a simple analytic form of the potential), we approximate the above density profile by a combination of three \cite{Miyamoto.Nagai.75} disk profiles \citep[][]{Smith.etal.15}, i.e., the 3MN profile as implemented in the {\tt Gala} Python package \citep[][]{gala, gala_code_adrian}. The corresponding disk potential is given by
\begin{align}
\Phi_\rmd(R,z) = -\sum_{i=1}^{3} \frac{G M_i}{\sqrt{R^2+{\left(a_i+\sqrt{z^2+b^2_i}\right)}^2}},
\label{Phi0_disk_app}
\end{align}
where $M_i$, $a_i$ and $b_i$, with $i=1,2,3$, are the mass, scale radius and scale height corresponding to each of the MN profiles.

The MW disk is believed to be embedded in a much more extended DM halo, which we model using a spherical NFW \citep[][]{Navarro.etal.97} profile with potential

\begin{align}
\Phi_\rmh(R,z)=-\frac{G \Mvir}{\Rvir} \, \frac{c}{f(c)} \,
\frac{\ln(1+r/r_\rms)}{r/r_\rms}.
\label{Phi0_halo_app}
\end{align}
Here $\Mvir$ is the virial mass of the halo, $r_s$ is the scale radius, $c=R_{\rm vir}/r_s$ is the concentration ($R_{\rm vir}$ is the virial radius), and $f(c)=\ln{\left(1+c\right)}-c/(1+c)$. The combined potential experienced by the disk stars is thus given by
\begin{align}
\Phi_0(R,z)=\Phi_\rmd(R,z)+\Phi_\rmh(R,z).
\end{align}

\section{Fourier coefficients of spiral arm or bar perturbing potential}
\label{App:fourier_spiral}

An essential ingredient of the disk response to spiral arm or bar perturbations is the Fourier component of the perturber potential, $\Phi_{nlm}$. This can be computed as follows:
\begin{align}
\Phi_{nlm}(\bI,t) &= \frac{1}{{\left(2\pi\right)}^3} \int_0^{2\pi} \rmd w_z \int_0^{2\pi} \rmd w_R \int_0^{2\pi} \rmd w_\phi\, \exp{\left[-i(n w_z + l w_R + m w_\phi)\right]}\, \Phi_\rmP \left(\br,t\right).
\end{align}
To evaluate this first we need to calculate $\br=(z,R,\phi)$ as a function of $\left(\bw,\bI\right)=(w_z,w_\phi,w_R,\bI)$ where $I_\phi=L_z$, the angular momentum. Under the epicyclic approximation, $R$ can be expressed as a sum of the guiding radius and an oscillating epicyclic term, i.e.,

\begin{align}
R\approx \Rc(L_z)+\sqrt{\frac{2 I_R}{\kappa}}\sin{w_R},
\label{R_epi_app}
\end{align}
and the azimuthal angle, $w_\phi$, is given by

\begin{align}
w_\phi \approx \phi - \frac{2\,\Omega_\phi}{\Rc \kappa} \sqrt{\frac{2 I_R}{\kappa}}\cos{\theta_R}.
\label{wphi_phi_app}
\end{align}

The vertical distance $z$ from the mid-plane is related to $\Rc(L_z)$ and $\left(w_z,I_z\right)$, according to
\begin{align}
w_z=\Omega_z(\Rc,I_z)\int_0^z \frac{\rmd z'}{\sqrt{2\left[E_z(\Rc,I_z)-\Phi_z(\Rc,z')\right]}},
\label{z_wz_Iz_app}
\end{align}
where $\Omega_z(\Rc,I_z)=2\pi/T_z(\Rc,I_z)$, with $T_z(\Rc,I_z)$ given by Equation~(\ref{T_z}). The above equation can be numerically inverted to obtain $z(\Rc,w_z,I_z)$. 

Upon substituting the above expressions for $R$, $\phi$ and $z$ in terms of $(\bw,\bI)$ in the expression for $\Phi_\rmP$ given in equation~(\ref{Phip_spiral}), we obtain

\begin{align}
\Phi_{nlm} \left(\bI,t\right) &= -\frac{2\pi G\Sigma_\rmP}{k_R}\left(\sum_{m_\phi=0,2,-2}\delta_{m,m_\phi}\right) \frac{\sgn(m)\exp{\left[i\,{\rm sgn}(m) k_R \Rc(I_\phi)\right]}}{2i}\nonumber \\
&\times \exp{\left[i\, l \tan^{-1}{\frac{2 m \Omega_\phi}{\Rc \kappa}\sqrt{\frac{2 I_R}{\kappa}}}\right]} \, J_l\left(\sqrt{k_R^2+{\left(\frac{2 m \Omega_\phi}{\Rc \kappa}\right)}^2} \sqrt{\frac{2 I_R}{\kappa}}\right) \nonumber \\
&\times\left[\alpha\,\calM_\rmo(t)\Phi_n^{(\rmo)}(I_z)+\calM_\rme(t)\Phi_n^{(\rme)}(I_z)\right] \exp{\left[-i m \Omega_\rmP t\right]},
\label{spiral_fourier_app}
\end{align}
where $J_l$ is the $l^{\rm th}$ order Bessel function of the first kind,
\begin{align}
\sgn(m) &=
\begin{cases}
1, & m\geq 0, \\
-1, & m<0,
\end{cases}
\end{align}
and $\Phi_n^{(\rmo)}(I_z)$ and $\Phi_n^{(\rme)}(I_z)$ are given by
\begin{align}
\Phi_n^{(\rmo)}(I_z) &= \frac{1}{2\pi} \int_0^{2\pi} \rmd w_z \sin{n w_z}\, \calF_\rmo\left(z,k_z^{(\rmo)}\right),\nonumber \\
\Phi_n^{(\rme)}(I_z) &= \frac{1}{2\pi} \int_0^{2\pi} \rmd w_z \cos{n w_z}\, \calF_\rme\left(z,k_z^{(\rme)}\right).
\label{wz_int_spiral_app}
\end{align}
In deriving equation~(\ref{spiral_fourier_app}) we have used the Hansen-Bessel formula which provides the following integral representation for Bessel functions of the first kind,
\begin{align}
\int_0^{2\pi} \rmd x \, \exp{\left[-i l x\right]} \, \exp{\left[i \alpha \sin{x}\right]} = 2\pi J_l\left(\alpha\right),
\label{Bessel_int_id}
\end{align}
and the identity for expansion in products of Bessel functions given in equation~(8.530.2) of \citet[][see also section~6.1 of \cite{Binney.Lacey.88}]{Gradshteyn.Ryzhik.65}. We have also used the identity,
\begin{align}
\int_0^{2\pi}\rmd \phi\, \exp{\left[-i m\phi\right]} = 2\pi\,\delta_{m,0}.
\end{align}

\section{Perturbation by encounter with satellite galaxy}

\subsection{Computation of the disk response}
\label{App:sat_disk_Resp}

To evaluate the disk response to satellite encounters using equation~(\ref{f1nk_gensol_f0}) we first evaluate the $\tau$ integral (with $t_\rmi\to -\infty$) of the satellite potential given in equation~(\ref{sat_pot}) and then compute the Fourier transform of the result. This yields the expression for the response in equation~(\ref{f1nk_gensol_f0}) with
\begin{align}
&\calI_{nlm}(\bI,t)=\exp{\left[-i\Omega t\right]}\int_{-\infty}^{t}\rmd \tau\, \exp{\left[i\Omega \tau\right]}\, \Phi_{nlm}\left(\bI,\tau\right)\nonumber \\
&=\frac{\exp{\left[-i\Omega t\right]}}{{\left(2\pi\right)}^3}\int_0^{2\pi}\rmd w_z \exp{\left[-inw_z\right]}\int_0^{2\pi}\rmd w_R \exp{\left[-ilw_R\right]}\int_0^{2\pi}\rmd w_\phi \exp{\left[-imw_\phi\right]}\int_{-\infty}^{t}\rmd \tau\, \exp{\left[i\Omega \tau\right]}\, \Phi_\rmP(z,R,\phi,\tau),
\label{Inlm_FT_app}
\end{align}
where
\begin{align}
\Omega = n\Omega_z + l\Omega_R + m\Omega_\phi.
\end{align}
We perform the inner $\tau$ integral of $\Phi_\rmP$ to obtain
\begin{align}
\int_{-\infty}^{t}\rmd \tau\, \exp{\left[i\Omega \tau\right]}\, \Phi_\rmP(z,R,\phi,\tau) &= -\frac{G M_\rmP}{\vp}\exp{\left[i\frac{\Omega\calS}{\vp}\right]} \int_{-\infty}^{t-\calS/\vp}\rmd\tau \frac{\exp{\left[i\Omega \tau\right]}}{\sqrt{\tau^2+{\left(\calR^2+\varepsilon^2\right)}/{v^2_\rmP}}}\nonumber \\
&=-\frac{G M_\rmP}{\vp}\exp{\left[i\frac{\Omega\calS}{\vp}\right]} \int_{-\infty}^{\left(\vp t-\calS\right)/\sqrt{\calR^2+\varepsilon^2}}\rmd x \frac{\exp{\left[i\left(\Omega\sqrt{\calR^2+\varepsilon^2}/\vp\right) x\right]}}{\sqrt{x^2+1}} \nonumber \\
&=-\frac{2\, G M_\rmP}{\vp}\exp{\left[i\frac{\Omega\calS}{\vp}\right]} K_{0i}\left(\frac{\Omega\sqrt{\calR^2+\varepsilon^2}}{\vp},\frac{\vp t-\calS}{\sqrt{\calR^2+\varepsilon^2}}\right).
\label{K0i_app}
\end{align}
Here $K_{0i}$ is defined as
\begin{align}
K_{0i}(\alpha,\beta) = \frac{1}{2}\int_{-\infty}^\beta\rmd x\, \frac{\exp{\left[i\alpha x\right]}}{\sqrt{x^2+1}},
\end{align}
which asymptotes to the zero-th order modified Bessel function of the second kind, $K_0\left(\left|\alpha\right|\right)$, in the limit $\beta \to \infty$. $\calR$ and $\calS$ are respectively the projections perpendicular and parallel to the direction of ${\bf \vp}$ of the vector connecting the point of observation, $(z,R,\phi)$, with the point of impact, and are given by
\begin{align}
\calR^2&= {\left[R\sin{(\phi-\phip)}+\rd\sin{\phip}\right]}^2 + {\left[(R\cos{(\phi-\phip)}-\rd\cos{\phip})\cos{\thetap}-z\sin{\thetap}\right]}^2\nonumber \\
\calS &= (R\cos{(\phi-\phip)}-\rd\cos{\phip})\sin{\thetap}+z\cos{\thetap}.
\label{sat_RS_app}
\end{align}
In deriving equation~(\ref{K0i_app}), we have only considered the direct term in the expression for $\Phi_\rmP$ given in equation~(\ref{sat_pot}); the indirect term turns out to be sub-dominant.

In the large time limit, i.e., $t\gg \calS/\vp$, $K_{0i}$ asymptotes to $K_0\left(\left|\Omega\right|\sqrt{\calR^2+\varepsilon^2}/\vp\right)$. We substitute the expressions for $R$ and $z$ in terms of $(\bw,\bI)$ given in equations~(\ref{R_epi_app}) and (\ref{z_wz_Iz_app}) in the above expressions for $\calR$ and $\calS$. Further substituting the resultant $\tau$ integral from equation~(\ref{K0i_app}) in equation~(\ref{Inlm_FT_app}), substituting $w_\phi$ in terms of $\phi$ using equation~(\ref{wphi_phi_app}), adopting the small $I_R$ limit, and performing the $w_R$ integral, we obtain 
\begin{align}
&\calI_{nlm}(\bI,t)\approx-\frac{2G M_\rmP}{\vp} \exp{\left[-i\Omega t\right]} \times \exp{\left[-i\frac{\Omega\sin{\thetap}\cos{\phip}}{\vp} \rd\right]} \times \exp{\left[i\, l \tan^{-1}{\frac{2 m \Omega_\phi}{\Rc \kappa}\sqrt{\frac{2 I_R}{\kappa}}}\right]} \nonumber \\ 
&\times \frac{1}{{\left(2\pi\right)}^2} \int_0^{2\pi} d w_z \exp{\left[-in w_z\right]} \exp{\left[i\frac{\Omega \cos{\thetap}}{\vp} z\right]} \int_0^{2\pi} d \phi\,\exp{\left[-im\phi\right]}\, \exp{\left[i\frac{\Omega \sin{\thetap} \cos{\left(\phi-\phip\right)}}{\vp} \Rc\right]} \nonumber \\
&\times J_l\left(\sqrt{{\left(\frac{\Omega \sin{\thetap}}{\vp}\right)}^2\cos^2{\left(\phi-\phip\right)} + {\left(\frac{2 m \Omega_\phi}{\Rc \kappa}\right)}^2} \sqrt{\frac{2 I_R}{\kappa}}\right) K_{0i}\left(\frac{\Omega\sqrt{\calR^2_\rmc+\varepsilon^2}}{\vp},\frac{\vp t-\calS_\rmc}{\sqrt{\calR^2_\rmc+\varepsilon^2}}\right),
\label{sat_gen_app}
\end{align}
where $\calR_\rmc=\calR(R=\Rc)$ and $\calS_\rmc=\calS(R=\Rc)$. Here we have used the integral representation of Bessel functions of the first kind given in equation~(\ref{Bessel_int_id}) and the identity given in equation~(8.530.2) of \cite{Gradshteyn.Ryzhik.65}.

The expression for $\calI_{nlm}$ given in equation~(\ref{sat_gen_app}) consists of the leading order expansion in $\sqrt{2 I_R/\kappa}$. A more precise expression that is accurate up to second order in $\sqrt{2 I_R/\kappa}$ is given, in the large time limit, as
\begin{align}
&\calI_{nlm}(\bI,t)\overset{\mathrm{t\to\infty}}{\approx}-\frac{2G M_\rmP}{\vp} \exp{\left[-i\Omega t\right]} \times \exp{\left[-i\frac{\Omega\sin{\thetap}\cos{\phip}}{\vp} \rd\right]} \nonumber \\ 
&\times \frac{1}{{\left(2\pi\right)}^2} \int_0^{2\pi} d w_z \exp{\left[-in w_z\right]} \exp{\left[i\frac{\Omega \cos{\thetap}}{\vp} z\right]} \int_0^{2\pi} d \phi\,\exp{\left[-im\phi\right]}\, \exp{\left[i\frac{\Omega \sin{\thetap} \cos{\left(\phi-\phip\right)}}{\vp} \Rc\right]} \nonumber \\
&\times \exp{\left[i\, l \tan^{-1}{\frac{2 m \Omega_\phi}{\Rc \kappa}\sqrt{\frac{2 I_R}{\kappa}}}\right]} \, \left[\zeta^{(0)} J_l\left(\chi\right) - i\zeta^{(1)} J'_l\left(\chi\right) -\frac{1}{2}\zeta^{(2)} J''_l\left(\chi\right) \right],
\label{sat_gen_IR_app}
\end{align}
where
\begin{align}
\chi = \sqrt{{\left(\frac{\Omega \sin{\thetap}}{\vp}\right)}^2\cos^2{\left(\phi-\phip\right)} + {\left(\frac{2 m \Omega_\phi}{\Rc \kappa}\right)}^2} \sqrt{\frac{2 I_R}{\kappa}},
\end{align}
and
\begin{align}
\zeta^{(0)} &= K_0\left(\eta\right), \nonumber \\
\zeta^{(1)} &= \sqrt{\frac{2 I_R}{\kappa}}\, \frac{\partial \calR_\rmc}{\partial \Rc} \frac{\calR_\rmc}{\sqrt{\calR^2_\rmc+\varepsilon^2}} \frac{\left|\Omega\right|}{\vp} K'_0\left(\eta\right), \nonumber \\
\zeta^{(2)} &= \frac{2 I_R}{\kappa}\, \left[{\left(\frac{\partial \calR_\rmc}{\partial \Rc}\right)}^2\frac{\calR^2_\rmc}{\calR^2_\rmc+\varepsilon^2}\frac{\Omega^2}{v^2_\rmP}K''_0(\eta) + \left\{\frac{\partial^2 \calR_\rmc}{\partial R^2_c}\frac{\calR_\rmc}{\sqrt{\calR^2_\rmc+\varepsilon^2}}+{\left(\frac{\partial \calR_\rmc}{\partial \Rc}\right)}^2\frac{\varepsilon^2}{{\left(\calR^2_\rmc+\varepsilon^2\right)}^{3/2}}\right\} \frac{\left|\Omega\right|}{\vp}K'_0(\eta)\right],
\end{align}
with
\begin{align}
\eta &= \frac{\left|\Omega\right|\sqrt{\calR^2_\rmc+\varepsilon^2}}{\vp}.
\label{eta_app}
\end{align}
Here each prime denotes a single derivative of the function with respect to its argument.

Substituting the expression for $\calI_{nlm}$ given in equation~(\ref{sat_gen_IR_app}) in the expression for the disk response given in equation~(\ref{f1nk_gensol_f0}), and adopting the fiducial parameters for the MW galaxy and those corresponding to satellite encounters (as detailed in section~\ref{sec:MW_disk_resp_sat}), we compute the response of the MW disk to past and future encounters with its satellite galaxies. Results for the steady state disk response (in the collisionless limit) of the $(n,l,m)=(1,0,0)$ mode, corresponding to $I_z=I_{z,\odot}=h_z\sigma_{z,\odot}$ and $R_c(L_z)=8\kpc$ and marginalized over $I_R$, are summarised in Table~\ref{tab:MW_sat_resp}.

\begin{table*}
\centering
\hspace{-3cm}
\tabcolsep=0.2 cm
%\begin{tabular}{lllllll}
\begin{tabular}{c|c|cc|cc|cc}
 \hline
MW satellite & Mass & $f_{1,n=1}/f_0$ & $t_{\rm cross}$ & $f_{1,n=1}/f_0$ & $t_{\rm cross}$ & $f_{1,n=1}/f_0$ & $t_{\rm cross}$ \\
name & $(\Msun)$ & & $(\Gyr)$ & & $(\Gyr)$ & & $(\Gyr)$ \\
 & & Penultimate & Penultimate & Last & Last & Next & Next \\
 (1) & (2) & (3) & (4) & (5) & (6) & (7) & (8) \\
 \hline
 Sagittarius & $10^9$ & $2.7\times 10^{-1}$ & $-1.01$ & $4.9\times 10^{-8}$ & $-0.35$ & $1.3\times 10^{-1}$ & $0.03$ \\
 Hercules & $7.1\times 10^6$ & $8.4\times 10^{-8}$ & $-3.78$ & $2.4\times 10^{-3}$ & $-0.5$ & $2.4\times 10^{-3}$ & $3.18$ \\
 Leo II & $8.2\times 10^6$ & -- & $-3.86$ & $1.6\times 10^{-3}$ & $-1.78$ & $3.2\times 10^{-3}$ & $2.31$ \\
 Segue 2 & $5.5\times 10^5$ & $6.2\times 10^{-4}$ & $-0.84$ & $8.5\times 10^{-4}$ & $-0.25$ & $6.3\times 10^{-5}$ & $0.28$ \\
 LMC & $1.4\times 10^{11}$ & $5.1\times 10^{-2}$ & $-7.63$ & -- & $-2.67$ & $2.3\times 10^{-2}$ & $0.11$ \\
 SMC & $6.5\times 10^9$ & $2.8\times 10^{-5}$ & $-3.32$ & -- & $-1.44$ & $8.7\times 10^{-6}$ & $0.22$ \\
 Draco I & $2.2\times 10^7$ & -- & $-2.46$ & $9.9\times 10^{-5}$ & $-1.24$ & $7.1\times 10^{-6}$ & $0.24$ \\
 Bootes I & $10^7$ & $1.8\times 10^{-7}$ & $-1.67$ & $3.7\times 10^{-5}$ & $-0.35$ & -- & $0.88$ \\
 Willman I & $4\times 10^5$ & $1.6\times 10^{-8}$ & $-0.66$ & $1.2\times 10^{-6}$ & $-0.21$ & $9.3\times 10^{-6}$ & $0.41$ \\
 Ursa Minor & $2\times 10^7$ & -- & $-2.28$ & $1.7\times 10^{-5}$ & $-1.17$ & $2.6\times 10^{-6}$ & $0.29$ \\
 Ursa Major II & $4.9\times 10^6$ & $5.8\times 10^{-6}$ & $-2.12$ & $2.5\times 10^{-6}$ & $-0.09$ & -- & $0.97$ \\
 Coma Berenices I & $1.2\times 10^6$ & $9.2\times 10^{-7}$ & $-2.58$ & $3.7\times 10^{-8}$ & $-0.25$ & -- & $0.71$ \\
 Sculptor & $3.1\times 10^7$ & -- & $-2.74$ & $3.4\times 10^{-8}$ & $-0.46$ & -- & $1.48$\\
 \hline
\end{tabular}
\caption{Steady state response of the MW disk to encounters with satellites in the collisionless limit, for the $(n,l,m)=(1,0,0)$ mode and for stars with $I_z = I_{z,\odot} = h_z\sigma_{z,\odot}$ in the Solar neighborhood. We have marginalized the response over $I_R$. Columns (1) and (2) list the name and dynamical mass of each satellite. The latter is taken from the literature \citep[][]{Simon.Geha.07,Bekki.Stanimirovic.09,Lokas.09,Erkal.etal.19,Vasiliev.Belokurov.20}, except for Sagittarius for which we adopt a mass of $10^9\Msun$. Note that there is a discrepancy between its estimated mass of $\sim 4\times 10^8\Msun$ \citep[][]{Vasiliev.Belokurov.20} and the mass required ($10^9-10^{10}\Msun$) to produce detectable phase-spiral signatures in N-body simulations \citep[see for example][]{Bennett.etal.22}. Columns (3) and (4) respectively denote the bending mode response assuming our fiducial MW parameters and the penultimate disk-crossing time. Columns (5) and (6) indicate the same for the last disk-crossing, while columns (7) and (8) show it for the next one. Only satellites that induce a bending mode response, $f_{1,n=1}/f_0\geq 10^{-8}$, in at least one of the three cases are shown. Any response weaker than $10^{-8}$ is considered negligible and is indicated with a horizontal dash.}
\label{tab:MW_sat_resp}
\end{table*}

\subsection{Special case: disk response for face-on impulsive encounters}
\label{App:special}

The disk response in the general case, expressed by equation~(\ref{sat_gen}), depends on several encounter parameters: $\rd$, $\thetap$, $\phip$, and is complicated to evaluate. Therefore, as a sanity check, here we compute the response as well as corresponding energy change for the special case of a satellite undergoing an impulsive, perpendicular passage through the center of the disk. 

As shown in \cite{vdBosch.etal.18a} \citep[see also][]{Banik.vdBosch.21b}, the total energy change due to a head-on encounter of velocity $v_\rmP$ with a Plummer sphere of mass $M_\rmP$ and size $\varepsilon$ is given by:
\begin{align}
 \Delta E = 4 \pi \left({G M_\rmp \over v_\rmP}\right)^2 \int_0^{\infty} I^2_0(R) \Sigma(R) {\rmd R \over R}
\end{align}
where
\begin{align}
 I_0(R) = \int_1^{\infty} {M_\rmP(\zeta R) \over M_\rmp} \, {\rmd \zeta \over \zeta^2 (\zeta^2 - 1)^{1/2}}
\end{align}
Using that the enclosed mass profile of a Plummer sphere is given by $M_\rmP(R) = M_\rmP R^3 (R^2 + \varepsilon^2)^{-3/2}$, we have that $I_0(R)=R^2/(R^2+\varepsilon^2)$, which yields
\begin{align}
 \Delta E = 4 \pi \left({G M_\rmp \over v_\rmP}\right)^2 \int_0^{\infty} \Sigma(R) {R^3 \rmd R \over (R^2 + \varepsilon^2)^2}.
\label{deltaE_impulse}
\end{align}

Now we compute the disk response to the face-on satellite encounter using equations~(\ref{f1nk_gensol_f0}) and (\ref{sat_gen_IR_app}-\ref{eta_app}). For a perpendicular face-on impact through the center of the disk we have $\rd=0$ and $\thetap=0$, implying that $\calR_\rmc$ becomes $\Rc$. The corresponding response is greatly simplified. In the large time and small $I_R$ limit, it is given by equation~(\ref{f1nk_gensol_f0}) with

\begin{align}
\calI_{nlm}(\bI,t) &\approx -\frac{2GM_\rmP}{\vp} \exp{\left[-i\Omega t\right]}\, \delta_{m,0}\times \frac{1}{2\pi}\int_0^{2\pi} d w_z \exp{\left[-in w_z\right]} \exp{\left[i\frac{\Omega z}{\vp}\right]} \nonumber \\
&\times \frac{1}{2\pi} \int_0^{2\pi} \rmd w_R\, \exp{\left[-ilw_R\right]}\, K_0\left[\frac{\left|\Omega\right|}{\vp}\sqrt{\varepsilon^2+{\left(\Rc+\sqrt{\frac{2I_R}{\kappa}}\sin{w_R}\right)}^2}\right],
\end{align}
where the $\phi$ integral only leaves contribution from the axisymmetric $m=0$ mode. The $w_R$ integrand can be expanded as a Taylor series and the $w_R$ integral can be performed to yield the following leading order expression for $\calI_{nlm}$:
\begin{align}
\calI_{nlm}(\bI,t) &\approx i\frac{GM_\rmP}{\vp} \exp{\left[-i\Omega t\right]}\, \delta_{m,0}\,\left(\delta_{l,1}-\delta_{l,-1}\right) \times \frac{1}{2\pi}\int_0^{2\pi} d w_z \exp{\left[-in w_z\right]} \exp{\left[i\frac{\Omega z}{\vp}\right]} \nonumber \\
& \times \sqrt{\frac{2 I_R}{\kappa}} \frac{\Rc}{\sqrt{\varepsilon^2+R^2_c}}\, \frac{\left|\Omega\right|}{\vp} K'_0\left[\frac{\left|\Omega\right|}{\vp}\sqrt{\varepsilon^2+R^2_c}\right].
\end{align}
In the impulsive limit, $\vp\to \infty$, this becomes
\begin{align}
\calI_{nlm}(\bI,t) &\approx i\,\delta_{n,0} \delta_{m,0} \left(\delta_{l,1}-\delta_{l,-1}\right) \frac{GM_\rmP}{\vp} \sqrt{\frac{2 I_R}{\kappa}} \frac{\Rc}{\varepsilon^2+R^2_c} \exp{\left[-i\, l\kappa\, t\right]},
\end{align}
which can be substituted in equation~(\ref{f1nk_gensol_f0}) to yield
\begin{align}
&f_{1nlm}\left(\bI,t\right) = f_0(\bI) \times \delta_{n,0} \delta_{m,0} \left(\delta_{l,1}-\delta_{l,-1}\right) \frac{G M_\rmP}{\vp} \frac{l\kappa}{\sigma^2_R} \sqrt{\frac{2 I_R}{\kappa}} \frac{\Rc}{\varepsilon^2+R^2_c} \exp{\left[-i\, l\kappa\, t\right]},
\label{f1nlm_sat_app}
\end{align}
with $f_0$ given by equation~(\ref{DF_MW}). Hence, the response is given by
\begin{align}
f_1\left(\bw,\bI,t\right)&=\sum_{n=-\infty}^{\infty} \sum_{l=-\infty}^{\infty} \sum_{m=-\infty}^{\infty} \exp{\left[i (n w_z + l w_R + m w_\phi)\right]}\, f_{1nlm}(\bI,t) \nonumber \\ &= f_0(\bI) \times \frac{2 G M_\rmP}{\vp} \frac{\sqrt{2 \kappa I_R}}{\sigma^2_R} \frac{\Rc}{\varepsilon^2+R^2_c} \cos{\left(w_R-\kappa t\right)}\,,
\label{f1_sat_spl}
\end{align}
which shows that the satellite passage introduces a relative overdensity, $f_1\left(\bw,\bI,t\right)/f_0(\bI)$, that scales as $\sim \Rc/\left(\varepsilon^2+R^2_c\right)$, which increases from zero at the center, peaks at $\Rc=\varepsilon$, and asymptotes to zero again at large $\Rc$. The $\cos(w_R - \kappa t)$-term describes the radial epicyclic oscillations in the response.

To compute the energy change due to the impact, we note that $\rmd E/\rmd t = \partial E/\partial \bI \cdot \rmd \bI/\rmd t$, where $\partial E/\partial \bI = {\bf\Omega}=(\Omega_z,\Omega_R,\Omega_\phi)$ and $\rmd \bI/\rmd t = -\partial \Phi_\rmP/\partial \bw$ from Hamilton's equations of motion. Thus the total phase-averaged energy injected per unit phase-space can be obtained as follows:
\begin{align}
\left<\Delta E \left(\bI\right)\right> &= \frac{1}{{\left(2\pi\right)}^3} \int \rmd \bw \int_{-\infty}^{\infty} \rmd t\, \frac{\rmd E}{\rmd t} f_1(\bI,t) = -\frac{1}{{\left(2\pi\right)}^3} \int \rmd \bw \int_{-\infty}^{\infty} \rmd t\; {\bf\Omega} \cdot \frac{\partial \Phi_\rmP}{\partial \bw} f_1(\bI,t).
\end{align}
We can substitute the Fourier series expansions of $\Phi_\rmP$ and $f_1$ given in equations~(\ref{fourier_series_gen}) in the above expression and integrate over $\bw$ to obtain \citep[][]{Weinberg.94a,Weinberg.94b}
\begin{align}
\left<\Delta E \left(\bI\right)\right> &= i\sum_{nlm} \left(n\Omega_z+l\kappa+m\Omega_\phi\right) \int_{-\infty}^{\infty} \rmd t\, \Phi^*_{nlm}(\bI,t) f_{1nlm}(\bI,t).
\label{delE_sat_app}
\end{align}
We can now substitute the form of $\Phi_\rmP$ for a Plummer perturber given in equation~(\ref{sat_pot}), with $\brp$ and $\br$ given by equations~(\ref{rp_sat}) and (\ref{r_sat}). The time integral can thus be written as
\begin{align}
\int_{-\infty}^{\infty} \rmd t\, \Phi^*_{nlm}(\bI,t) f_{1nlm}(\bI,t) &= - \frac{1}{{\left(2\pi\right)}^3} \int_0^{2\pi}\rmd w_z \exp{\left[inw_z\right]}\int_0^{2\pi}\rmd w_R \exp{\left[ilw_R\right]}\int_0^{2\pi}\rmd w_\phi \exp{\left[imw_\phi\right]} \nonumber \\
&\times \int_{-\infty}^{\infty} \rmd t \frac{G M_\rmP}{\sqrt{{\left(\vp t - z\right)}^2+R^2+\varepsilon^2}} f_{1nlm}(\bI,t).
\end{align}
Using equations~(\ref{R_epi_app}) and (\ref{z_wz_Iz_app}) to express $R$ and $z$ in terms of $(\bw,\bI)$, and substituting the form for $f_{1nlm}(\bI,t)$ from equation~(\ref{f1nlm_sat_app}), we can perform the above integrals over $\bw$ and $t$. Substituting the result in equation~(\ref{delE_sat_app}) we obtain
\begin{align}
\left<\Delta E \left(\bI\right)\right> &= {\left(\frac{G M_\rmP}{\vp}\right)}^2 f_0(\bI)\, \frac{2\kappa I_R}{\sigma^2_R}\, \frac{R^2_c}{{\left(\varepsilon^2+R^2_c\right)}^2}.
\end{align}

The total energy, $\Delta E_{\rm tot}$, imparted into the disk by the impulsive satellite passage can be computed by integrating the above expression over $\bI$ and $\bw$ (which simply introduces a factor of ${\left(2\pi\right)}^3$ since $\left<\Delta E \left(\bI\right)\right>$ is already phase-averaged), using equation~(\ref{DF_MW}) and transforming from $L_z$ to $\Rc$ using the Jacobian $\rmd L_z/\rmd \Rc = \Rc \kappa^2/2 \Omega_\phi$. This yields
\begin{align}
\Delta E_{\rm tot} = 4\pi{\left(\frac{GM_\rmP}{\vp}\right)}^2 \int_0^{\infty}\rmd \Rc\, \Rc\, \Sigma(\Rc) \frac{\Rc^2}{{\left(\varepsilon^2+\Rc^2\right)}^2}.
\end{align}
This is indeed the expression for $\Delta E_{\rm tot}$ derived under the impulse approximation given by equation~(\ref{deltaE_impulse}).

%%
%\begin{figure}
%\centering
%\hspace{-1mm}
%\includegraphics[width=1\textwidth]{f1max_vs_t_gaussian_spiral.jpeg}
%\caption{}
%\label{fig:f1max}
%\end{figure}
%

%%%%%%%%%%%%%%%%%%%%%%%%%%%%%%%%%%%%%%%%%%%%%%%%%%
% Don't change these lines
%\bsp	% typesetting comment
\label{lastpage}



\end{document}